\documentclass[final,12pt]{elsarticle}
\usepackage{amssymb}
\usepackage{multimedia}
\usepackage{amsmath,amsthm}
\usepackage{amssymb}
\usepackage{epstopdf}
\usepackage{times}
\usepackage{mathrsfs}
\usepackage{color}
\usepackage{gensymb}
\usepackage{appendix}
\usepackage{subfigure}

\journal{International Journal of Engineering Science}

\begin{document}

\begin{frontmatter}
\title{On the critical nature of plastic flow: one and two dimensional models}
\author{ O. U. Salman$^{\rm a,}$$^{\rm b,}$\thanks{$^\ast$Corresponding author. Email: umut.salman@polytechnique.edu} and L. Truskinovsky$^{\rm a,}$$^{\rm b}$ }
\address{$^{\rm a}${\em{LMS,  CNRS-UMR  7649,
Ecole Polytechnique, Route de Saclay, 91128 Palaiseau,  France}}\\
$^{\rm b}${\em{SEAS, Harvard University, 29 Oxford Street,
Cambridge, MA 02138, USA}}}
\date{\today}
\begin{abstract}

Steady state plastic flows have been compared to developed turbulence because the two phenomena share the inherent complexity of particle trajectories, the scale free spatial patterns and the power law statistics of fluctuations. The origin of the apparently chaotic and at the same time highly correlated microscopic response in plasticity remains hidden behind conventional engineering models which are based on smooth fitting functions. To regain access to fluctuations, we study in this paper a minimal mesoscopic model whose goal is to elucidate the origin of scale free behavior in plasticity.  We limit our description to fcc type crystals and leave out both temperature and rate effects. We provide simple illustrations of the fact that complexity in rate independent athermal plastic flows is due to marginal stability of the underlying elastic system. Our conclusions are based on a reduction of an over-damped visco-elasticity problem for a system with a rugged elastic energy landscape to an integer valued automaton. We start with an overdamped one dimensional model and show that it reproduces the main macroscopic phenomenology of rate independent plastic behavior but falls short of generating self similar structure of fluctuations.  We then provide evidence that a two dimensional model is already adequate for describing power law statistics of avalanches and fractal character of dislocation patterning. In addition to capturing experimentally measured critical exponents, the proposed  minimal model shows finite size scaling collapse and generates realistic shape functions in the scaling laws.

\end{abstract}
\end{frontmatter}

\section{Introduction}

It is known that crystalline and amorphous solids flow plastically when macroscopic stresses exceed certain thresholds. In plasticity, thermal relaxation may often be neglected while the driving can still be viewed as quasi-static. In this case the flow is an athermal process with rate independent dissipation. The  singularity of this dissipative mechanism distinguishes plasticity from the regular dissipative phenomena such as viscosity or heat conductivity,  and places it instead in the  class of dry friction and fracture \cite{Cottrell:1953fk,NADGORNY:1988dq, Lubliner:1990cr,Kachanov:2004nx,Falk:2011fk}.

At the macroscale rate independent plasticity can be perceived as a smooth process, whereas
at  microscale the plastic flow is known to exhibit fluctuations revealing isolated dissipative events \cite{Brechet.1994,Puglisi:2005fk}. In crystals these fluctuations can be linked to nucleation and collective depinning of dislocations interacting
through long range elastic stress fields \cite{Nabarro:1967uq,Hirth:1982kx}. In amorphous and random granular systems the nature of the 'quantum' of plastic  strain  is still debated \cite{Argon:1979kx,PhysRevE.74.016118, Liu:2010uq,Barrat:2011vn, PhysRevE.84.016115} and therefore in what follows we mostly limit our discussion to crystals.

The smooth description of crystal plasticity adopted in engineering theories presumes spatial homogenization and time averaging. In some cases (bcc metals, tetrahedral covalent crystals, etc.) the obstacles are strong, the dislocation interaction is weak and the plastic flow can be viewed as a sum of uncorrelated events. The collective effects can then be neglected making classical homogenization an appropriate tool. Here we are interested in an alternative case (fcc metals, hcp crystals with basal glide, etc.) when hardening can be neglected, dislocational mobility	
 is high and the  elastic interaction among distant dislocations is significant. Then a specific collective behavior at the macroscale emerges as a manifestation of many correlated
 events at the microscale \cite{Zaiser:2008sb,Seeger:1957fk,Kubin:2002ly,anantha_rep}.  In particular,  such  plastic flows exhibit in the steady state irregular isolated bursts and reveal apparently random localized active slip volumes, with both spatial and temporal fluctuations spanning all scales. The temporal intermittency manifests itself through acoustic emission with power law statistics of avalanches \cite{Weiss:1997vn,Weiss:2000kx,zapperi_exp,Richeton2006190} while spatial fluctuations   take  the form
  of fractal dislocational cell structures \cite{PhysRevLett.81.2470, Zaiser1999299,doi:10.1080/01418618608245293}.

The ensuing macroscopic dynamics is usually characterized as critical because of the reasons that will become more clear in what follows. In physical terms the resulting behavior can be described as glassy.  Since dislocational motion proceeds in a self similar manner in both space and time the importance of such notions as average dislocation density and average velocity in such regimes become questionable. Those notions, however, remain relevant in non-critical plastic regimes where dislocational microstructures are regular and are characterized by particular scales which depend on the level of loading, e.g. \cite{kubin:1993}.

Despite the over-schematic description of dislocational activity in classical engineering theory, it has been remarkably successful in capturing the most important plasticity phenomenology such as yield stress, work hardening and shakedown \cite{Hill:1998vn,Rice:1975ys}. However, the fractal patterning and the peculiar scale free structure of the temporal fluctuations remain invisible at the macroscale and are missed by the approaches based on  'intuitive' spatial and temporal homogenization. As a result, a quantitative link between the fitting functions used in phenomenological plasticity and the microscopic picture of a crystalline lattice with moving line defects remains elusive. This gap has been a major obstacle on the way of quantitative estimate of plastic response for artificially designed materials.

The emergence of power laws  suggests that the relation between  microscopic and macroscopic pictures of plastic flow is rather complex and more akin to turbulence \cite{Cottrell:2002zr,Kubin:2002ly, Choi:2011ys} than to elasticity where classical homogenization of lattice models is usually sufficient to predict macroscopic response \cite{Born:1998oq,Blanc2006627}. It is also clear that powerful methods of equilibrium statistical mechanics implying ergodicity  and absence of correlations are not applicable for the description of small scale plasticity which appear to be highly nonequilibrium and strongly correlated phenomenon. The main problem with applying the classical continuum approach is that spatial renormalization symmetry (self-similarity)  excludes separation of scales while non gaussian statistics prevents local description. More generally, we still lack analytic tools allowing quantitative description of evolutionary phenomena that depend on rare events spanning the whole domain and have convergence problems handling the uncertainty which grows algebraically with time which qualifies such systems as located at the 'edge of chaos' \cite{Bak:1991}.

The fine structure of long range correlations in plasticity has been extensively studied in recent experiments which unambiguously established the power law statistics  of avalanches and produced numerous examples of the scale-free nature of dislocation patterns \cite{Weiss:1997vn,Weiss:2000kx,zapperi_exp,Richeton2006190, Dimiduk:2006ys,RichetonBreakdown,PhysRevLett.87.165508,Bharathi:2002C,Zaiser:2008sb,Richeton20054463,PhysRevB.76.224110,
CreepNG,PhysRevLett.100.155502}.   Similar phenomena of self organization towards criticality have  been also observed in other mechanical systems operating in nonequilibrium steady regimes including friction, fracture, porous and granular flows and martensitic transformations \cite{0034-4885-62-10-201,Sornette:2000ve,Jensen:1998qf,Vives_1994,Vives_1998}.

While the general mechanisms for the formation of scale free correlations remain a major subject of research \cite{Henkel:2008kl}, several important ingredients of dynamic criticality, such as marginal stability, quasi-static driving, threshold type nonlinearity and nonlocal feedback, have been identified \cite{Papanikolaou:2011qf,ISI:000086325400004}. All these components are present in crystal plasticity. For instance, marginal stability reveals itself through the fact that
 infinitesimal perturbations not only trigger small scale rearrangements of the ensemble of  locked dislocations but can also induce a global transition between the jammed and the flowing regimes. The long range interactions responsible for the feedback have elastic nature, suggesting that, somewhat paradoxically, elasticity plays a fundamental role even in 'ideal' plasticity.

 The experimental findings of plastic criticality in crystals have been supported
by a number of numerical simulations reviewed in \cite{ISI:000237065700003,springerlink:10.1023/A:1023293117309}.
The two main approaches are: the discrete dislocation dynamics (DDD) accounting for dislocation interaction on
different slip planes   \cite{Miguel:2001dk,PhysRevLett.89.165501,Miguel:2001dk,Csikor12102007, Beato:2011dq}  and the
pinning-depinning  models considering dislocations dynamics on a single slip plane
  \cite{PhysRevB.69.214103, PhysRevLett.93.265503,PhysRevLett.93.125502}. Discrete mesoscopic
 models of crystal and amorphous plasticity implying some averaging and providing an effective description of the pinning-depinning  and jamming-unjamming processes  have also been shown to
generate power law statistics of avalanches with realistic exponents
\cite{MichaelRandomness,Zaisermoretti2005,  PhysRevLett.102.175501,PhysRevLett.89.195506,PhysRevLett.105.015502, PhysRevE.84.016115}.
 Among those we particular emphasize the models of earthquakes aimed at reproducing the Gutenber-Richter law because the rapture-healing phenomena on a preexisting fault are very similar to 2D plasticity showing almost the same critical exponents \cite{PhysRevA.43.625,Ben-Zion:1993kx,PhysRevLett.78.4885}. In view of the implied analogy with turbulence, of considerable  interest are also the  models dealing with dynamics of continuously distributed dislocations and capturing the scale free effects already in a PDE setting \cite{Amit2001761,Roy:2005fk, Limkumnerd20081450, PhysRevLett.96.095503,PhysRevB.79.014108,Choi:2011ys,PhysRevLett.105.105501}.

Despite their ability to reproduce experimental findings, the above models cannot provide theoretical insight into the origin of criticality because they lack analytical transparency and depend on large scale numerical experiments. On the other hand, the physical adequacy of the proposed numerical schemes may still be disputed in some basic details. For instance, in case of DDD models, the fast microscopic topological changes (nucleation, annihilation, interaction with obstacles, etc) are treated by auxiliary hypotheses  formulated phenomenologically in terms of local stresses without addressing the mechanism of barrier crossing \cite{PhysRevLett.93.016001, Tanguy:2006zr}. Likewise, the theory of continuously distributed dislocations have difficulty representing  kinks, locks, jogs and other individual entanglements with nontrivial topology. The numerical models can be made more adequate by involving various quasi-continuum schemes, however, it is clear that in order to understand plastic criticality mathematically, the existing models have to be  simplified  rather than further complicated.

The anticipation of a simple and transparent theory is based on the idea that truly scale free behavior is independent of either microscopic or macroscopic details of the system. The statistics of dislocation avalanches should then  be an intrinsic feature of a particular crystal class described in terms of symmetry and dimensionality and not affected by specific material parameters and details of the loading geometry. It is then of interest to formulate  simple  prototypical models representing particular universality classes of plastic flows.  The simplicity is understood here in the sense that the model is amenable to rigorous mathematical study while still capturing not only the observed exponents but also the characteristic shape functions in scaling relations \cite{Papanikolaou:2011qf, PhysRevE.84.061103}.

 At present, the only analytically tractable  approaches to criticality in driven distributed systems are the models based on branching processes  \cite{Sornette:2000ve}, the mean field model \cite{dharMajumdar1990}, the elastic depinning model amenable to FRG  methods \cite{PhysRevE.79.051106} and the Abelian sand pile type automata with integer valued fields \cite{Dhar19994,Redig:2005fk}. The challenge is  to reduce a realistic model of plasticity to one of these analytically transparent types.

 Since the relation between the prototypical branching processes and the dynamical phenomena in distributed systems remains rather imprecise, we will not pursue this line of modeling here.  In contrast, the observed critical exponents for crystal plasticity are so close to the mean field values that claims have been made that 3D plasticity is in the mean field universality class. This hypothesis is supported by the estimates of the critical dimension in various systems with long range elastic interactions \cite{Papanikolaou:2011qf,PhysRevB.58.6353,colaiori-2008-57,Dahmen:2011ly}. However, the question of validity of this approximation remains open because the match of the measured exponents is not perfect and the observed avalanche shapes appear to be less symmetric than the theoretical predictions \cite{Sethna:2001ud}. Also,  the mean field values of exponents have not been supported by numerical experiments with colloidal crystals \cite{PhysRevLett.106.175503} and within phase field crystal model \cite{ChanThesis}.

 The depinning models have a considerable relevance for plasticity because they study quasi-statically driven elastic objects interacting with a set of randomly distributed obstacles. The corresponding theoretical critical exponents for the case of dislocations have been computed with high precision \cite{Fisher:1998fk, Kardar:301ly, ISI:000237065700003, Pierre201049} {and confirmed by molecular dynamics studies \cite{PhysRevB.84.174101}. These models, however, may not be in the same universality class as crystal plasticity because the critical exponents are different from those observed in plasticity experiments. Most probably, the origin of divergence is the neglect in depinning models of important interactions among dislocations on different slip planes.

Left with the only remaining option, we study in this paper the possibility to reduce the conventional dislocation mediated plasticity model to an integer valued Abelian automaton.  The Abelian symmetry means that the order, in which the eligible 'spin variables' are updated during the avalanche, is irrelevant for the avalanche outcome. Some Abelian automata are amenable to rigorous analysis due to partial ergodicity on a set of recurrent sequences which were fully characterized in several cases \cite{Dhar19994,Redig:2005fk,Jarai:2005ve, Dhar:1999qf,PhysRevE.82.031121}. The aim of this paper is to show that a reduction of a multi-slip-plane dislocational dynamics to an integer automaton with some form of Abelian symmetry allows one to reproduce the experimentally observed statistics of avalanches. An analytical study of the ensuing automaton  will be reported elsewhere.

\section{Summary of the main results}

 In search for the simplest representation of plasticity universality class(es) we begin with a series of one dimensional models allowing one to reproduce rate independent dissipation and hysteresis. Then we move to two dimensional models capturing also intermittency and power law statistics.

  Our starting point is an assumption that the micro-scale dynamics is overdamped  and that the rate of loading is much slower than the rate of viscous relaxation but much faster than the rate of thermal relaxation. Such models belong to the class of AQS (athermal quasi-static) systems that have been found relevant for the description of many rate independent phenomena from wetting to magnetism \cite{Maeda:1978ve,Kobayashi19801641,doi:10.1080/01418618108236167, PhysRevLett.81.5576, PhysRevE.58.3515,Vives_2001}.

The simplest 1D lattice model of this type,  representing a chain of interacting particles (or a deck of interacting rigid cards),  can be developed for transformational plasticity  of shape memory alloys. To obtain in this setting a plastic response at the macro-level we must assume that the interaction potential has a double well structure which fully characterizes the local 'mechanism' of the  transformation. The presence  of local bi-stability, which is also relevant for the description of the 'easy glide' stage in crystal plasticity, places such mechanical system into the class of snap-spring lattices. The non-convexity of local interactions may be equally relevant for amorphous plasticity operating with a concept of STZ (shear transformation zones), which implies a multi-well structure of an effective potential \cite{Falk:2011fk, PhysRevE.74.016118}. We emphasize that the discreteness at this stage has to be understand broadly as describing a generic structural inhomogeneity, for instance, a grain structure.

In the simplest chain with nearest neighbor (NN) interactions the elastic elements are subjected to common stress field. The resulting mean field model  captures the basic mechanism of rate independent dissipation and can be used to illustrate the idea of marginal stability  of the underlying elastic system  which is ultimately responsible for the  plastic yield \cite{Puglisi:2005fk,A.:2011kx}.  A  more realistic model must incorporate short range interactions and  the simplest extension in this direction is a chain  with next to nearest neighbors (NNN). The latter may be of ferromagnetic or anti-ferromagnetic type \cite{Truskinovsky:2004qf, Braides:2008bh} and in both cases the resulting mechanical system can be viewed as a soft spin  generalization of the classical Ising model.

In the case of ferromagnetic interactions penalizing gradients we show  that  that the NNN model captures the important difference between the nucleation and  the propagation thresholds and describes the characteristic nucleation peak \cite{PhysRevE.68.061502}. In the case of anti-ferromagnetic interactions, favoring lattice scale oscillations, such model can reproduce periodically distributed shear bands. We find, however, that even in the presence of gaussian quenched disorder, the internal dynamics in all these 1D models is still too simple and none of them can generate scale free correlations in the overdamped setting. The situation does not improve if the double-well NN potential is replaced by a periodic potential.

Despite their failure to capture criticality, the 1D models are very important because they allow one to show  that in the limit of quasi-static driving a viscous evolution in a rugged energy landscape takes the form of a stick slip dynamics adequately described by a discrete  automaton.  The replacement of the fast stages of dynamics by jumps leads to considerable simplification of the computational problem. It also shows that in quasi-static setting all dissipative mechanisms producing the same set of jumps are equivalent.

To simplify the system even further we replace a smooth multi-well potential by its piece-wise parabolic analog allowing one to solve the elastic problem analytically when the 'phase configuration' is known. Since the phase configuration is prescribed by an integer valued field, such 'condensation' of elastic problem allows one to reduce the original smooth dynamical system to an integer valued automaton of a sand pile (threshold) type.  Because of the long range nature of elastic interactions, the discharge rules in this automaton are nonlocal which makes its rigorous analysis challenging.

While minimizing out the elastic fields in plasticity problems is a rather common approach leading to a reduced description  in terms of either plastic distortion or dislocation density \cite{Ortiz20002077,Koslowski:2002uq,berd:2006,Limkumnerd20081450,Choi:2011ys,Chen:2011uq}, the proposed integer automaton representation appear to be new. Its salient feature is that in place of straightforward time discretization of continuum dynamics \cite{PhysRevB.48.7030} we utilize  inherent  temporal discreteness  hidden behind the conventional gradient flow dynamics (see also \cite{Perez-Reche:2007et,Puglisi:2005fk,A.:2011kx}).

The next level of complexity is exemplified by 2D lattice models. Here we abandon the double-well framework  with its prototypical defects and consider the simplest setting allowing one to represent actual dislocations which interact through realistic long range elastic fields and are capable of generating collective effects.  As it has been repeatedly proposed in the literature,  the two dimensional models of plasticity may be sufficient to capture the corresponding universality class at least for FCC and hexagonal crystals \cite{Miguel:2001dk,Zaisermoretti2005,PhysRevLett.105.085503}. For similar models of criticality associated with transformational plasticity, see \cite{Perez-Reche:2007et,PhysRevLett.101.230601,J.:2009kx}).

In general terms, our 2D model represents a cross section of a crystal with a single slip geometry. More precisely, we consider a set of parallel  edge  dislocations which can move only in horizontal direction.  This implies that the crystal is highly anisotropic with the deformation in one direction strongly constrained.  As a partial justification we  mention that intermittency has been mostly observed under single slip conditions \cite{Zaisermoretti2005, PhysRevLett.105.085503}.

The ensuing discrete model with continuous  dynamics can  be viewed as a parallel set of coupled overdamped FK chains \cite{ISI:A1982QE67200035,ISI:A1993LF66200007, ISI:A1993MF12500011,A.-I.-Landau:1994vn,PhysRevLett.90.135502}.  It has been shown before that this  model allows for generation and annihilation of dislocations and that it describes adequately their long range elastic interactions \cite{PhysRevB.71.134105, Plans_homo}. Most importantly, both the propagation and the nucleation/anihillation mechanisms in this model are associated with reaching the  same strain thresholds. In particular, this model correctly accounts for Peierls stress associated with underlying periodicity, describes adequately sufficiently slow dislocational kinetics and does not require special rules governing the interaction of the dislocations with quenched disorder \cite{PhysRevLett.90.135502,PhysRevB.71.134105}. At the same time, this model still neglects tensorial nature of elasticity and ignores such important physical effects as cross-slip and climbing which may affect critical exponents.

In our numerical simulations we assume periodic boundary conditions and submit the 2D lattice to quasi-static cycling in the hard device by imposing periodic shear deformation. We show that after several loading-unloading cycles the  system reaches a non-equilibrium steady state (plastic shakedown) where it exhibits critical behavior characterized by the power law statistics of avalanches. The associated  spatial fractality is revealed through the study of spatial correlations. A study of  the power spectrum of the energy fluctuations shows the characteristic $1/f$ noise which is another signature of  scale free behavior.  Overall, the critical manifestations of the plastic flow revealed by our simplified modeling are in full agreement with experimental observations and with previous more elaborate numerical studies in 3D.

In an attempt to  understand the origin of criticality in analytical terms we performed a  reduction of the system of ordinary differential equations describing our 2D visco-elastic lattice model to a threshold automaton with conservative dynamics.  As in 1D case, we minimized out elastic fields and obtained an automaton model for a discrete variable which serves as an indicator of the number of dislocations passing a given point. Despite partial linearization implied by the use of piece-wise quadratic potential and the replacement of time-continuous dynamics by a sequence of jumps, the automaton model exhibits the  same  critical behavior as the original ODE model. Our numerical study suggests that the discrete automaton behind the FK model has a statistical (weak)  Abelian property which makes it amenable in principle to rigorous mathematical treatment despite the nonlocal nature of the discharge rules.

A short announcement of some of the results of this paper can be found in \cite{PhysRevLett.106.175503}.

\section{The model}

Classical phenomenological theory of plasticity is applicable in both crystal and amorphous settings because it circumvents an explicit reference to the nature of  the defects carrying inelastic deformation. Instead, it operates with a concept of an averaged  non affine  deformation represented locally by a tensorial measure of plastic strain. The evolution of the plastic strain is governed by a dissipative potential which is  assumed to be a homogeneous function of degree one   making the dissipation rate independent. Other internal variables characterizing hardening, softening, additional inelastic rearrangements, etc. are often added  as well, each equipped with its own  dynamics. For instance, in crystal plasticity the effective plastic strain rate is parameterized by slip measures associated with different slip directions and evolved by  coupled rate independent mechanisms. Likewise, in transformational plasticity of shape memory alloys the evolution of the phase/twin fractions is usually assumed to be governed by (coupled) rate independent dynamic equations.

Despite their success in fitting the observed memory behavior, the phenomenological models  of plasticity do not describe adequately the internal dynamics revealed by endogenous fluctuations because they do not deal explicitly with microscopic time and length scales. In particular they miss temporal intermittency and spatial fractality.

An alternative  program of describing plasticity directly in terms of defects responsible for the flow is far from being complete despite many important recent successes \cite{Motz20091744, Csikor12102007, Zhou:2010vn,PhysRevLett.105.015502,Needleman:2006,Limkumnerd:2008, Shishvan:2011}. First of all, in order to handle microscopic dynamics of dislocations one needs to go back to the elastic framework. Since  the strain field of a dislocation  decays rapidly  away from the core region, the corresponding long-range distortions can be adequately described within the framework of linear elasticity. Therefore dislocations are modeled as (line) singularities of linear displacement fields  bounding surfaces that carry  the slip. The latter is represented in the elastic models as a quantized displacement discontinuity and this is the only place where the inherent discreteness  enters the picture.

The localization of dislocations in space  and time requires  constitutive assumptions which are outside the realm of linear elasticity. Such assumptions are usually presented in the form of additional kinetic relations of phenomenological nature describing the associated mesoscopic,  rate dependent  dissipation \cite{O.:2003fk}.  An additional problem in this approach is that independent phenomenological assumptions are also required to describe nucleation, annihilation and other topological transitions. Notwithstanding all this uncertainty, the simulation of realistic 3D flows requires tracking of a huge number of interacting defects which remains prohibitively expensive in terms of computational time.

In search for an alternative description we first recall that a simple  hybrid discrete-continuous dislocational model can be formulated by using the Peierls-Nabarro framework \cite{Hirth:1982kx,Nabarro:1951ad,Koslowski:2002uq}. Suppose that the bulk elastic energy is defined by
$$
\Phi_V(u)=\int_V \phi(\nabla u)dv,
$$
where   $u(x)$ is the displacement field and the function $\phi$ is  quadratic.  The  field $u(x)$  is allowed to have  discontinuities $[u]$ which are penalized by surface energy of the form
$$
\Phi_{S(t)}(u)=\int_{S(t)}\varphi([u])ds.
$$
In Peierls-Nabarro model the function $\phi$ is assumed to be periodic and the evolution of the field $u$  is governed by an appropriate gradient flow dynamics.
 The main technical difficulty in this approach is to track the \textit{a priori} unknown surfaces of displacement discontinuity $S(t)$ with arbitrary complex geometry.

In order to avoid the explicit 'tracking' of the slip surfaces one can consider their 'capturing' by
  incorporating discreteness already into the bulk energy term. To this end one can introduce an internal length scale $a$ and
  consider a mesoscopic energy which we schematically represent as
\begin{equation}
\label{eq_toy}
\Phi(u)= a\sum f \left(\frac{[u]}{a}\right ).
\end{equation}
Now $u$ is the lattice field and the function $f$ is periodic with  infinite symmetry group \cite{ISI:000222353800003,J.:2009kx} . The discrete energy $f$ contains information about both the bulk energy $\phi(\nabla u)$ and the surface energy $\varphi([u])$; the  mechanism of such  splitting in the continuum limit was discussed in the context of fracture in \cite{Truskinovsky:1996tg} and in the context of plasticity in \cite{pellegrini2010}. It is important to keep in mind that the discrete energy $\Phi(u)$ describes multi-particle elastic units and is therefore different from the conventional molecular dynamics potentials dealing directly with interatomic interactions.

After the energy density in (\ref{eq_toy}) is specified, one can formulate the rate dependent dynamics for the elastic fields carrying dislocations. Under the assumption that such dynamics is viscous and overdamped we obtain the following  system of ODEs

\begin{equation}
\label{eq_mov}
\nu \dot{u}=- \partial \Phi(u)/\partial u,
\end{equation}
where $\nu$ is the ratio of the internal time scale and the time scale of the driving \cite{ Perez-Reche:2007et,Puglisi:2005fk}. Notice that instead of classical visco-elasticity we consider here an 'environmental' viscosity which allows us to avoid certain dynamic degeneracies in the system
 without quenched disorder  \cite{A.:2011kx}. Notice also that the system (\ref{eq_mov}) is not autonomous because time enters through the equations for boundary units as, for instance,  in the case of cyclic driving in a hard device.

The numerical solution of (\ref{eq_mov}) may be rather challenging under generic loading. However, in the limit of quasistatic driving
$
\nu \rightarrow 0,
$
which can also be interpreted as the limit of zero viscosity, the dynamic system (\ref{eq_mov})  can be replaced almost everywhere (in time) by the equilibrium system \cite{Puglisi:2005fk}
\begin{equation}
\label{eq_mov1}
  \partial \Phi(u)/\partial u =0.
\end{equation}
The dynamics is then projected onto the branches of metastable equilibria  which are defined as  sequences of local minima (of the elastic energy) parameterized by the loading parameter. The crucial observation is that similar to the case of relaxation oscillations, the dissipation-free phases of such dynamics are necessarily interrupted by the branch switching events which are instantaneous at the time scale of the quasi-static driving. Such reduction of an original dynamic problem opens the way towards minimizing out the elastic variables and reformulating the problem in terms of a discrete integer valued variable labeling individual metastable branches  \cite{Perez-Reche:2007et,Puglisi:2005fk,A.:2011kx}.

In the next section we illustrate this general approach  by using the simplest one dimensional setting where our goal will be modest: to make the idea of marginal stability fully transparent and to obtain the simplest examples of fluctuating plastic flow. Then we formally expand the model to the second dimension and show that the ensuing minimal geometrical complexity is already sufficient to generate realistic fluctuational behavior.

\section{One dimensional setting}

The very first model of plasticity accounting for fluctuations was probably Prandtl's zero-dimensional model describing a particle driven through a spring in a periodic landscape \cite{Prandtl:1928fk}. This model involves marginal stability, captures hysteresis, and reproduces rate independent dissipation \cite{Puglisi:2005fk} but misses the crucial interactions among the defects. To obtain criticality in this model one needs to considerably complicate the effective landscape, replacing, for instance, a periodic potential by a Brownian motion type potential \cite{colaiori-2008-57}. Such fine 'tuning' of the interaction potential places this model in the class of phenomenological models  representing a natural development of the classical rheological models used in engineering plasticity.

The simplest nontrivial 'first principle type' model with the energy (\ref{eq_toy})  describes the transformational plasticity of shape memory alloys \cite{springerlink:10.1007/BF00289355,springerlink:10.1007/BF01208928,Puglisi:2000uq,springerlink:10.1007/s001610200083,Puglisi:2005fk}.  In this model each elastic element interacts with its nearest neighbors (NN) only through the total stress which then plays the role of a mean field. Having the advantage of utmost  simplicity, this model possesses  a nonphysical permutational invariance which introduces undesirable degeneracy. Therefore we subsequently augment this model by  introducing  linear interactions between next to nearest neighbors (NNN) and allowing them to be either  attractive and repulsive \cite {Truskinovsky:2004qf,Braides:2008bh}. Finally, we replace a bi-stable NN potential by a periodic one and introduce quenched disorder.

\subsection{NN model}

Consider a chain of particles forming in the reference configuration a regular 1D lattice.  Denote the reference particle positions by $x_i^0=i$ where we imply that the atomic spacing is taken as the length scale ($a=1$).  Introduce discrete displacement field $u_i(t)$ and the discrete strain field $\epsilon_i(t)= u_{i+1}(t)-u_{i}(t).$   Our Fig.\ref{chain}  shows the simplest representation of this mechanical system.  In plasticity framework one should rather
view the chain as a deck of rigid cards interacting through  nonlinear shear springs and view the displacement field $u_i(t)$ as being perpendicular to the direction of the coordinate axis.
\begin{figure}[h!]
 \begin{center}
 \includegraphics[scale=0.35]{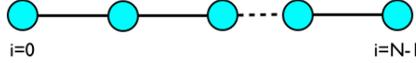}
 \end{center}
 \caption{\label{chain} A schematic representation of a one dimensional NN chain.}
 \end{figure}
\begin{figure}
\begin{center}
\subfigure[]{\label{udot_ss}\includegraphics[scale=0.25 ]{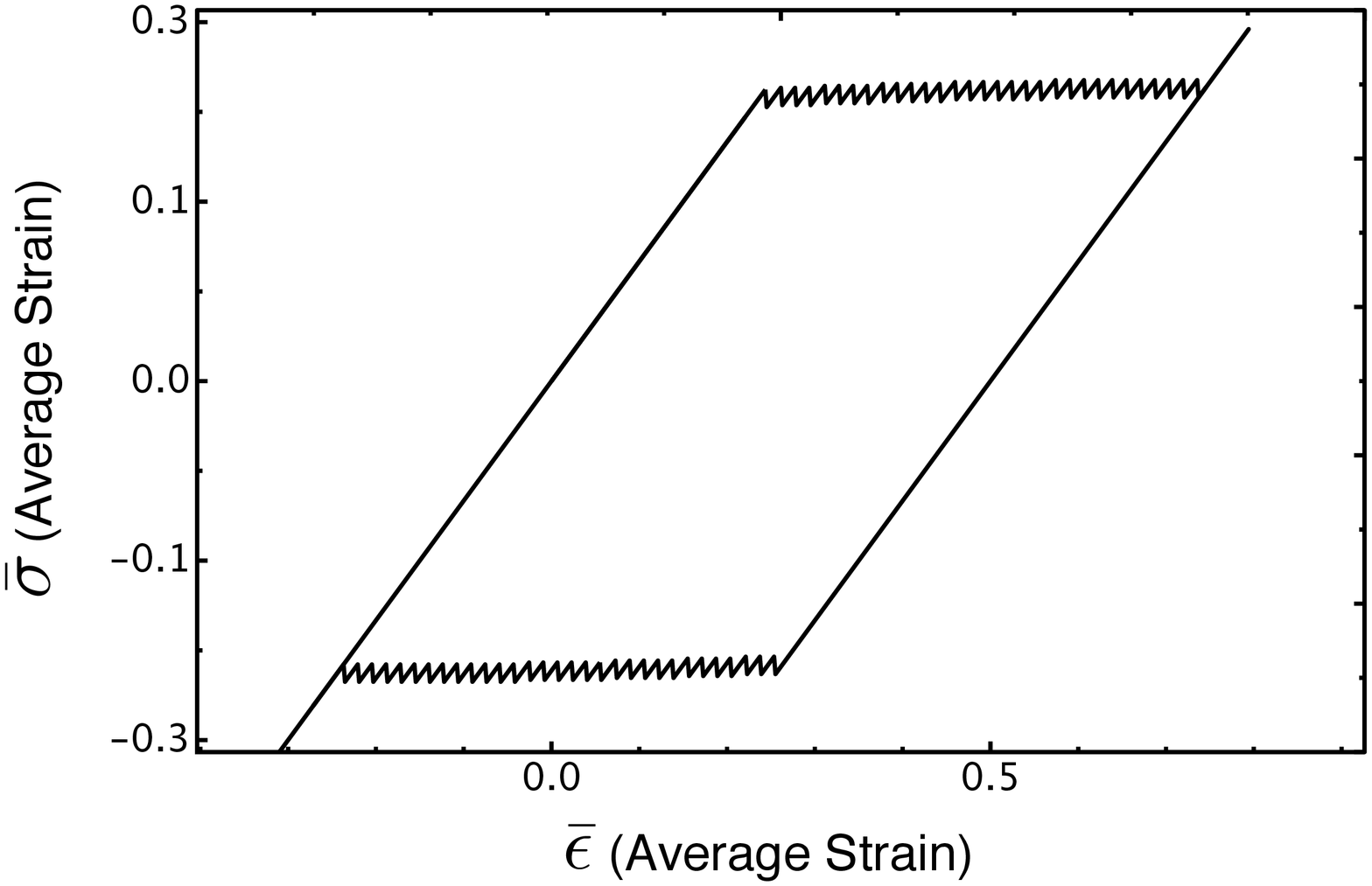}}
\subfigure[]{\label{particle1}\includegraphics[scale=0.25]{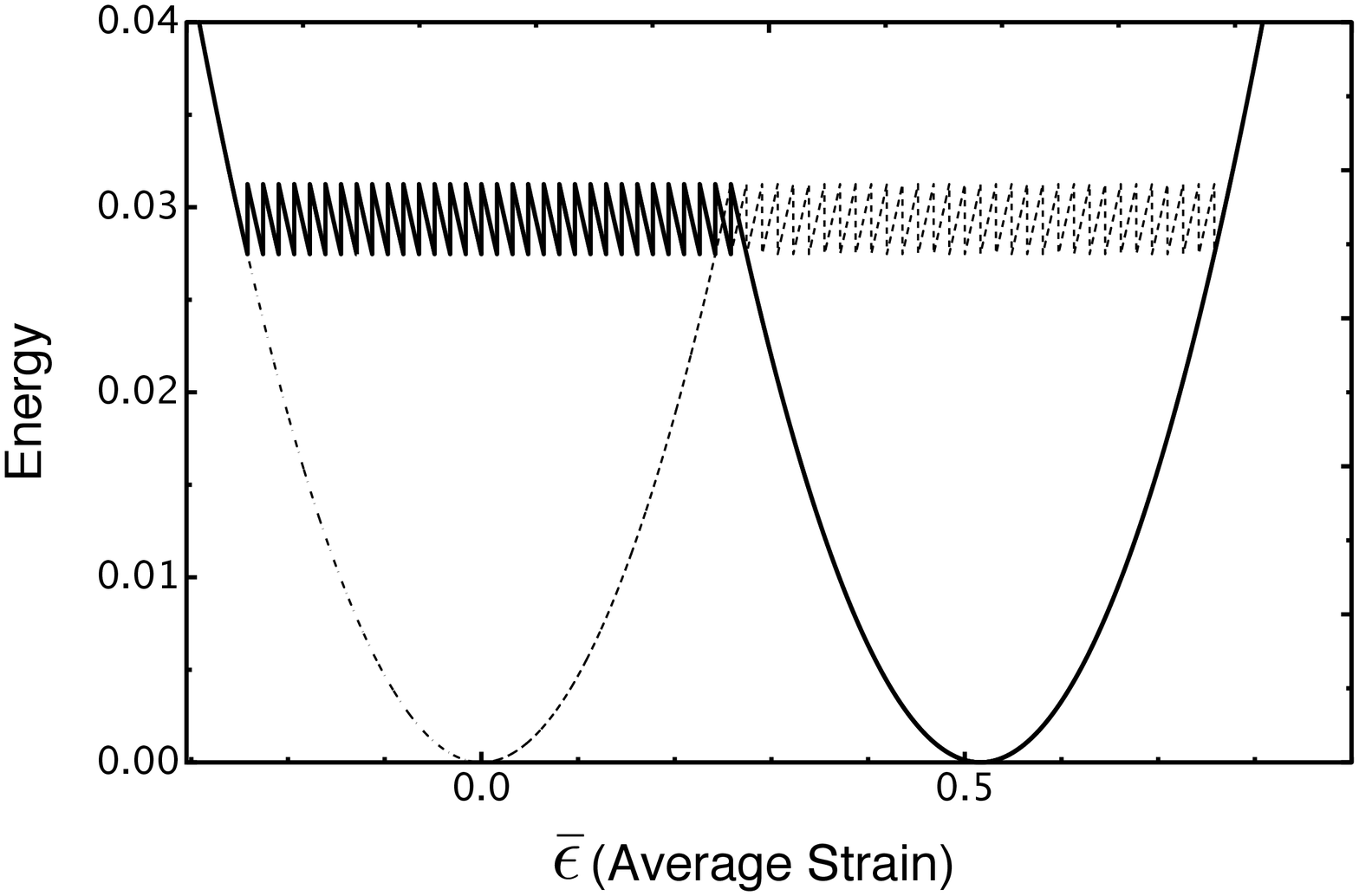}}
\end{center}
\caption{ Stress-strain  (a) and energy-strain  (b) relations in the NN model with $N=32$, $\kappa=1$, $\nu=10^{-4}$.
In (b) dashed line represents loading, solid line-unloading.}
\label{fig1}
\end{figure}

For the NN model the  equations of motion (\ref{eq_mov}) take the form
\begin{equation}
\label{eq:1}
 \nu \dot{u}_i  + \sigma(u_{i+1}-u_{i}) - \sigma(u_{i}-u_{i-1})=0,
\end{equation}
where   $\sigma (\epsilon)=  \partial f/ \partial \epsilon$ is the elastic force resulting from stretching of a spring.
 We assume that the system is placed in a hard device  with  $$u_0(t)=0, u_N(t)=t$$
where $t$ is the dimensionless time playing the role of the loading parameter.

One can see that the total energy in the NN model is the sum of the energies of individual springs. This means that the springs interact only through the constraint imposed by the hard device. Indeed,  in equilibrium
 $\sigma(\epsilon_i)=\bar{\sigma}, $   where  $\bar{\sigma}$  is the stress common to all springs; such infinite range of interactions is characteristic of paramagnetism.

To facilitate the subsequent elimination of elastic variables we assume that the  double well potential describing individual springs is  piece-wise quadratic:
 \begin{equation}
 \label{pot}
   f(\epsilon_i)  =\left\{
  \begin{aligned}
&{}  \frac{\kappa}{2}\epsilon_i^2,\hspace{1mm}\text{if} \hspace{1mm}\epsilon_i <0.25\\
&{}  \frac{\kappa}{2}(\epsilon_i -0.5)^2, \hspace{1mm}\text{if}\hspace{1mm}\epsilon_i > 0.25.
\end{aligned}
\right.
 \end{equation}
Here  $\kappa$ is  the elastic modulus which we assume to be the same in both phases. The interval $\epsilon_i <0.25$ will corresponds to Phase 1 with equilibrium strain $\epsilon=0$ and the interval $\epsilon_i >0.25$,   to Phase 2 with equilibrium strain $\epsilon=0.5$.

Our Fig. \ref{udot_ss} shows the macroscopic strain-stress curve  obtained from the numerical solution of (\ref{eq:1}) with $\nu$ sufficiently small. As we load the system from a homogeneous state in Phase 1, the phase transition starts at the critical strain $\bar \epsilon = 0.25$, where only one spring changes phase. The transformation in one spring corresponds  to a stress drop as the other springs relax.  Each subsequent stress drop is also associated with only one spring changing phase during each jump. One can see that the resulting macroscopic stress strain curve forms a wiggly plateau with a rudimentary intermittency manifesting itself through equally spaced oscillations. This behavior was studied in \cite{springerlink:10.1007/s001610200083,Puglisi:2000uq} where it was shown that the individual springs  can change phase only consequently.

\begin{figure}[h!]
\centering
\includegraphics[scale=0.35]{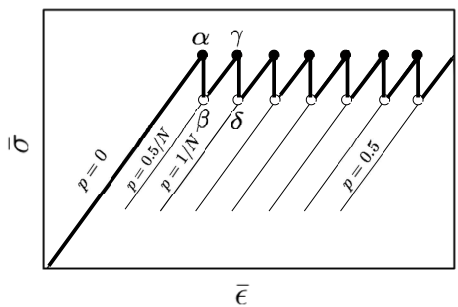}
\caption{Blow up of a schematic strain-stress relation for the NN model in the limit $\nu\rightarrow0$  ($N=8$). Solid lines show the path followed by the system during quasi-static loading. Thin lines are metastable branches. The black dots mark marginally stable states  while the white dots correspond to local minima where the system stabilizes after a jump.}
\label{marginal}
\end{figure}
As we mentioned in the Introduction, our model can be simplified further in the case of quasi-static loading  which for the the dynamical system (\ref{eq:1}) corresponds to a zero viscosity limit .  We begin with rewriting the  piece-wise quadratic potential (\ref{pot}) as
 \begin{equation}
 \label{pwN}
   f(\epsilon_i)  =
  \begin{aligned}
&{}  \frac{\kappa}{2}(\epsilon_i-m_i)^2.
\end{aligned}
 \end{equation}
Here, we introduced a double-valued spin variable $m_i$ describing the phase state of a given spring with $m_i=0$ for  $\epsilon_i <0.25$ and $m_i=0.5$ for $\epsilon_i > 0.25$.  At a given spin field, the displacement field $u_i$ must satisfy the following set of linear  equations
\begin{equation}
\label{eq:at1D}
  (u_{i+1}+u_{i-1} - 2u_i)  = (m_{i}-m_{i-1}),
\end{equation}
where the right hand side plays the role of 'dislocation density'\cite{berd:2006}. If the phase field is known, the tri-diagonal matrix on the left may
be inverted and the strain field can be obtained explicitly. Moreover, due to permutational invariance of the problem,
one can always assume that there is only one defect and therefore the equilibrium branches can be parameterized by a single  variable $p=(1/N)\sum m_{i}$ which changes in the interval $0 \leq p \leq 0.5$ and describes the 'phase fraction' in the mixture (see \cite{Puglisi:2000uq}).

Several   metastable  branches parameterized by $p$ are shown schematically in Fig. \ref{marginal}. The first branch with all springs in the first energy well (Phase 1) corresponds to $p=0$. It ends at point $\alpha$ where this homogeneous state becomes unstable.  Suppose that we drive the system by quasi-statically changing  the total strain. Then at  point $\alpha$ one spring changes phase and the system undergoes a dynamic transition to the new local minimum identified in Fig. \ref{marginal} by point $\beta$. This fast process  is accompanied by a jump in stress; as a consequence of this jump a finite amount of energy dissipates into heat. 

From point $\beta$  the driven chain evolves along the second equilibrium branch corresponding to $p= 0.5/N $ which in turn  becomes unstable at point $\gamma$. Here again one spring switches phase, the stress drops and the system finds itself in a new local minimum identified as point $\delta$ on the branch with $p=1/N $. The process continues till all springs change phase and the system reaches the second homogeneous  branch corresponding to $p=0.5$. Interestingly, the
way we introduced viscosity ensures that the transformation proceeds as a single front which is not the case if we directly deal with the degenerate problem at $\nu=0$ or if we consider classical visco-elasticity \cite{A.:2011kx}.

A full analysis of the thermodynamics of the NN model including the computation of the 'heat to work ratio' can be found in \cite{Puglisi:2005fk}.

\subsection{Automaton model}

After the elastic fields are minimized out (either by the inversion of the matrix in (\ref{eq:at1D}) or by the Fourier transform as we show below), the ensuing problem in terms of the variables $m_{i}$  can be reformulated as a spin-valued automaton.

To be more precise suppose again that initially all the units are in the Phase 1, meaning that the phase variable vanishes everywhere. Since the field $m_{i}$  is known, one can compute $e_{i}$  by solving (\ref{eq:at1D}). Then one can find $\bar{\sigma}$ and locate the system on the  equilibrium branch with $p=0$. The stability condition in this case reduces to an inequality $\bar{\sigma}(p,t)\leq \sigma^c,$
where  $\sigma^c=0.25$ is the upper spinodal stress \cite{Puglisi:2005fk}. Suppose that the stability condition is violated at time $t_1$. Then we need to update the phase configuration
 $ \tilde{p}= p+0.5/N$
which leads to a stress drop  from $\bar{\sigma} (p,t_1)$ to $\bar{\sigma}(\tilde{p},t_1)$.  Notice that the choice of the flipping element $m_i$ remains arbitrary in the automaton why in the ODE framework it is uniquely prescribed by the dynamics. To overcome this (artificial) non-uniqueness in the automaton setting we later consider two regularizing mechanisms: NNN interactions and/or quenched disorder.

Along the new metastable branch $ p=\tilde{p}$ one can continue computing  $\bar{\sigma} (\tilde{p},t)$ till the spinodal stress is reached again and then it is necessary to make the next update. The stress oscillations continue till we reach the homogeneous Phase 2 with $p=0.5$. During  unloading the stability condition changes into $\bar{\sigma}(p,t)\geq \sigma_c,$ where $\sigma_c=0$ is the lower spinodal stress and the update rule becomes $p\rightarrow p -0.5/N$. The homogeneous branch corresponding to Phase 1 is reached again when  $p=0$.
We do not present here the strain-stress curve obtained from the automaton-based simulations because it is practically indistinguishable from the results obtained by solving directly the dynamic equations (\ref{eq:1}) with $\nu=10^{-4}$ shown in Fig. \ref{fig1}. 

In order to show that on the yielding plateau the system evolves through a sequence of marginally stable states,  we plot in Fig. \ref{particle1} the energy of the system  against the applied strain for one loading-unloading cycle. When the system is in one of the pure phases (affine or Cauchy-Born state) it resides in a global minimum of the energy until  the Maxwell threshold $\sigma_M=0$  is crossed (during either loading or unloading). Then the global minimum becomes a local minimum  and eventually the homogeneous state  becomes unstable as the system reaches either upper or lower spinodal stress. After that the system evolves through the sequence of
non-affine states where different springs occupy different phases.  Notice, however, that in this non-affine stage the stress in the chain never deviates considerably from one of the spinodal stresses (see points $\alpha$, $\beta$, $\gamma$, $\delta$, etc. in Fig. \ref{marginal}). It is not hard to see that in the continuum/thermodynamic limit $N \rightarrow \infty$ the jumps become infinitesimal and the ensuing yielding plateau collapses the spinodal \cite{Puglisi:2005fk}. Therefore, in the continumm limit the yielding system is  marginally stable on the whole yielding plateau.

In summary, plasticity in this simplest model is associated with elastic states of minimal stability. This peculiar feature of plastic flow may be considered as a first manifestation of criticality. In addition, the model exhibits stress drops that can be interpreted as avalanches. The problem is that these avalanches are all of the same size and the corresponding microscopic transformation events are either un-correlated (zero viscosity limit) or over-correlated (finite viscosity case).

\subsection{NNN model with 'ferromagnetic' interactions}

As we have seen, the NN model lacks the short range interactions capable of generating extended dislocation cores.  To bring the missing internal length scale into the theory one can add to the NN model an additional 'ferromagnetic' interactions favoring local homogeneity of the strain field \cite{Rogers1997370}.

The simplest way to introduce short range interactions is to consider  an NNN model with the energy density \cite{Truskinovsky:2004qf}
 \begin{equation}
   f(\epsilon_i)  =
  \begin{aligned}
&{}  \frac{\kappa}{2}(\epsilon_i-m_i)^2+ \frac{\lambda}{2}(\epsilon_{i}+\epsilon_{i-1})^2.
\end{aligned}
\label{eq:nnne}
 \end{equation}
Here $m_i$ is the spin variable introduced in the previous subsection and  $\lambda$ is the strength of short range interactions (we consider only the automaton version of the model).   The 'ferromagnetic'  interactions  ensuring the localization of phase boundaries corresponds to $\lambda<0$; the case of 'anti-ferromagnetic' interactions $\lambda>0$,  favoring the formation of fine mixtures, will be considered in the next subsection (see also \cite{doi:10.1080/14786430500363270}).

In the quasi-static limit discussed above, the equilibrium equations for the NNN model can be written in the form
 \begin{equation}
\label{eq:at1DNNN}
  \begin{aligned}
&{} [ u_{i+1}+u_{i-1} - 2u_{i}  ]+(\lambda/\kappa) [u_{i+2}+u_{i-2} - 2u_{i}] = m_{i}-m_{i-1},
 \end{aligned}
\end{equation}
where again the 'dislocation density' in the right hand side plays the role of a source of elastic strain. Due to non locality of the NNN model, we need to complement the hard device boundary conditions used in the NN model by two additional conditions. One way to avoid surface boundary layers is to add  two fictitious springs just outside the boundaries of the chain ensuring the periodicity conditions $\epsilon(0)=\epsilon(1)$ and $\epsilon(N+1)=\epsilon(N)$. This leads to additional boundary terms in the energy, however, the elastic problem can be again solved explicitly \cite{Truskinovsky:2004qf}.
The remaining problem for the spin field $m_i(t)$ can be formulated as an automaton which is similar to the one described in the previous section and we omit the details.

\begin{figure}[h!]
\begin{center}
\subfigure[]{\label{NNNss1}\includegraphics[scale=0.25]{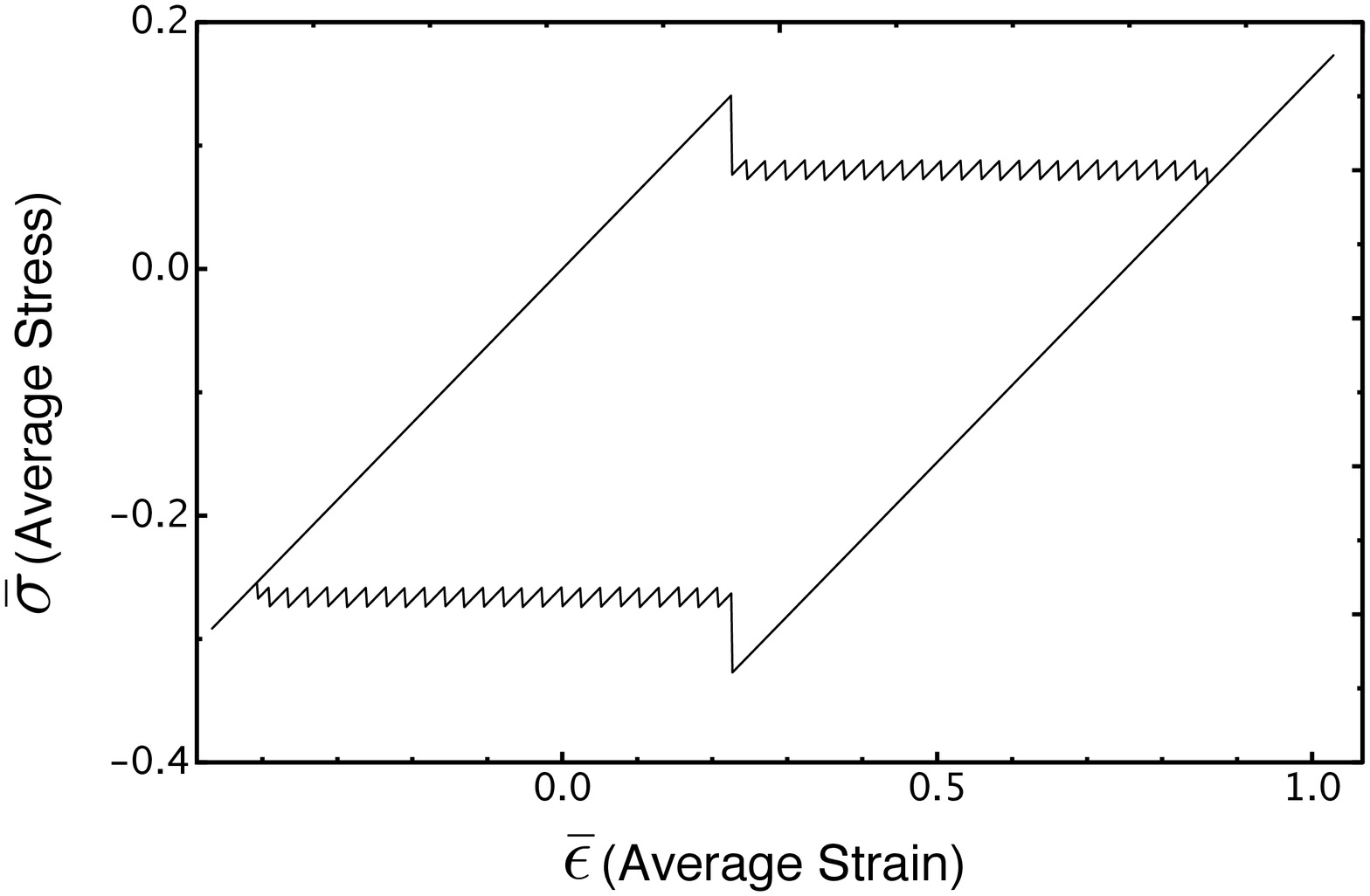}}
\subfigure[]{\label{NNNss2}\includegraphics[scale=0.25]{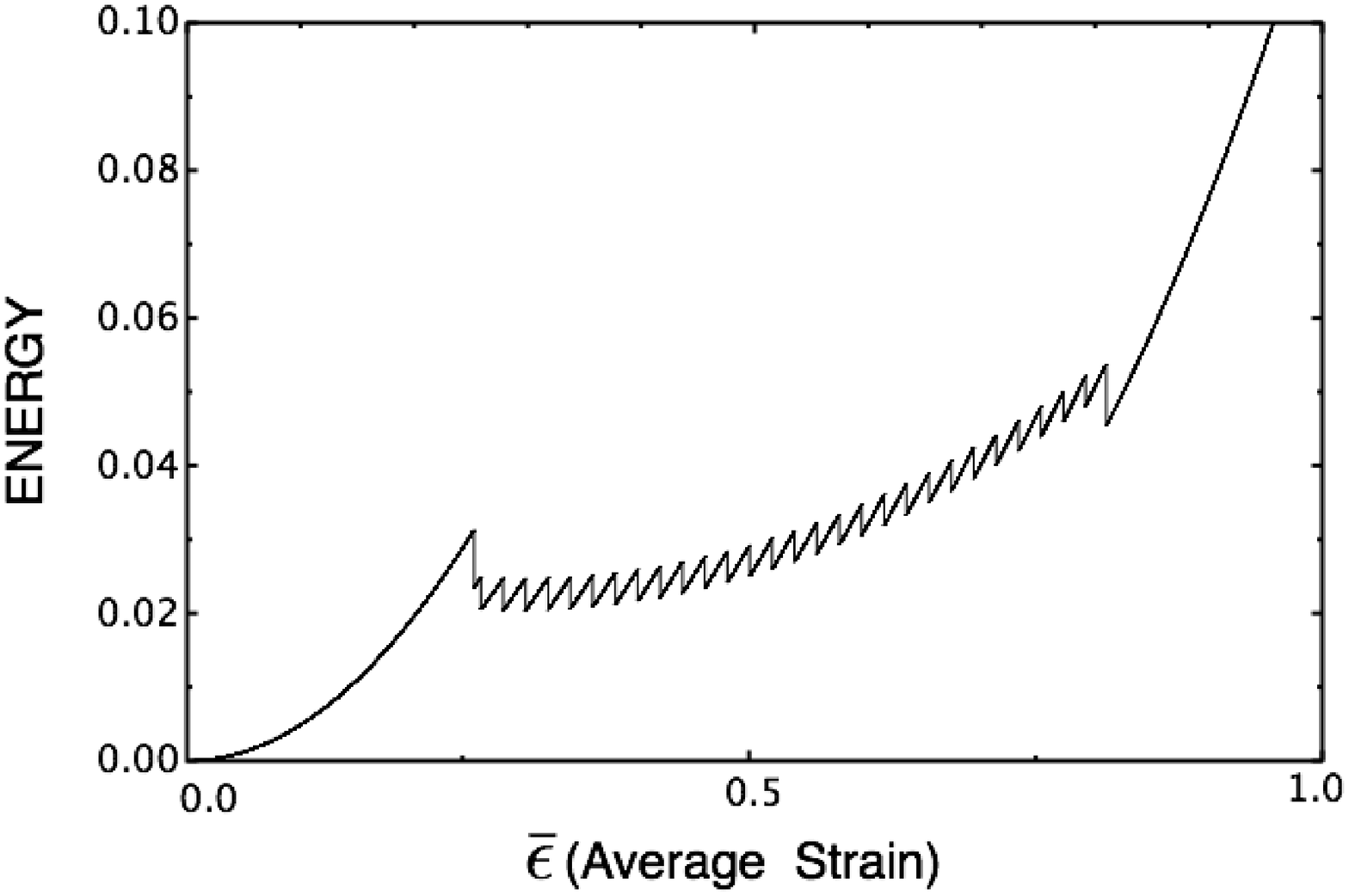}}
\end{center}
\caption{\label{NNN_aut} (a) Strain-stress relation for the NNN automaton. (b) Evolution of the total energy. Here and in other similar plots $N=32$,
$\kappa = 1$ and $\lambda =-0.01$ }
\end{figure}

The macroscopic strain-stress curves for the NNN model with $\lambda <0 $ are shown in Fig.\ref{NNNss1}.  In addition to the horizontal plateaus, that again correspond to the regimes when an isolated  phase boundary sweeps through the chain (propagation), one can also see two peaks at the onset of the direct and inverse transformation (nucleation)  and two smaller anti-peak  describing  the disappearance the existing phase boundaries (annihilation). In the computational experiment shown in  Fig.\ref{NNNss1}
 four springs out of 24 participate in the nucleation event  while the propagation involves successive jumps of individual springs. The necessity of the peak in the NNN model and the nature of the nucleation event in the continuum limit were discussed in \cite{Truskinovsky:2004qf}.

In Fig. \ref{E2_f}) we show the macroscopic energy-strain relation for this model  (loading only). In contrast to what we have seen in the NN model, here the energy  increases in the systematic manner as the system evolves along the plastic plateau. This means that plastic dissipation is accompanied by an additional energy build up which can be vaguely qualified as the 'cold work' \cite{Puglisi:2005fk}. The origin of this effect in our model is the particular form of the NNN term in the energy (\ref{eq:nnne}) which, in addition to penalizing gradients, also contributes to homogeneous elasticity. Interestingly, similar continuing growth of the energy along the flat-stress yielding plateau have been observed in several microscopic (MD-type) models of amorphous plasticity \cite{Procaccia:2009}.

 \begin{figure}[h!]
\begin{center}
\subfigure[]{\label{E1_f}\includegraphics[scale=0.25]{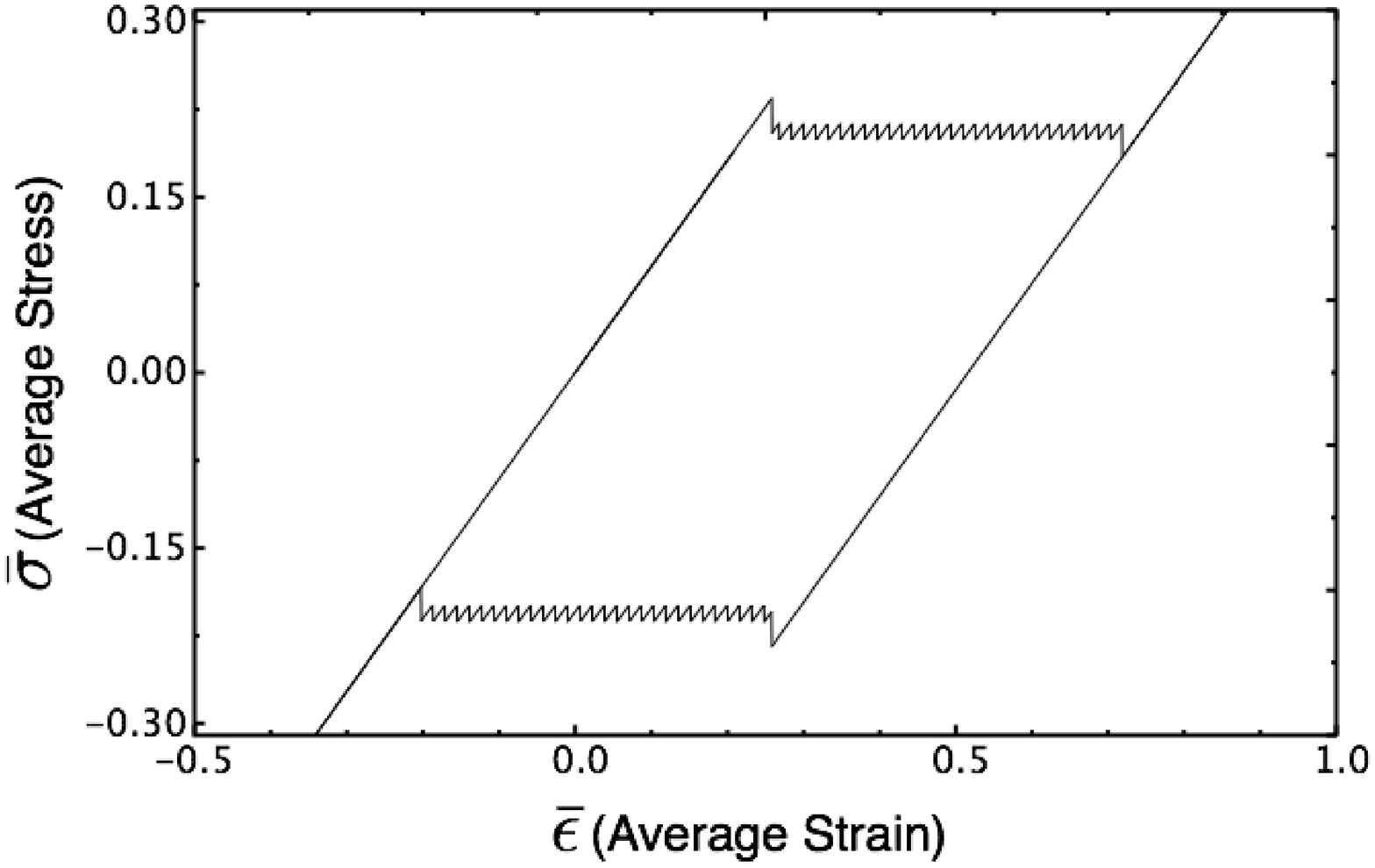}}
\subfigure[]{\label{E2_f}\includegraphics[scale=0.25]{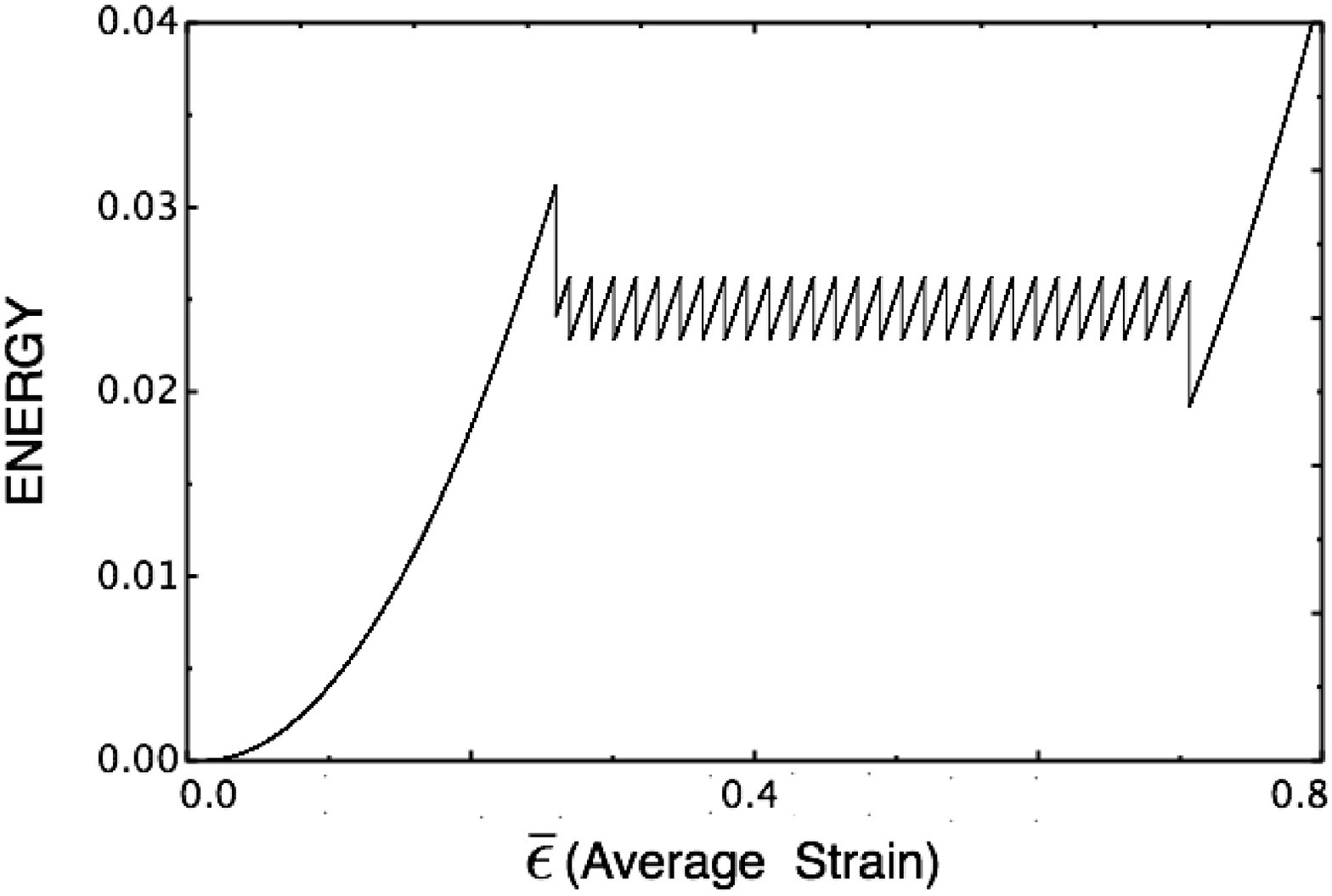}}
\end{center}
\caption{\label{sg_nn} Stress-strain (a) and  energy-strain (b) relations in the  SG model with $\beta=0.2$ (loading only).}
\end{figure}

In order to avoid such  hidden contributions to bulk elasticity, one can introduce short range interactions in a different form. The simplest way to keep the ferromagnetic coupling intact while removing the NNN contribution to homogeneous elasticity  is to introduce a three-particle interaction mimicking the strain gradient (SG) term in the discrete setting
 \begin{equation}
 \label{pwN}
   f(\epsilon_i)  =
  \begin{aligned}
&{}  \frac{\kappa}{2}(\epsilon_i-m_i)^2+ \frac{\beta}{2}\bigr(\epsilon_{i}-\epsilon_{i-1}\bigl)^2.
\end{aligned}
 \end{equation}
 It is not hard to see that in the case $\beta>0$ this energy is equivalent to the NNN energy (\ref{eq:nnne}) with $\lambda <0$ up to a  'local' term depending on  $\epsilon_{i}$ only. This means that the 'cold work' is eliminated in the SG formulation which is confirmed by our Fig. \ref{sg_nn}(b).  Notice that outside  the nucleation (annihilation) events the behavior of the  NN and SG models are quite similar with isolated phase boundaries propagating along the chain as the total strain increases. The main difference is that in the NN model a phase boundary (mimicking dislocation) is atomically sharp, while in the SG model its core has a finite size.
\begin{figure}
\begin{center}
\subfigure[]{\label{SS3_f}\includegraphics[scale=0.25]{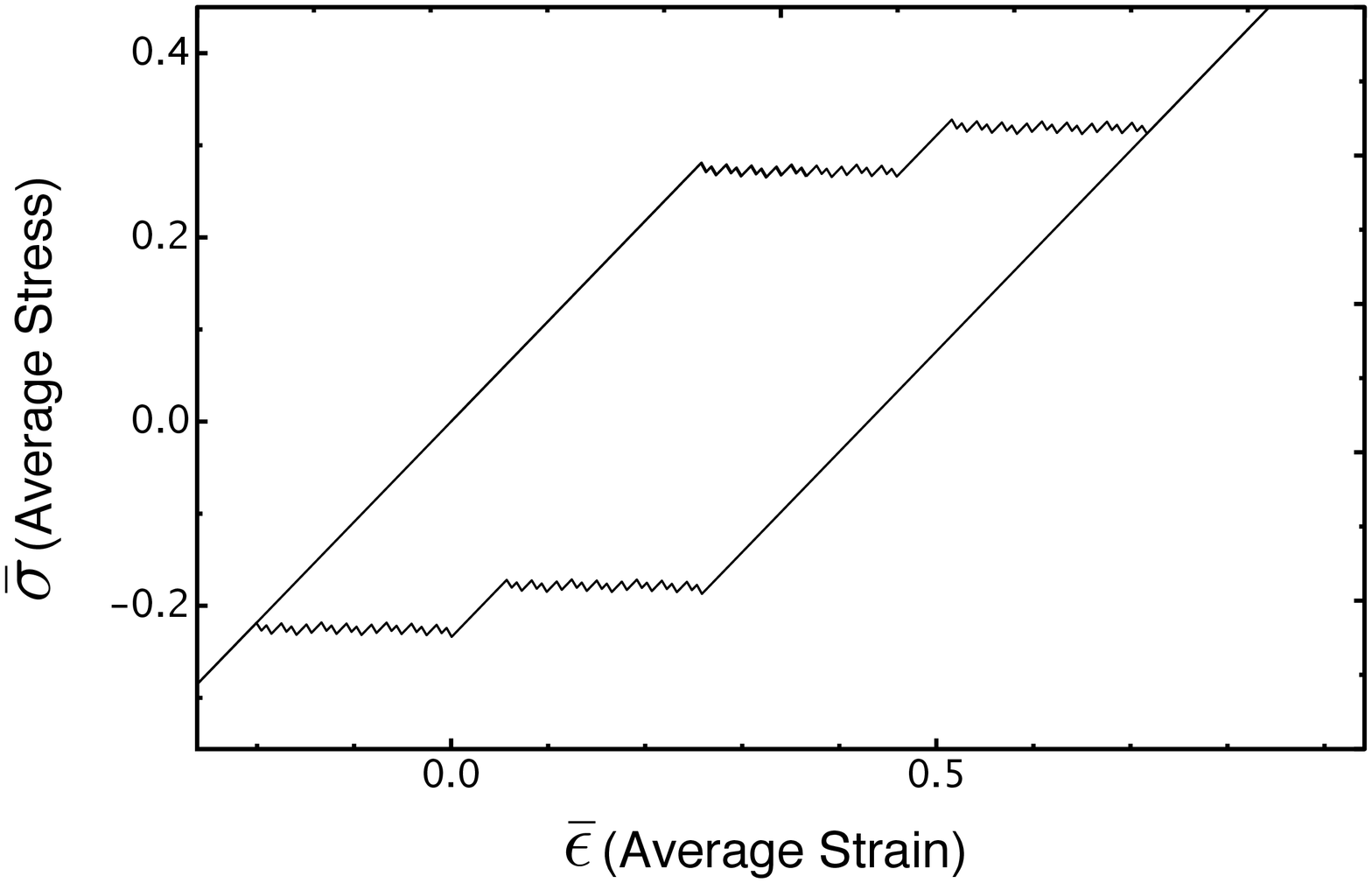}}
\subfigure[]{\label{SS2_f}\includegraphics[scale=0.25]{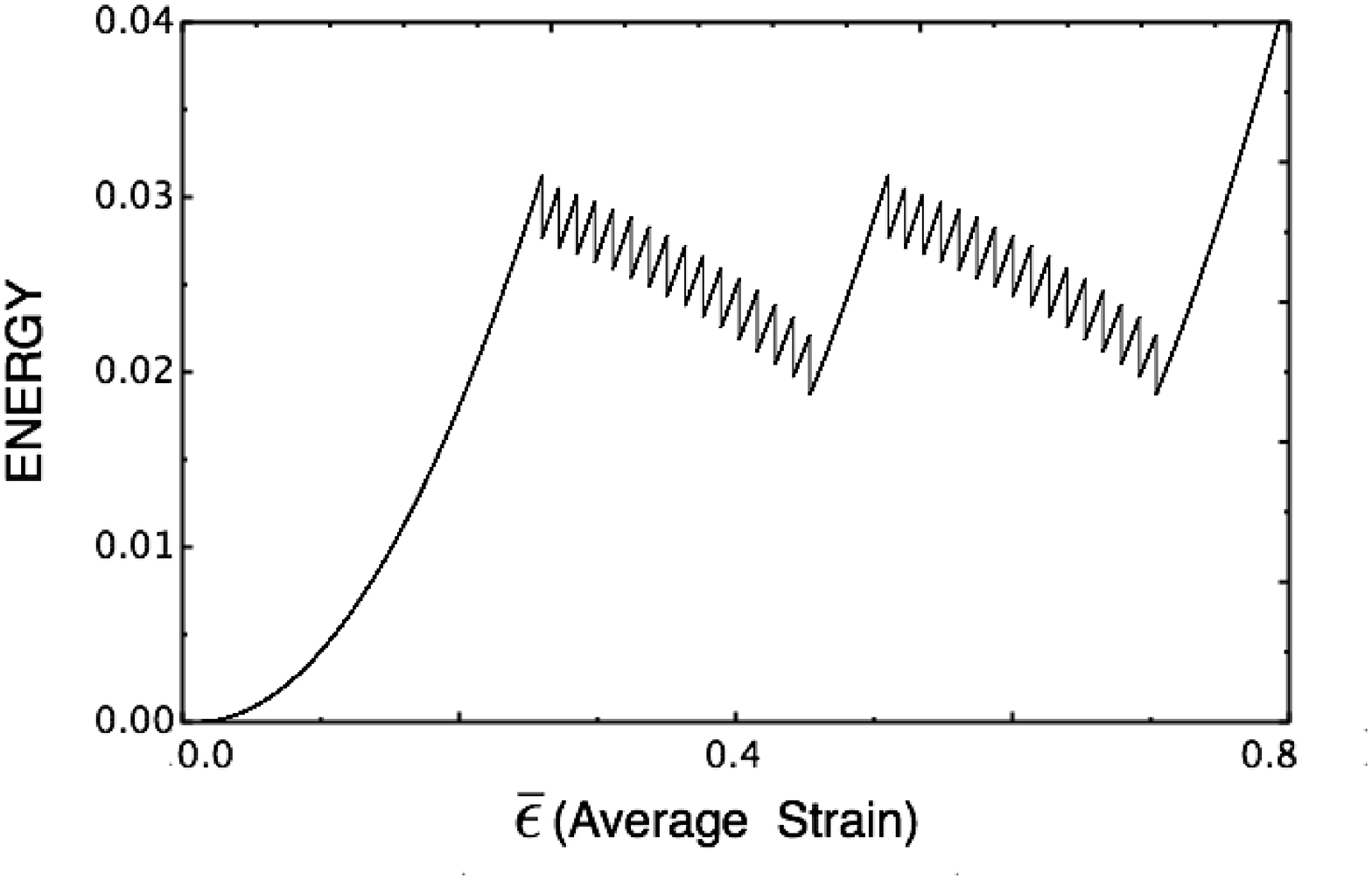}}
\end{center}
\caption{\label{SS1_f_all} (a) Strain-stress relations for the NNN  model  with  $\lambda=0.01$, other parameters are the same as in the other similar graphs. A step in the center of the plateau corresponds to elastic deformation of a 'binary' phase with lattice scale oscillations; (b) Evolution of the total energy.}
\end{figure}

 Despite all these new features, neither NNN nor SG  model are successful in predicting the realistic statistical structure of avalanches. Indeed, both models exhibit one correlated 'snap' event (nucleation) and a succession of equally sized 'pop' events (propagation). It is clear that this fluctuation/avalanche structure is still too simple to be compared with realistic experiments.

 \subsection{NNN model with 'anti-ferromagnetic' interactions}

As it is well known, the nonlocal kernel describing linear elastic interactions in   dimensions 2 and 3 has both ferromagnetic and anti-ferromagnetic contributions \cite{PhysRevB.52.803,Ren:2000uq, PhysRevB.67.024114}. Therefore it is of interest to see how  plastic behavior in our toy model changes if  'ferromagnetic' NNN interactions with  $\lambda<0$  are replaced by  'anti-ferromagnetic' NNN interactions with  $\lambda>0$. In the 1D setting, anti-ferromagnetic interactions favoring formation of small scale microstructures, should be understood as a poor man's attempt to capture the gradient nature of the strain (elastic compatibility) and the effect of the constraints placed by higher dimensional boundary conditions \cite{Ren:2000uq}.

Our numerical results for the anti-ferromagnetic NNN model driven in a hard device (automaton version) are summarized in Fig. \ref{SS1_f_all}. The most important feature is that in addition to two homogeneous phases the model exhibits the third phase representing binary  lattice scale mixture which is homogeneous only in average (see Fig. \ref{Sp_f}). This new phase shows binary oscillations with neighboring springs occupying different phases; the energetic preference of such non Cauchy Born (non-affine) phase arrangements follows from the studies of the global minimum of the elastic energy, e.g.  \cite{Cahn19994627,Pagano_asimple,Braides:2006zr,Braides:2008bh,Charlotte02linearelastic,doi:10.1080/14786430500363270}.

\begin{figure}[h!]
\begin{center}
\subfigure[]{\label{Sp2_f}\includegraphics[scale=0.25]{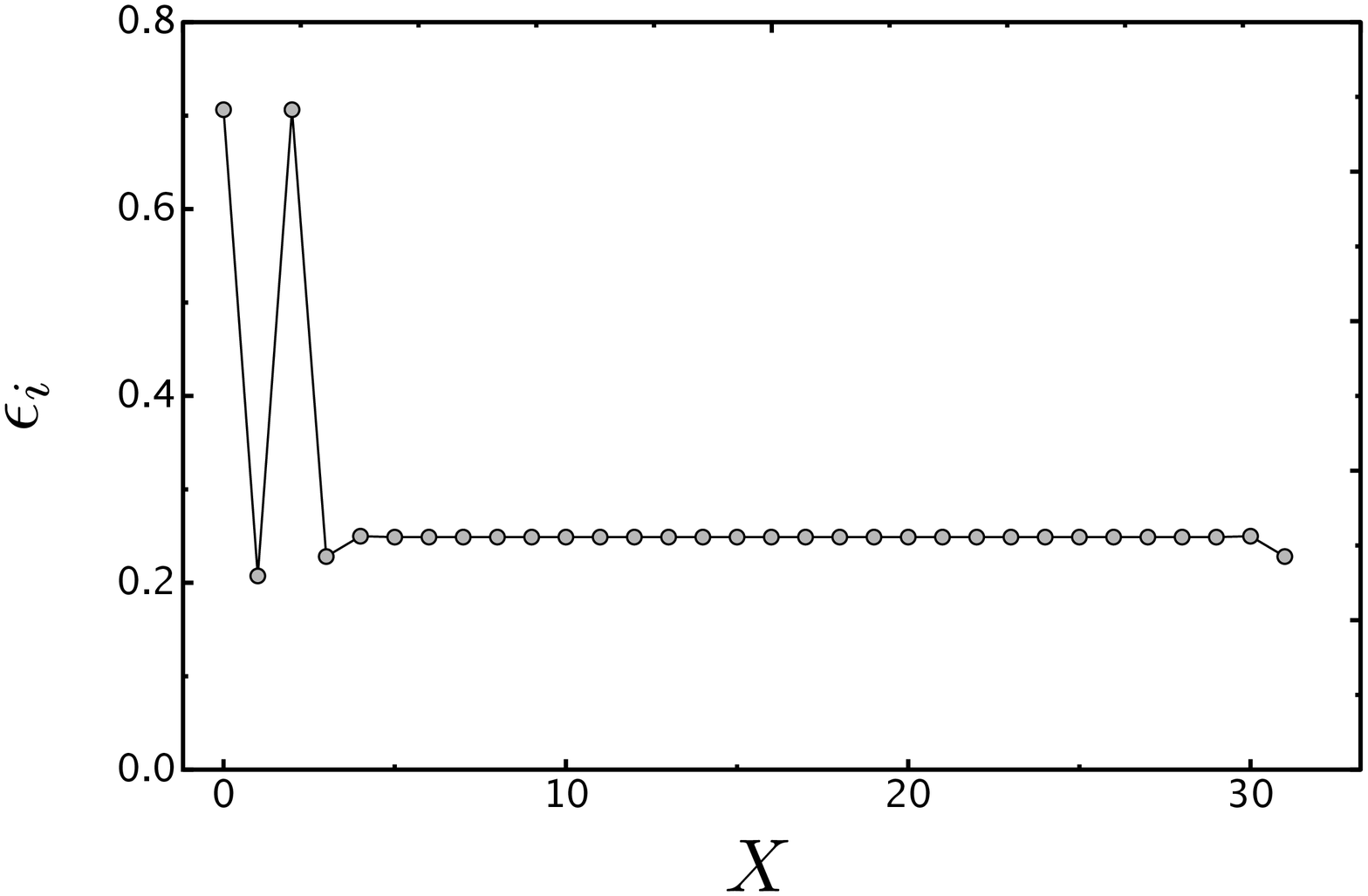}}
\subfigure[]{\label{Sp3_f}\includegraphics[scale=0.25]{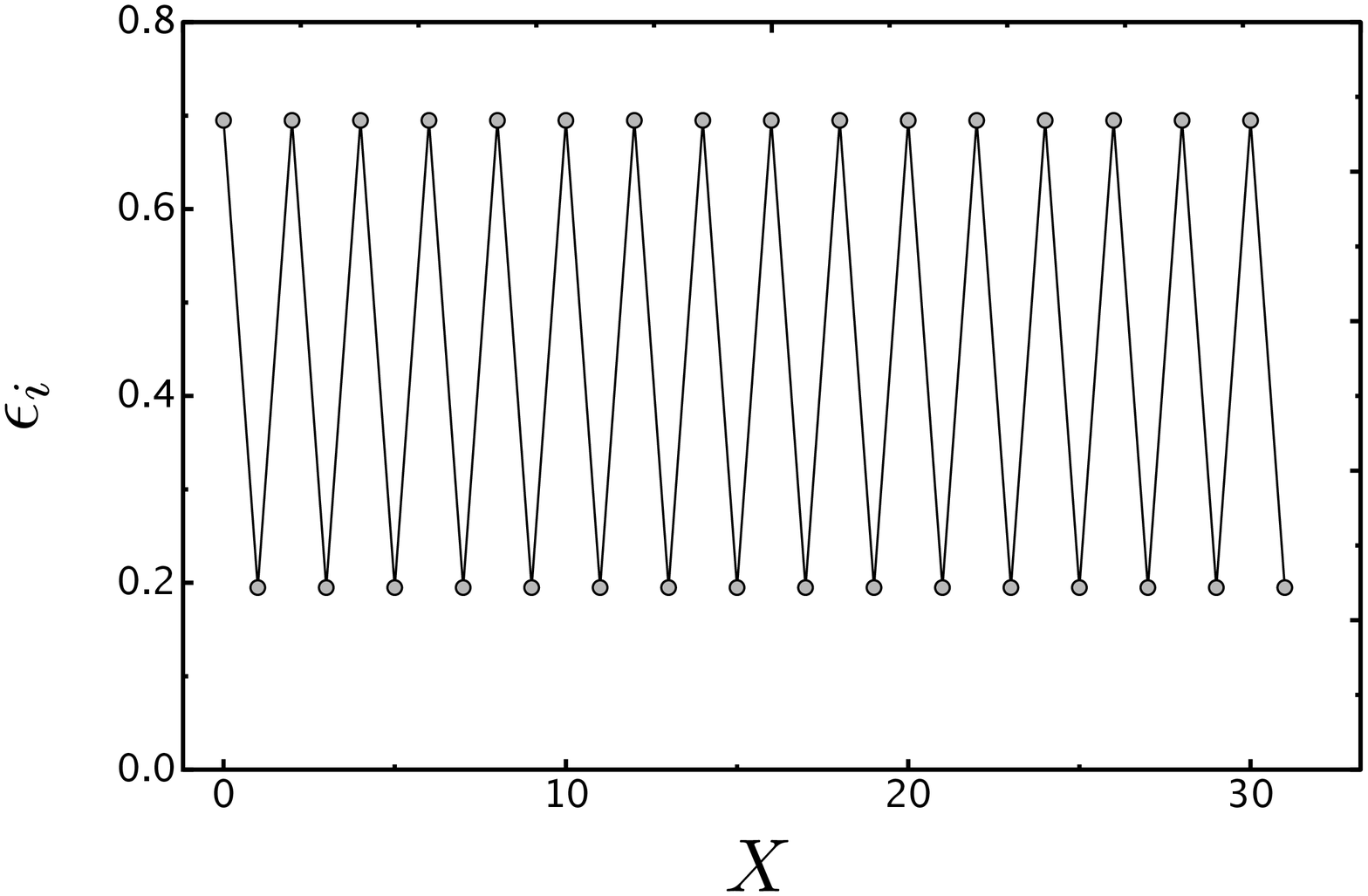}}
\end{center}
\caption{\label{Sp_f}Strain profile evolution in the driven  NNN model  with  $\lambda=0.01$. (a) Propagation of the binary phase along the first half of the yielding plateau; (b) homogeneous binary phase with lattice scale oscillations on the elastic step separating  two yielding regimes in Fig. \ref{SS1_f_all}.}
\end{figure}
The salient feature of the macroscopic response in the anti-ferromagnetic NNN model is the disappearance of the nucleation peak. One can see that on the first half-plateau of the strain-stress curve (Fig. \ref{SS3_f}), the  homogeneous phase is progressively replaced by the growing 'binary phase' and the corresponding front is shown in Fig. \ref{Sp2_f}. Once again, each drop in stress profile corresponds to one spring changing phase. Notice that the thresholds for the nucleation of the binary phase and for its propagation are similar.  By the end of the first half-plateau the binary phase occupies the whole domain and the step in the middle of the strain-stress curve (Fig. \ref{Sp3_f}) describes purely elastic deformation of the binary mixture. Along the second half-plateau the mixture phase is gradually replaced by the second phase and we again observe a propagating front. The dynamics of such fronts has been studied in \cite{Cahn:1999kx,PhysRevB.79.144123}.

From Fig. \ref{SS2_f} we see that 'ferromagnetic' model, the bulk contribution to the energy in the 'anti-ferromagnetic' model changes its sign. The main plastic phenomenology, however, remains the same even though the microscopic spatial arrangement of phases changes from phase separation to fine scale mixing.

For completeness, we also present here the results of the  simulations for the anti-ferromagnetic SG model with  $\beta<0$, see Fig. \ref{SS1_f1_all}. The strain profiles in this model are basically the same as in the NNN model with anti-ferromagnetic interactions ($\lambda < 0$), however, the energy profile shows a new feature: a negative slope in the first segment and a positive slope in the second.  This implies that the negative contribution to the cold work along the first half-plateau is compensated by the equal in magnitude positive contribution on the second half-plateau.
\begin{figure}
\begin{center}
\subfigure[]{\label{SS5_f}\includegraphics[scale=0.25]{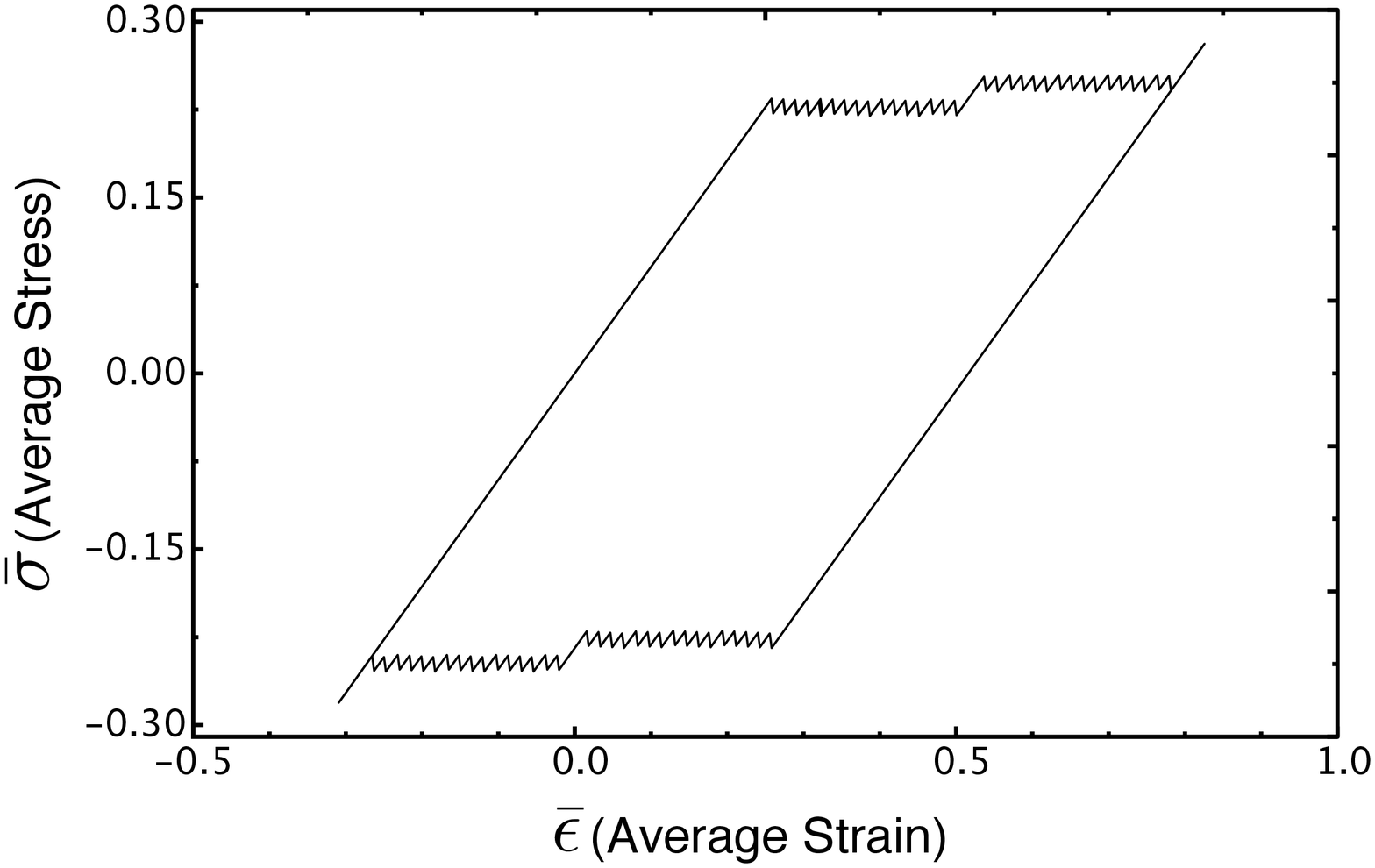}}
\subfigure[]{\label{E3_f}\includegraphics[scale=0.25]{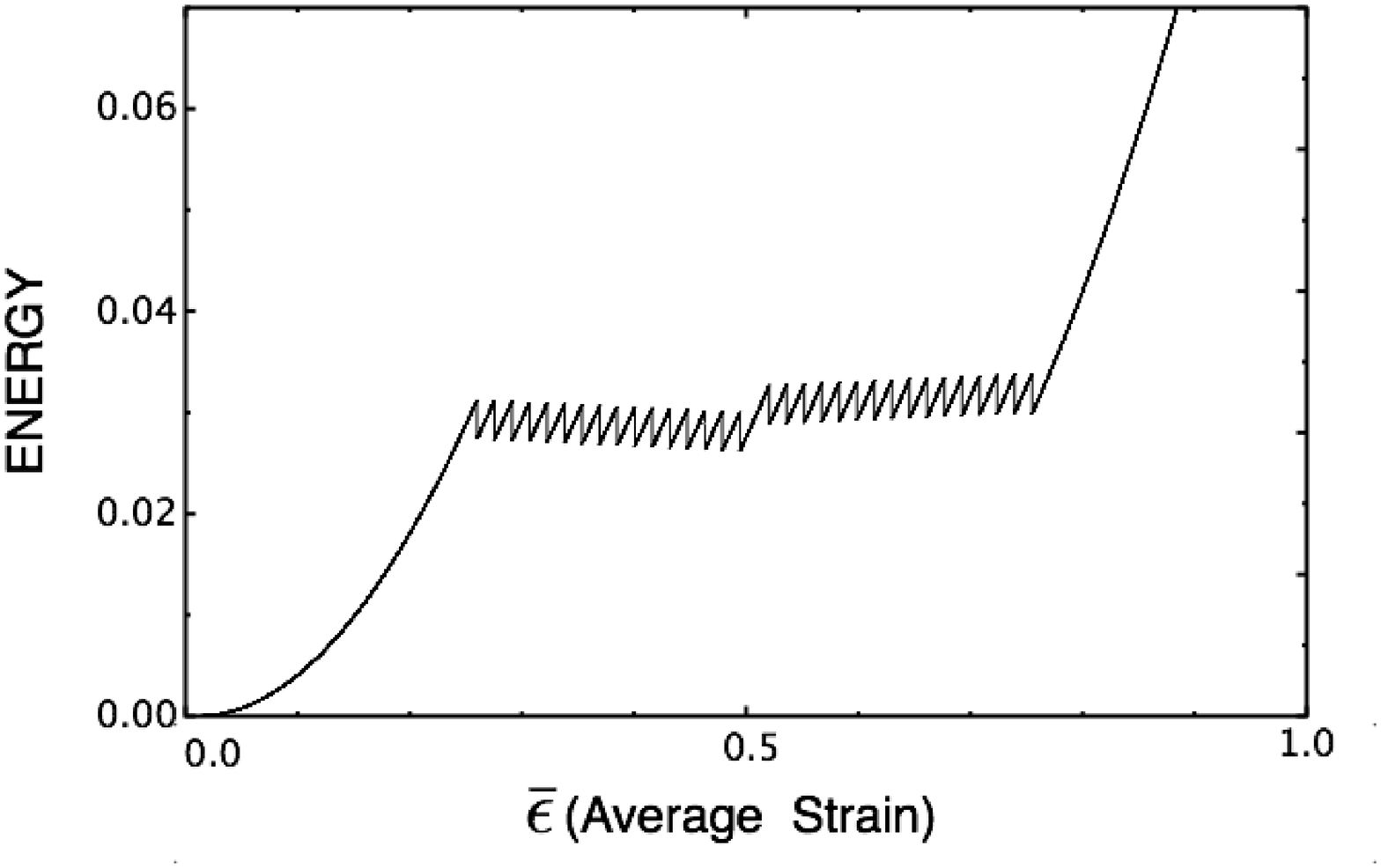}}
\end{center}
\caption{\label{SS1_f1_all}(a) Strain-stress relation in the driven  SG model with  $\beta=-0.2$, (b) evolution of the total energy at varying applied strain.}
\end{figure}

To summarize, the 1D anti-ferromagnetic NNN model and its SG analog describe rate independent dissipation and show a new feature which can be interpreted as a propagation of regular slip band microstructures. Still, both models predict highly coherent fluctuations structure and therefore none of them captures the complexity of the experimentally observed spatial and temporal correlations.

\subsection{Quenched disorder}

In modeling of realistic material response one cannot neglect disorder due to various chemical and mechanical imperfections at the level of a lattice. Such disorder can influence the macroscopic behavior of the system by facilitating inhomogeneous nucleation of defects on one side and by creating a variety of pinning obstacles for the propagating defects and defect microstructures, on the other side. In this subsection we show how the quenched disorder influences the macroscopic response in the basic NNN model with ferromagnetic interactions.

A  disorder can be incorporated into the energy (\ref{pwN}) in different ways, for instance, by randomizing the size of the barriers separating individual energy wells \cite{springerlink:10.1007/s001610200083}. In this paper, due to technical reasons, we follow a different approach originating from the RFIM model \cite{PhysRevLett.70.3347} where disorder is represented by a random field biasing different phases in different spatial points. The ensuing RSSM model (random soft/snap spring model \cite{PhysRevLett.101.230601}) is characterized by  the energy

 \begin{equation}
 \label{eq:nnne1}
  f(\epsilon_i)  =
  \begin{aligned}
&{}  \frac{\kappa}{2}(\epsilon_i-m_i)^2+ \frac{\lambda}{2}(\epsilon_i+\epsilon_{i-1})^2
-h_i\epsilon_i,
\end{aligned}
 \end{equation}
where $\lambda<0$ and $h_i$ are independent random numbers  distributed according to $$ P(x)=(\sqrt{2\pi\sigma^2})^{-1} \exp( -(x)^2 / (2\sigma^2) ).$$  In all our numerical illustrations we put  $\sigma^2=0.01$; we have  checked that the value of dispersion does not affect our statistical results even though it affects some  macroscopic features, for instance, the degree of hardening (see \cite{A.:2011kx}).
\begin{figure}[h!]
\begin{center}
\includegraphics[scale=0.3]{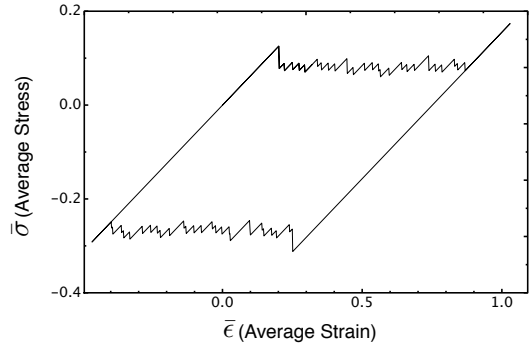}
\end{center}
\caption{\label{NNN_ss_dis} Strain-stress relation for the NNN ferromagnetic model with quenched disorder. Here $N=2048$, other parameters are the same is in similar graphs shown before.}
\end{figure}

A typical strain-stress relation for a particular realization of disorder is shown in Fig. \ref{NNN_ss_dis}. We observe that the nucleation peak persists, however the successive stress jumps due to phase boundary advances have  random amplitudes and each avalanche now involves different number of springs. Between the jumps the system is trapped/pinned in metastable configurations where it evolves purely elastically. The instabilities can be again associated with a succession of minimally stable states where at least one element have reached the spinodal state. Notice that the disorder in this model does not evolve and therefore the shakedown state is reached already during the first cycle.

\begin{figure}[h!]
\label{fig:dissip1d}
\begin{center}
\subfigure[]{\label{Di_fa}\includegraphics[scale=0.28]{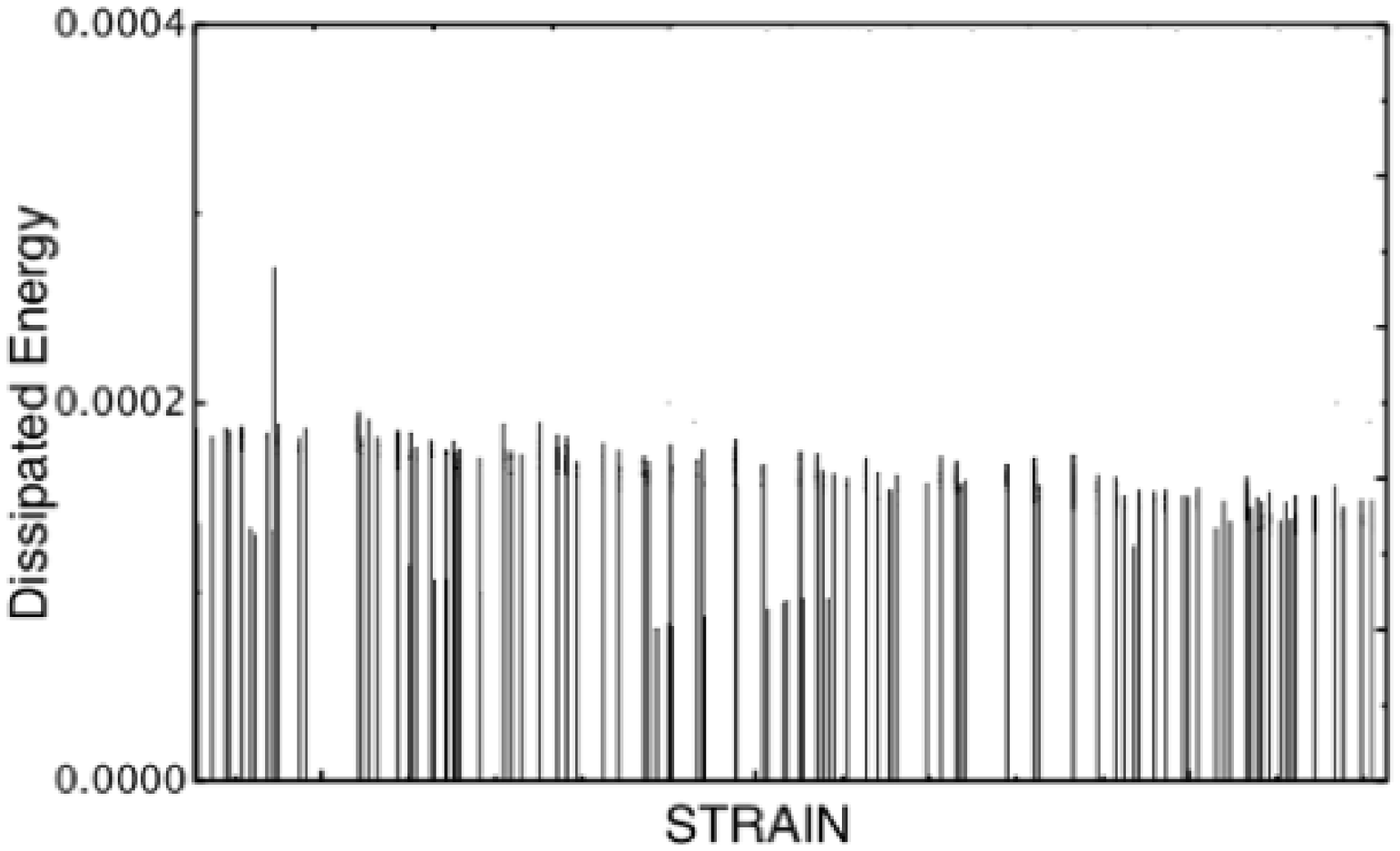}}\hspace{10mm}
\subfigure[]{\label{Di_fb}\includegraphics[scale=0.24]{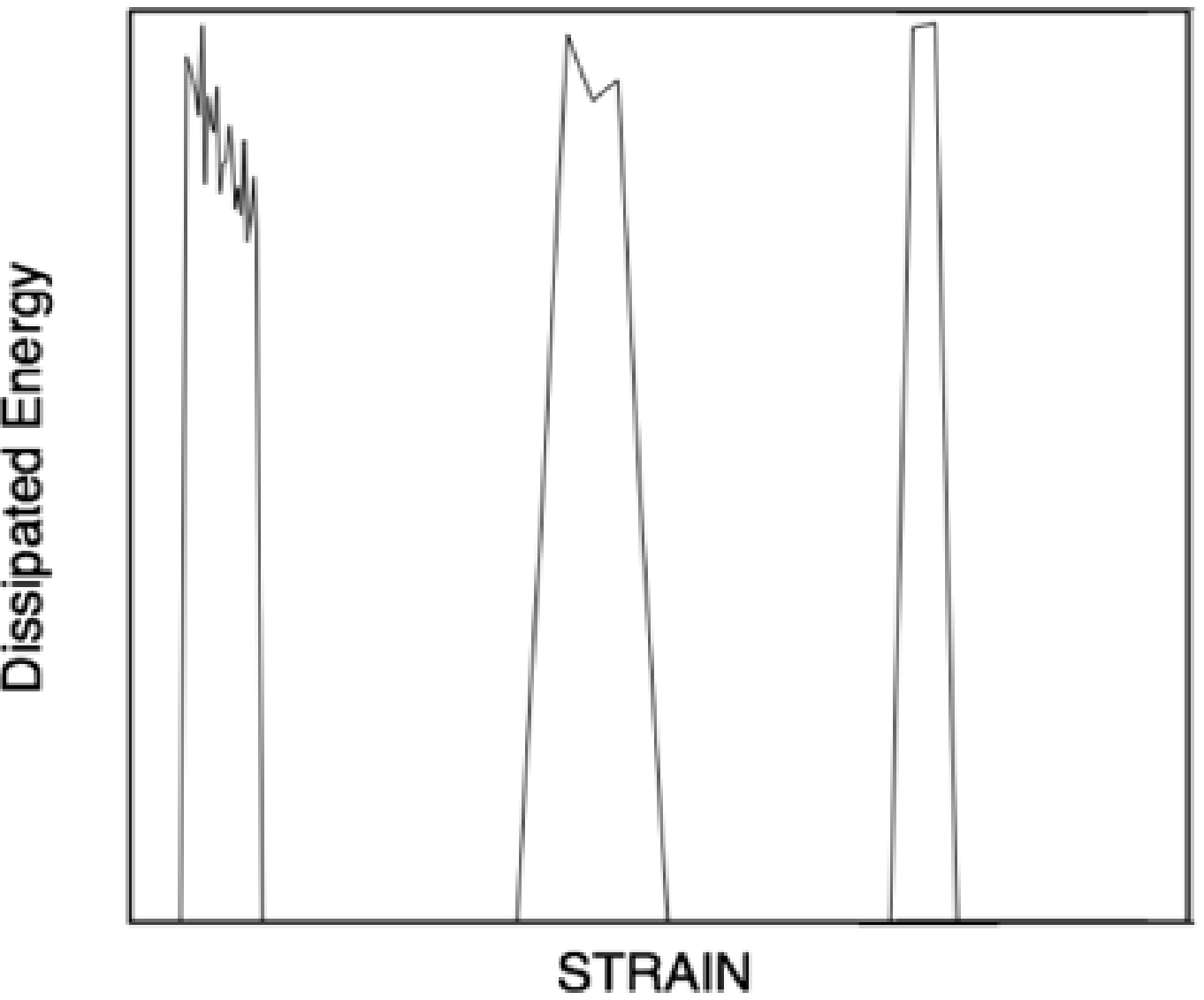}}
\end{center}
\caption{(a) Time series for the dissipated energy $E(t)$ during one loading cycle in the NNN model with quenched disorder shown in Fig. \ref{NNN_ss_dis} ; (b) blow up of three successive avalanches.}
\end{figure}

We observe that as in the case without disorder, each jump at time $t$ is associated with a finite dissipation which can be computed as the energy difference before and after the jump
\begin{equation}
\label{eq:dissip1dg}
E(t) = \sum_i (f_i (t+)- f_i (t-)).
\end{equation}
At a given $t$ the value of $E(t)$ may be either zero (elastic deformation) or positive (irreversible jump).  In a system with disorder a single jump, originating from a phase change in one unstable element, does not necessary lead to a stable state and therefore
an avalanche may  contain several elementary jumps. We have already encountered this phenomenon while dealing with nucleation requiring simultaneous transition in several elastic elements. In the case of nucleation, however,  we had to deal with massive transformation at one value of the loading parameter. In a disordered  system typically only one element becomes unstable at a given value of the load, however, several such jumps may take place before the next elastic stage is reached.

A typical time series for the dissipated energy  $E(t)$ is shown in Fig. \ref{Di_fa}. We observe  that jumps are separated by quiescent time intervals where $ E(t)=0$. The shape of several typical avalanches is shown in Fig. \ref{Di_fb} where we do not see any significant variation of time and length scales. Notice also that  avalanches  are very short, corresponding to 10-20 time steps. All this indicates that despite marginal stability of the system we record a succession of local events which fail to trigger  global rearrangements.
\begin{figure}[h!]
\begin{center}
{\includegraphics[scale=0.4]{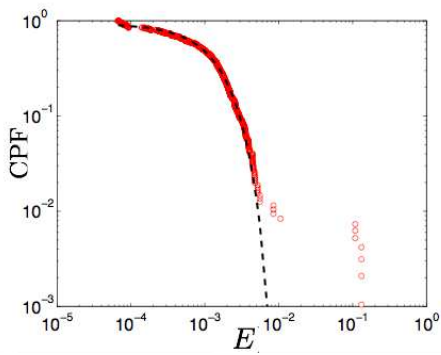}}
\end{center}
\caption{Log-log plot of cumulative probability function (CPF) for the dissipated energy  $E$  in the NNN  model shown in Fig.\ref{NNN_ss_dis} and Fig. \ref{fig:dissip1d} .  Dashed line is  a gaussian fit: $
P(x)=A \exp( -(x-\mu)^2 / (2\sigma^2) ),
$ with  $A=8.9078$, $\mu=-0.0071$ and $\sigma=0.0034$.}
\label{gauss}
\end{figure}

The total dissipated energy during an avalanche separated by two silent intervals can be written as $$  E = \sum_A E(t_A),$$ where  summation is over the avalanche duration. The cumulative probability distribution \cite{clauset:661}  of the variable $E$  is shown in Fig.\ref{gauss}. One can see that the distribution is close to Gaussian  if we exclude few large scale  events corresponding to macroscopic nucleation/annihilation. The fact that we do not record a  power law signal implies that this 1D system does not exhibit criticality and that outside highly synchronized nucleation events the avalanches remain largely uncorrelated.

\subsection{Explicit elimination of elasticity}

In the case when the elastic potential is piece-wise quadratic, one can  solve  the linear elastic problem analytically  and present the automaton explicitly in terms of spin variables. Given our periodic boundary conditions, it is natural to invert the linear elasticity operator in the Fourier space.
The discrete Fourier transform of a lattice field  $X_k$ is defined by
$
\hat X(q) =  1/N \sum_k X_k e^{-iqk} ,
$
where $ q$ is the 1D wave vector with  values ${q=2\pi j/N} $ , $j=1,...N$.  The inverse Fourier transform is given by
$
 X_k = \sum_{q} \hat X(  q) e^{iqk}.
$

Consider first the NN model. By transforming (\ref{eq:at1D}) with added quenched disorder into Fourier space we obtain
\begin{equation}
  \kappa  c(   q)\hat u (   q,t)  - s_x^-(  q) (\kappa\hat m (  q,t)+ \hat h (  q))  )   = 0,
\label{inh_prob}
\end{equation}
where $$  c(  q) = 2[\cos(  q)-1],$$ $$s^{-}_x (  q) =  (1-\cos (   q) + i\sin (  q)).$$
The field  $\hat m(  q,t)$  is the Fourier transform of the discrete spin field $m_i$ and the field  $\hat h ( q)$ is the transform of the random field $h_i$; both fields represent sources of elastic strains. The difference between these two sources is that the field $\hat h ( q)$ is fixed (quenched disorder), while the field $\hat m(  q,t)$  is evolving in accordance with the automaton rules which we formulate below.

First, we decompose our linear elastic field into a homogeneous part associated with prescribed time dependent affine deformation on the boundary and an inhomogeneous part which is a solution of a problem with periodic boundary conditions. More precisely, we assume that the $  q = 0$ component of the strain field is controlled by the loading while the $  q \ne 0$ component can be obtained  by solving a periodic problem with given sources.

The expression for the non affine  component of  the displacement field can be obtained from (\ref{inh_prob})
\begin{equation}
\label{Fourierdisp}
\hat u (  q,t) = \frac{s_x^-(  q) (\kappa\hat m (  q,t)+ \hat h (  q))  }{\kappa c(  q)}.
\end{equation}
The corresponding strain field $\hat \epsilon^{NN}(  q)  = s_x^+ (  q)\hat u (  q)$   where 
$$s^{+}_x (  q) = - (1-\cos (   q) - i\sin (  q)),$$ can be written as
\begin{equation}
\label{Fourierstrain}
\hat \epsilon^{NN}  (  q) = \frac{s_x^+(  q)s_x^-(  q)   }{\kappa c(  q)}(\kappa\hat m (  q)+ \hat h (  q)).
\end{equation}
The affine component of the deformation field  $\epsilon_0 = 1/N\sum_i\epsilon_i$,  is  fully defined  by the loading conditions
\begin{equation}
\hat \epsilon_0  (  q) = \delta(  q)t.
\label{loading}
\end{equation}
The total strain is then equal to $$\hat \epsilon  (  q)=\hat \epsilon_0  (  q) + \hat \epsilon^{NN} (  q).$$
The same method can  be applied to the  NNN model and the SG models with the replacement of $\hat \epsilon^{NN}  (  q)$ by either $\hat \epsilon^{NNN}  (  q)$ or $\hat \epsilon^{SG}  (  q)$. We  obtain
\begin{equation}
\label{FourierstrainNNN}
\hat \epsilon^{NNN} (  q) = \frac{ s_x^+(  q)s_x^-(  q)   }{\kappa c(  q)-\lambda\sin^2(  q)}(\kappa\hat m (  q)+ \hat h (  q)).
\end{equation}
and
\begin{equation}
\label{FourierstrainGinz}
\hat \epsilon^{SG} (  q) = \frac{ s_x^+(  q)s_x^-(  q)   }{\kappa c(  q)-\beta [c(  q)]^2}(\kappa\hat m (  q)+ \hat h (  q)).
\end{equation}
Such elimination of the elastic fields allows one to formulate the condensed automaton model for the spin variable $\hat m (  q)$ in  explicit form.

\subsection{SG model with periodic potential}

As an application of the reduction method presented in the previous section we consider here a discrete chain with  periodic NN potential. Once again we circumvent  dynamic simulations and move directly  to the study of the automaton model corresponding to the inviscid limit $\nu\rightarrow0$.
\begin{figure}[h!]
\begin{center}
\subfigure {\includegraphics[scale=0.2]{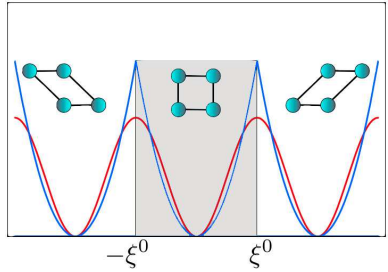}}
\end{center}
\caption{ \label{Fig_pw} Periodic potential and its piece-wise quadratic approximation. One elastic domain is shadowed.}
\label{atomic111}
\end{figure}

We chose the periodic potential to be  piece-wise quadratic (see Fig. \ref{atomic111}) and assume that in one period
 \begin{equation}
 \label{periodicN}
   f(\epsilon_i)  =
  \begin{aligned}
&{}  \frac{\kappa}{2}(\epsilon_i-(2\xi^0)d_i)^2,
\end{aligned}
 \end{equation}
where $d=0,\pm1,\pm2,...$ is a new integer-valued spin variable which describes quantized slip; each value of $d$ defines a quadratic energy well  with a period $(2\xi^0)$. At $d_i$ field given, the elastic strains $e_i$  must satisfy
\begin{equation}
 \label{periodic1}
 (d_i-1/2)(2\xi^0)\leq\epsilon_i\leq(d_i+1/2)(2\xi^0).
  \end{equation}
To regularize the NN model we add to the elastic energy the ferromagnetic SG term  as in (\ref{pwN}) and introduce a quenched disorder as in (\ref{eq:nnne1}).
The elastic field can now be decomposed into three components
\begin{equation}
\label{inverseFourierstrain}
\epsilon_i =  \epsilon_0 + \epsilon^h_i+ \epsilon^d_i = \{\hat \epsilon_0  ( q) \}_i^{-1}  + \{\hat \epsilon^{h}  ( q) \}_i^{-1}+\{\hat \epsilon^{d}  ( q) \}_i^{-1}.
\end{equation}
Here the affine component $\hat \epsilon_0  (  q)$ is given by (\ref{loading}); the Fourier transform of the residual strain due to quenched disorder can be written as
\begin{equation}
\label{FourierstrainGinzper}
\hat \epsilon^{h} (  q) =  \hat L( q)\hat h ( q).
\end{equation}
where the Green's function is
\begin{equation}
 \label{eq:kernel1d}
  \hat L( q) =\frac{\kappa (2\xi^0)}{\kappa + 2\beta -2\beta \cos( q)},
\end{equation}
and finally the Fourier representation of the strain field due to plastic slip is
\begin{equation}
\label{FourierstrainGinzper}
\hat \epsilon^{d} ( q) = \hat L( q)\hat d ( q).
\end{equation}
Now, according to  (\ref{periodic1}), a  metastable configuration with a given distribution $d_i$ is defined (is stable) within the following limits
\begin{equation}
 - \xi^0-t-  \{\hat \epsilon^{h} ( q)\}_i^{-1}  \leq  G_i(d)\leq \xi^0-t- \{\hat \epsilon^{h} ( q)\}_i^{-1} ,
\end{equation}
where
\begin{equation}
G_i(d)=   \{ \hat L(  q)\hat d ( q)\}_i^{-1}-(2\xi^0)d_i.
\end{equation}
\begin{figure}[h!]
\begin{center}
\subfigure[]{\label{EK1DA}\small\includegraphics[scale=0.6]{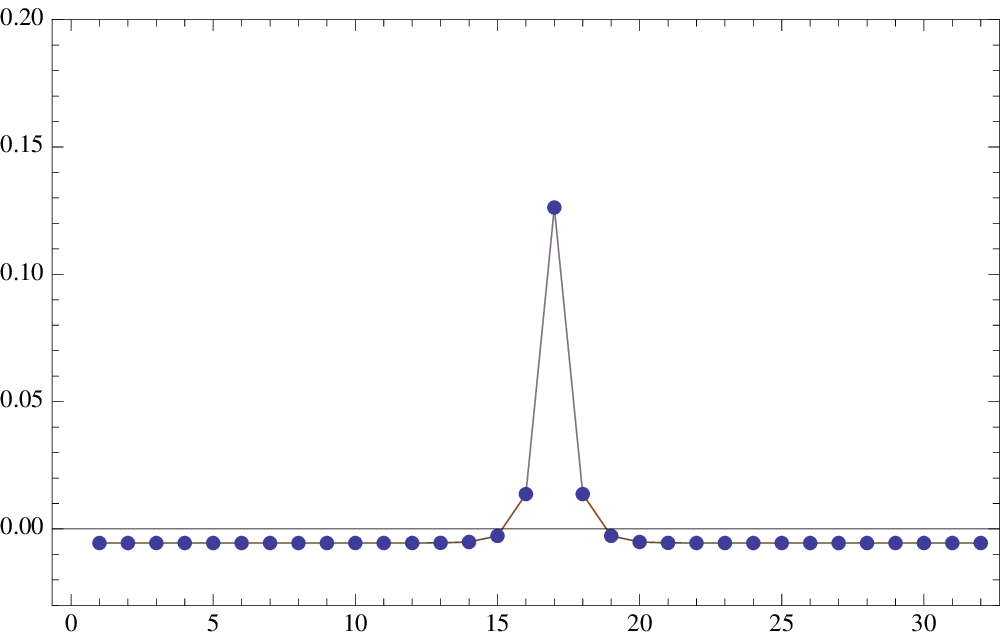}}
\subfigure[]{\label{EK1DB}\includegraphics[scale=0.6]{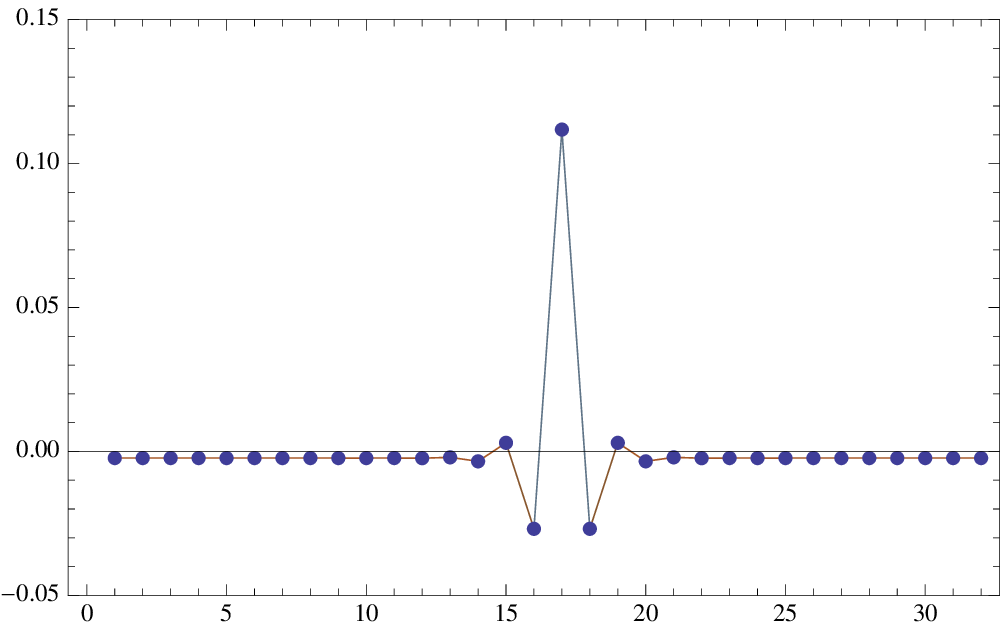}}
\end{center}
\caption{ Elastic kernel (\ref{eq:kernel1d}) with  (a) $\beta$=$ 0.2$ (ferromagnetic SG model) and (b) $\beta$=$-0.2 $ (anti-ferromagnetic SG model).}
\end{figure}
When $G_i(d)$ reaches one of the   thresholds, the integer parameter  $d_i$ must be updated
\begin{equation}
\label{M_field}
d_i\rightarrow d_i + M_i(d),
\end{equation}
where
\begin{equation}
\label{M_field}
   M_i( d ) =\left\{
  \begin{aligned}
&{}  +1,\hspace{1mm}\text{if} \hspace{1mm} G_i(d)> \xi^0-t- \{\hat \epsilon^{h} (  q)\}_i^{-1},\\
&{} -1, \hspace{1mm}\text{if}\hspace{1mm} G_i(d)< - \xi^0-t-  \{\hat \epsilon^{h} ( q)\}_i^{-1}, \\
&{}    \hspace{0.5cm} 0 \hspace{1mm}\text{otherwise.}
\end{aligned}
\right.
\end{equation}
\begin{figure}[h!]
\begin{center}
\subfigure[]{\label{sinnobeta}\small\includegraphics[scale=0.28]{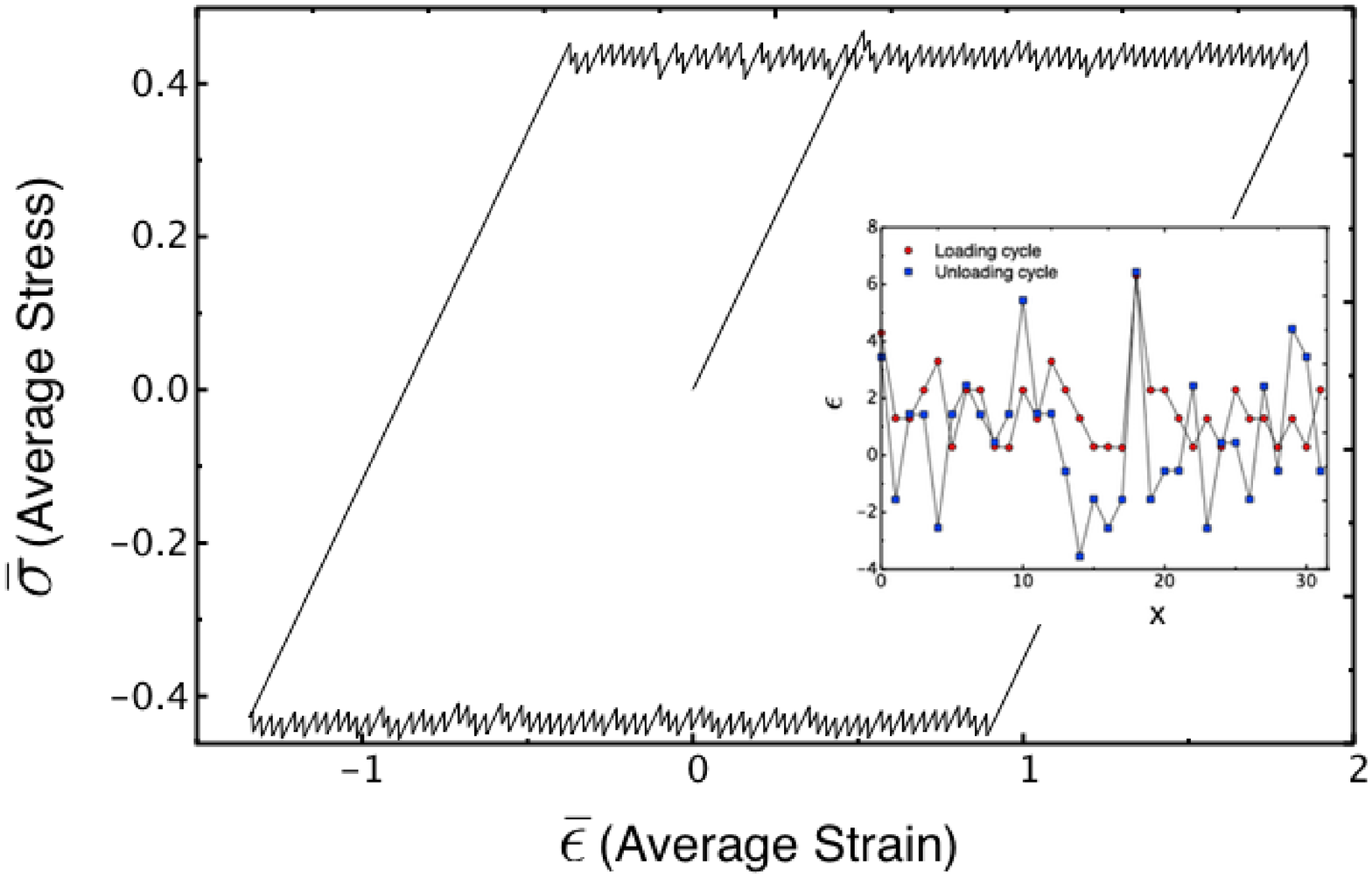}}
\subfigure[]{\label{sinbeta}\small \includegraphics[scale=0.28]{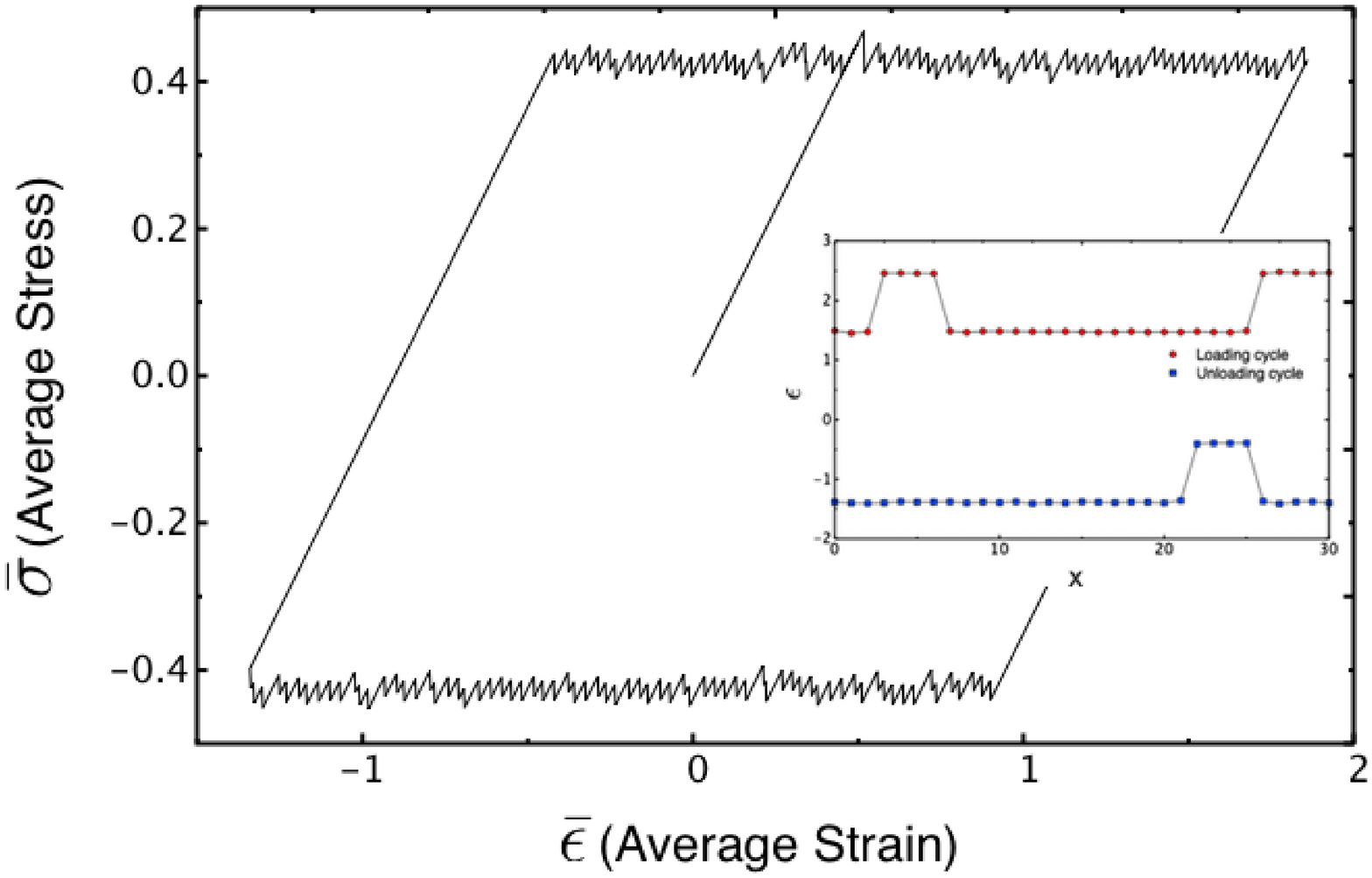}}
\end{center}
\caption{ Stress-Strain relations in two models with periodic potential and quenched disorder (a)$\beta$=$ 0$ (NN model)  and (b)$\beta= 0.01$ (SG ferromagnetic model).  Inserts show the corresponding strain profiles. }
\end{figure}
In a driven system an initially stable distribution of slip $d_i$  can get destabilized  because the thresholds evolve with time. The update of the slip $d_i$ given by
(\ref{M_field}) leads to a redistribution of the elastic strains  $ \epsilon_i^d$.   In Fourier space the corresponding 'discharge rules' take the form
$$
\hat \epsilon (q) \rightarrow  \hat  \epsilon (q) + \hat L(q) \hat M(q) .
$$
%
In Fig. \ref{EK1DA} we show the  physical image $L(x)$ of the discrete kernel $\hat L(q)$ for $\beta > 0$ (ferromagnetic SG model) and $\beta < 0$ (anti-ferromagnetic SG model).  One can see that if  an elementary positive slip takes place in an element, the elastic strain is increased in several neighboring elements and this 'influence domain'  increases with absolute value of $\beta$. In NN model  where $\beta = 0$  the strain of only the unstable unit increases  while the strain of all another units decreases  which is a typical mean field behavior. In the regularized model with $\beta \neq 0$ the local response is more complex as one can see in Fig. \ref{EK1DB}.

One of the crucial points in the automaton formulation is the choice of the update strategy when several elements become simultaneously unstable.  To avoid this problem we introduce slip dependent disorder $h_i(d)$ which signifies that random bias fields may be different in different energy wells. Physically this means that  imperfections may affect unevenly the repeated slips along the same slip direction. Similar assumption is usually made in the meso-scopic models of friction/amorphous plasticity where new random thresholds are assigned to the newly formed bonds, e.g. \cite{Bak:1991}.

%

In Figs. \ref{sinnobeta} and  \ref{sinbeta} we show the  strain-stress relations in the periodic problem with $\beta > 0$ (SG ferromagnetic model) and $\beta$=$0$ (NN model). In both cases, the macroscopic response is very similar: deformation is irreversible,  springs always remain close to  the marginal stability limit and the transformation proceeds intermittently through random avalanches.
However, the strain  profiles $\epsilon_{i}$ in the two models are different as  shown in the inserts in Figs. \ref{sinnobeta} and \ref{sinbeta}.  Thus, in the local model with $\beta=0$,
 the strain field is chaotic at the lattice scale due to the presence of disorder.  Instead,  in the nonlocal model which penalizes interfaces we see a more orderly (but still random) arrangement of the domains with different level of slip.  In both cases the statistical distribution of dissipated energy during
    avalanches remains Gaussian as in the case of transformational plasticity (a model with a double well potential).
\begin{figure}[!h]
\begin{center}
\label{PS1D1} \includegraphics[scale=0.30]{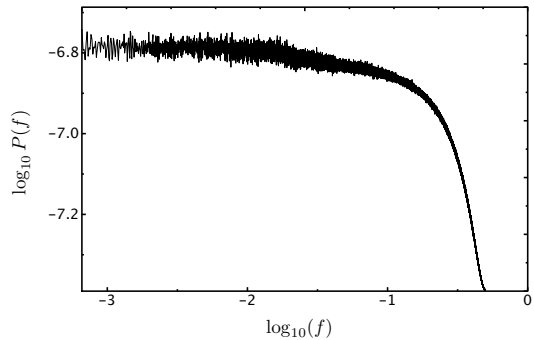}
\end{center}
\caption{ Power spectrum in the 1D  automaton model with periodic NN potential and SG ferromagnetic interactions.}
\end{figure}

It is also of interest to analyze the power spectrum of the signal $E(t)$ \cite{Jensen:1998qf,1742-5468-2005-11-L11001}
\begin{equation}
\label{eq:psf}
P(f) = \frac{1}{T}\biggr|\int_0^T dt  E(t)e^{-i2\pi f t}\biggl|^2,
\end{equation}
where $T$ is the total duration of the time series, $f$ is the frequency. The log-log plot of the power spectrum for the SG model with $\beta=0.01$  is shown in Fig. \ref{PS1D1}.
 We see the signature of a typical random process when the energy distribution is constant over a broad range in the low frequency domain.  The  rapidly decaying part of the spectrum at high frequencies is just a sign that the exerted power is finite.  This behavior suggests a short range of temporal correlations which is incompatible with criticality.  We recorded similar structure of the power spectrum in all other 1D models, with and without short range interactions.
\section{Two dimensional setting}

Even though our driven 1D models exhibited marginal stability, we could not detect in our numerical experiments any signs of scale free correlations. This may be associated with the simplicity of the geometrical structure of interaction which  prevents local events from triggering global rearrangements. As we have seen, this conclusion remain valid even in the presence of strong Gaussian disorder that is not sufficient in the 1D setting to generate avalanches with  power law statistics. It is then natural to try to augment the model by introducing an additional spatial dimension.

\subsection{The model}

The simplest 2D extension of the  models presented in the previous sections can be viewed as a series of 1D NN chains linked transversally by linear elastic springs (see Fig.\ref{atomic1111} ); the ensuing 2D lattice model can also be interpreted as an array of coupled FK chains \cite{suziki1988,Landau_dislo}.
\begin{figure}[h!]
\begin{center}
\subfigure {\includegraphics[scale=0.40]{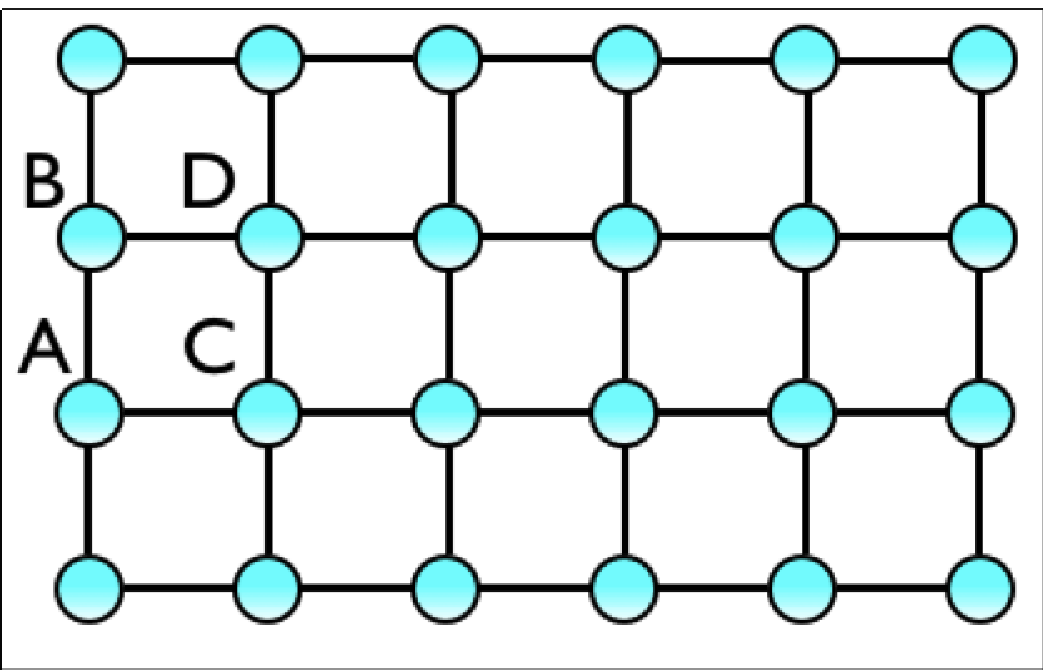}}
\end{center}
\caption{ The schematic representation of the 2D model. Shear springs AB and CD are described by periodic potentials, longitudinal linear springs AC and BD are described  by quadratic potential. Displacement field is horizontal.}
\label{atomic1111}
\end{figure}
Notice that in 2D the short range NNN or SG terms are not necessary because the nonlocal elastic interactions take place already in the NN model by virtue of an additional geometrical dimension. We also emphasize that while in  1D  we could discriminate only homogeneous slips (see Fig.\ref{atomic1}), in  2D  one can also describe the structure of the dislocation cores (see Fig.\ref{atomic2}).

We assign to each particle with coordinates $(i,j)$ defined on a $N \times N$ square lattice ($a=1$) a horizontal scalar displacement field $u_{i,j}$. We associate with linear longitudinal springs  the strain measure
 $\theta_{i,k}=u_{i+1,k}-u_{i,k}$
and with nonlinear shear springs - the strain measure
 $\xi_{i,k}=u_{i,k+1} - u_{i,k}.$
 The total energy of this highly anisotropic lattice can be written as
 \begin{equation}
\Phi(u)=\sum_{i,j} f(\theta_{i,j},\xi_{i,j}),
 \end{equation}
where the  potential $f$ is assumed to be quadratic in $\theta$ and periodic in $\xi$.  More specifically, we set
\begin{equation}
f(\theta_{i,j},\xi_{i,j})=  g (\xi_{i,j} )+ \frac{K}{2} (\theta_{i,j})^2 -h^1_{i,j} \xi_{i,j}-h^2_{i,j} \theta_{i,j}.
\end{equation}
where $g$ is a  periodic function (see Fig. \ref{atomic111}), in particular, in our dynamic (ODE based) model we use
\begin{equation}
g(\xi) = (2\pi)^{-2}(1-\cos(2\pi\xi)).
\label{periodic}
\end{equation}
The two lattice functions  $h^{1,2}_{i,j}$ are independent Gaussian random variables introducing quenched disorder.

 \begin{figure}[h!]
\begin{center}
\subfigure[]{\label{mm2}\includegraphics[scale=0.28]{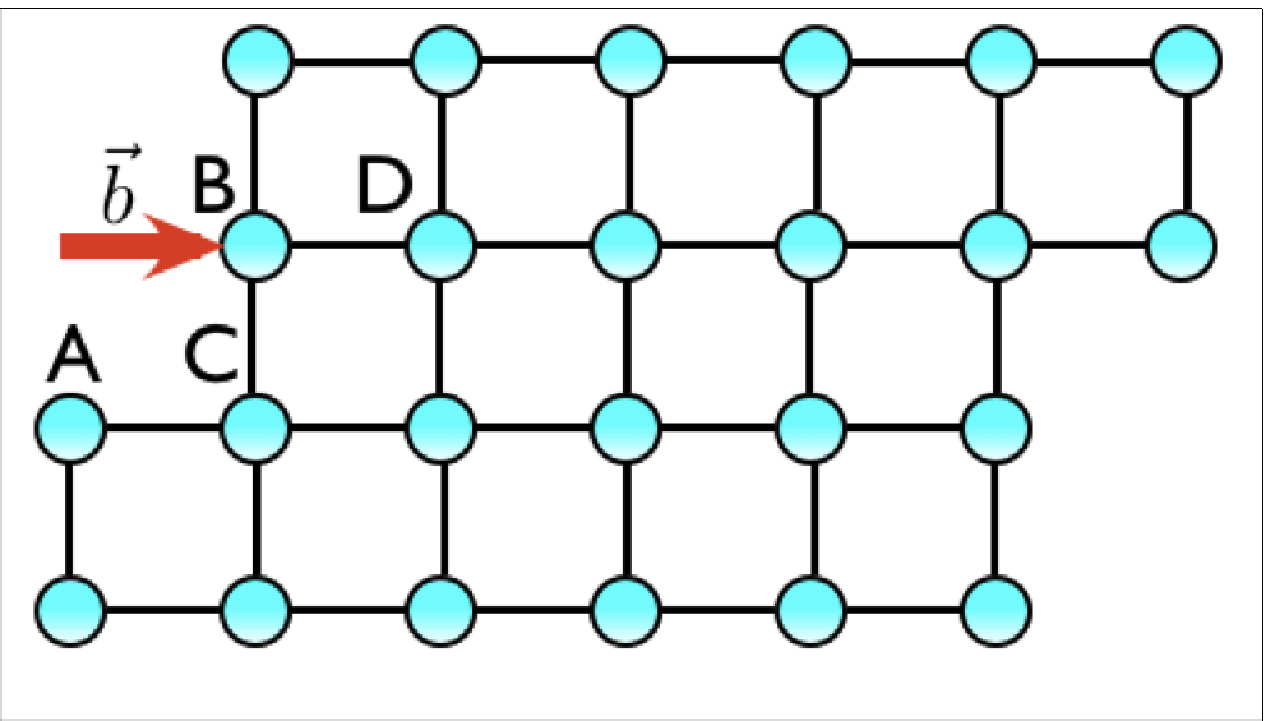}}
\subfigure[]{\label{mm3}\includegraphics[scale=0.29]{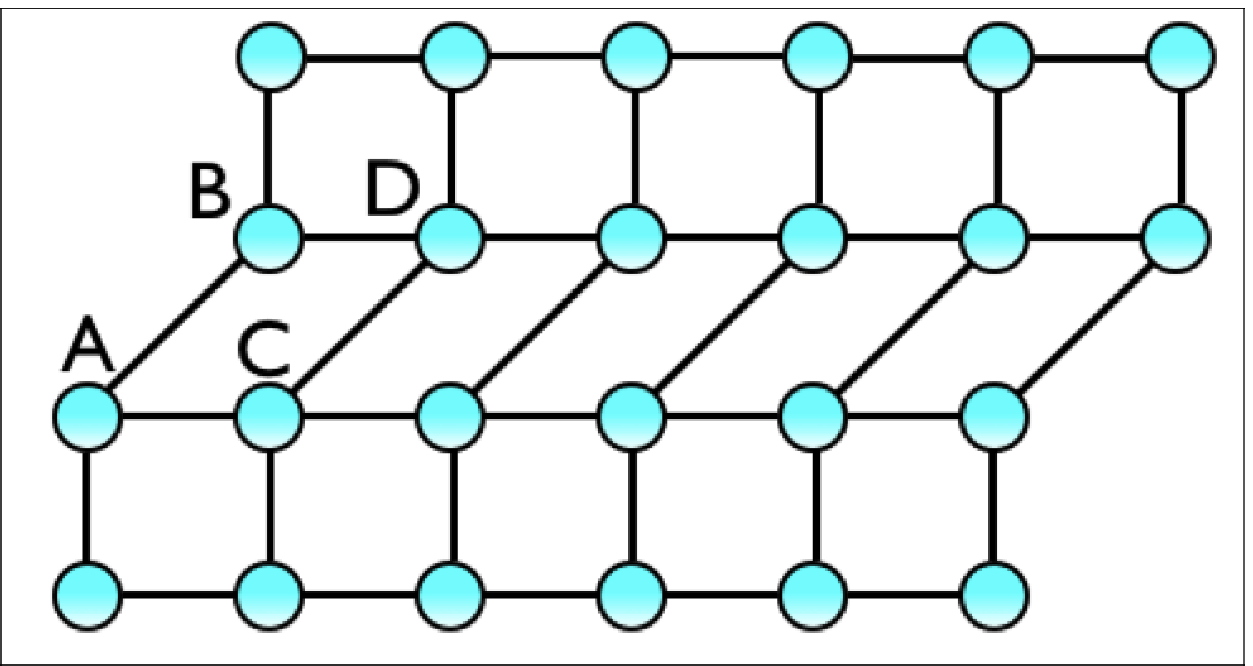}}
\end{center}
\caption{\label{atomic1} A quasi-one dimensional slip deformation in a 2D model: (a) with relabeling,  (b) without relabeling; $\textbf{b}$ is the Burgers vector}
\end{figure}

Observe that in the limit $K\rightarrow\infty$, we must have $\theta=0$ and our 2D model reduces to a one dimensional model with periodic potential which we have already discussed in the previous sections (see Fig. \ref{atomic1}). When $K<\infty$ the variables $\theta$ and $\xi$ are no longer independent and  the corresponding long-range interactions can be revealed by  minimizing out a linear variable $\theta$ which is lacking in the 1D model \cite{PhysRevB.52.803,Rasmussen:2001xi}.

An elementary slip in 2D can be illustrated already in the constrained model with $K=\infty$ if the adjacent atomic planes are displaced by one or several lattice spacings (Burgers vector) without changing the energy, see Fig. \ref{atomic1}. Within the classical interpretation,  one usually relabels  the nearest atoms after the slip takes place:  notice that the bond in Fig. \ref{m4} between the atoms $A$ and $B$  is broken, therefore,  a new bond has to be added between the atoms $B$ and $C$, which are now nearest neighbors. This approach gives rise to discontinuous displacement fields  and requires additional phenomenological hypotheses to deal with evolving discontinuities. In our  model we do not perform the relabeling and consider  compatible representation of the lattice before and after slip,  see Fig. \ref{mm3}. Notice that in this case, one cannot use linear elasticity and has to introduce a periodic potential which ensures lattice invariance with respect to finite shears.

If $K<\infty$ the  plastic flow does not occur by a slip of complete atomic planes. Instead the slipped atomic planes end  inside the crystal forming  dislocations. In   Figs. \ref{m4} and \ref{m5}  we illustrate the role of relabeling in the classical representation of dislocations and compare it with the compatible description of large shears adopted in this paper.
\begin{figure}[h!]
\begin{center}
\subfigure[]{\label{m4}\includegraphics[scale=0.22]{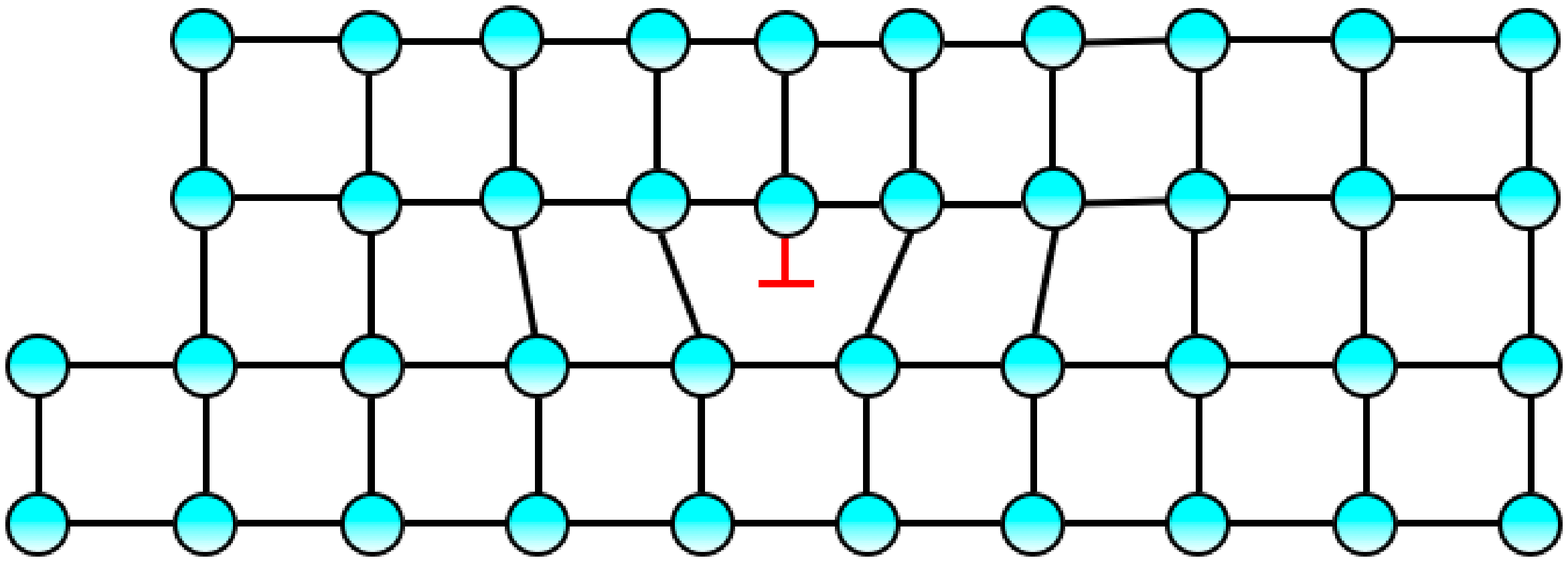}}
\subfigure[]{\label{m5}\includegraphics[scale=0.22]{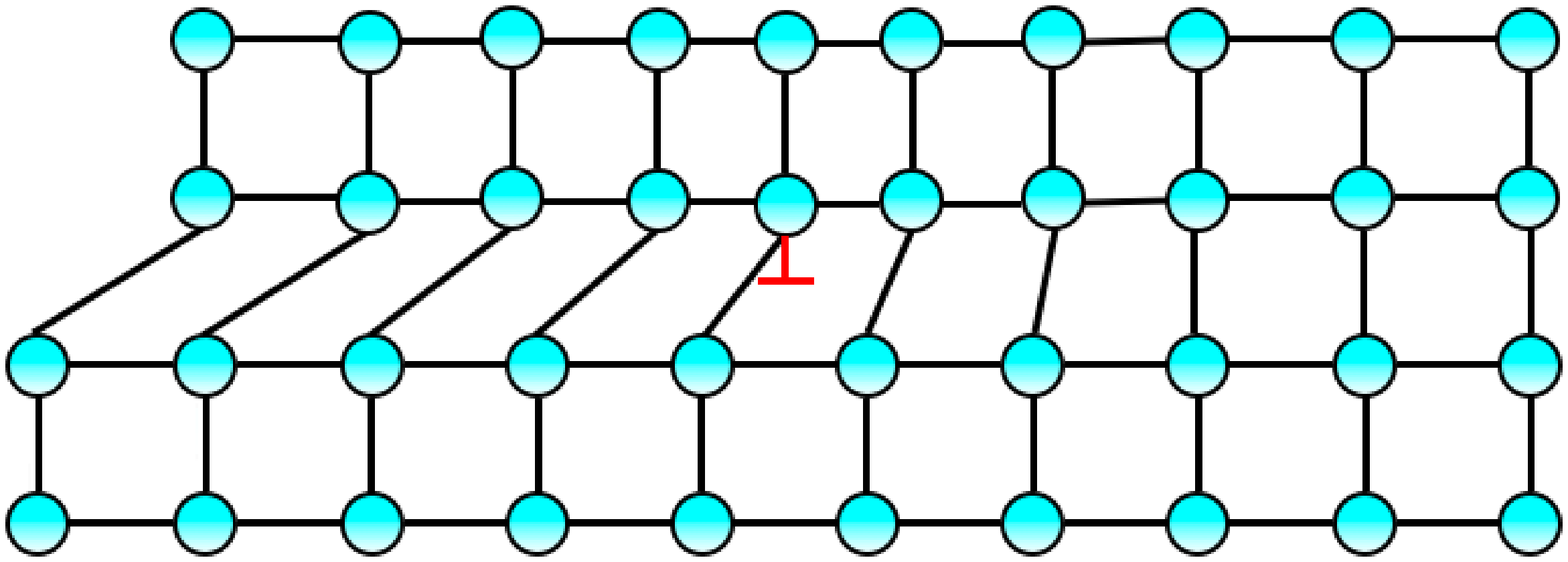}}
\end{center}
\caption{\label{atomic2}  A single dislocation in a 2D model: (a) classical model with relabeling or (b) compatible description without relabeling.}
\end{figure}

To complete the formulation of the problem we assume that the lattice is subjected to a time dependent shear in a hard device. In our framework driving with constant strain rate
can be presented as displacement controlled boundary condition $$u_{i,N-1} = u_{i,0}+t$$ for $i=0,N-1$. This boundary condition can also be rewritten in terms of the  overall shear strain as
$$\sum _{k=0}^{N-1} \xi_{i,k} =t,$$
where $t$ is again the slow time playing the role of loading parameter.
In longitudinal direction we assume the periodic boundary conditions given by $$u_{N-1,j}= u_{0,j}$$ for $j=0,N-1$. The overall  axial strain must then satisfy
 $$\sum_{i=0}^{N-1} \theta_{i,k} =0.$$

Observe that our elastic energy depends on two material parameters: $K$, representing the ratio of the bulk and shear moduli  \textit{and $\xi^0$, the ratio of the Burgers parameter} and the interatomic distance.  In addition, the behavior of the system may be affected  by the variance of the quenched disorder $\sigma$. The natural question is where in the space of parameters one can expect to see the critical regimes.  For large   $K$ the model becomes one dimensional, while for large $\xi^0$, the dislocations disappear. When $K$ is small the model again becomes one-dimensional and for small $\xi^0$ we lose lattice trapping. At very large disorder, dislocation mobility is very limited (POP regime, \cite{PhysRevLett.101.230601}) and the critical behavior can hardly be expected. On the other hand, at small disorder, one can expect synchronization (SNAP regime, \cite{PhysRevLett.101.230601}). The detailed reconstruction of the corresponding boundaries presents a formidable task which is outside the scope of this paper. A limited  parametric study shows robustness of the scale free behavior, in particular, we observed that the power law  statistics was disorder insensitive  in the interval $0.01-0.2$.

\subsection{\label{subsec1}The dynamic problem}

As we have seen in the 1D case, a quasistatic evolution of the driven system can be studied either in a dynamic setting with small but finite viscosity or in an automaton setting corresponding to the limit of infinitely slow driving. In the 2D setting we first formulate the dynamical model with small viscosity and smooth periodic potential (\ref{periodic}). Later we compare the results with the automaton model implying zero viscosity and operating with a piece-wise quadratic approximation of the periodic elastic potential.

To facilitate the Fourier representation we reintroduce the  finite difference operators $s_x^{+}$ and $s_y^{+}$
\begin{eqnarray}
\theta_{i,j}=s_x^{+}u_{i,j}=u_{i+1,j}-u_{i,j}, \,
\xi_{i,j}=s_y^{+}u_{i,j}=u_{i,j+1}-u_{i,j}.
\end{eqnarray}
Then the  axial and shear stresses  can be written as
\begin{eqnarray}
 \sigma^{xx}_{i,j} = Ks_x^{+}u_{i,j}, \,
 \sigma^{xy}_{i,j} = 2\pi^{-1}\sin(2\pi(s_y^{+}u_{i,j})),
\end{eqnarray}
and the equation of motion (\ref{eq_mov}) takes the form
\begin{equation}
\label{eq_mov_oper}
\nu \dot u_{i,j} = Ks_x^{-} \sigma^{xx}_{i,j} +  s_y^{-} \sigma^{xy}_{i,j},
\end{equation}
where
\begin{eqnarray}
s_x^{-} \sigma^{xx}_{i,j}  = \sigma^{xx}_{i,j}-\sigma^{xx}_{i-1,j}, \,
s_y^{-} \sigma^{xy}_{i,j}  = \sigma^{xy}_{i,j}-\sigma^{xy}_{i,j-1}.
\end{eqnarray}
In order to avoid numerically challenging non-local spatial derivative terms in (\ref{eq_mov_oper}),  we use the Fourier method.   The discrete Fourier transform of a 2D lattice function is defined by
\begin{equation}
\hat X(\bold{q}) = \frac{1}{N^2}\sum_k\sum_l X(k,l) e^{-i(q_k+q_l)},
\end{equation}
 where $\bold q=(q_x,q_y)=(\frac{2\pi k}{N},\frac{2\pi l}{N}) $ is the wave vector. By mapping our dynamic equation  (\ref{eq_mov_oper}) into Fourier space we obtain
\begin{equation}
\label{eq_mov_Fourier}
\nu \dot {\hat {u}}(\bold q ) = K\hat s_x^{-}(\bold q )  \hat \sigma^{xx}(\bold q ) +  \hat s_y^{-}(\bold q ) \hat \sigma^{xy}(\bold q )-\hat H(\bold q),
\end{equation}
where we defined
  $$\hat s^{-}_a (\bold q) =  (1-\cos (q_a ) + i\sin (q_a)), \, {a}=x,y,$$
and added an inhomogeneity due to quenched disorder
$$\hat H(\bold q) =   \hat s_x^{-}(\bold q )\hat h^1(\bold q)+ \hat s_y^{-}(\bold q )\hat h^2(\bold q).$$
To  solve (\ref{eq_mov_Fourier}) numerically we use a semi-explicit Euler time discretization. The axial strain $\sigma^{xx}$ is a linear term with respect to the displacement $u$ and we can treat  it implicitly in Fourier space. The non linear term $ \sigma^{xy} $ is first calculated in real space and then  transformed into Fourier space and treated explicitly. As a result, we obtain
\begin{equation}
\hat u^{t+1}(\bold q) = \frac{\hat u^{t}(\bold q) + \Delta t    \biggl\{ \hat s_y^{-}(\bold q ) \hat \sigma^{xy}(\bold q)-   \hat H(\bold q) \biggr\}}{1- \Delta t K \hat s_x^{-}(\bold{q})\hat s_x^{+}(\bold{q})},
\end{equation}
where 
 $$\hat s^{+}_a (\bold q) =  -(1-\cos (q_a ) - i\sin (q_a)), \, {a}=x,y,$$
and $\Delta t$ is the time step. After the field  $\hat u^t (\bold q)$ is known, the strains can be computed in Fourier space by simple multiplication.  To obtain the real space solution we need to invert the Fourier transform by using the formula
\begin{equation}
 X_{k,l} = \sum_k\sum_l \hat X(\bold{q}) e^{i(q_k+q_l)}.
\end{equation}

Notice that the FFT method automatically enforces periodic boundary conditions for the displacement field $u_{i,j}$. For instance, the   conditions  $\sum_{i=0}^{N-1} \theta_{i,k} =0$ are automatically satisfied. Driving in shear $\sum_{i=0}^{N-1} \xi_{i,k} =t$ can be again implemented by means of decoupling of the homogeneous shear strain component $\xi_0(\bold q)=t\delta(\bold q).$

\subsection{Numerical results in the dynamic problem}

In this subsection we present the results of our numerical experiments  with implicit-explicit FFT method described above. In our tests  we always take $K=2$ and  $N=512$ while our initial state is always dislocation free.
\begin{figure}[h!]
\begin{center}
\includegraphics[scale=0.22]{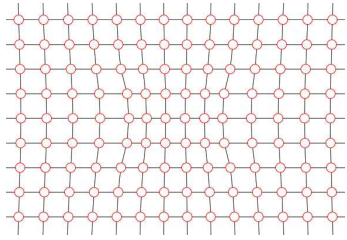}
\end{center}
\caption{\label{bc}  A  pre-strained zone in the form of an ellipse representing a local inhomogeneity and triggering dislocation nucleation.}
\end{figure}

As a first  illustration we show in  Fig. \ref{bc} the strain state due to internal prestress  $h^{1}_{i,j}=0.1$ inside an ellipsoidal inclusion.  In the driven system  such inhomogeneously stressed regions play the role of nucleation centers for dislocational dipoles (mimicking dislocation loops in 2D).
\begin{figure}[h!]
\begin{center}
\includegraphics[scale=0.22]{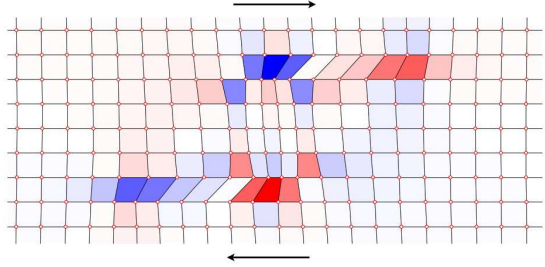}
\end{center}
\caption{ \label{atomica11}  A fragment of the deformed lattice with two dislocation dipoles nucleated under the applied shear strain around an imperfection shown in Fig. \ref{bc}).Red and blue colors indicate stress concentration corresponding to dislocations of different signs.  }
\end{figure}
In Fig. \ref{atomica11} we show  a fragment of the deformed lattice with two dislocation dipoles nucleated around an ellipsoidal imperfection as a result of the application of homogeneous shear strain on the boundary of the domain.   The local shear stress given by  $ \partial g(\xi)/\partial \xi$, is indicated by colors: red, for positive shear stress and blue for negative shear stress. One can see that the stress  reaches its maximum inside the four distinct dislocation cores. As the applied strain is increased the dislocations of different sign move in opposite directions. Eventually they reach the boundaries which means that two full atomic layer have slipped.
\begin{figure}[h!]
\begin{center}
\includegraphics[scale=0.50]{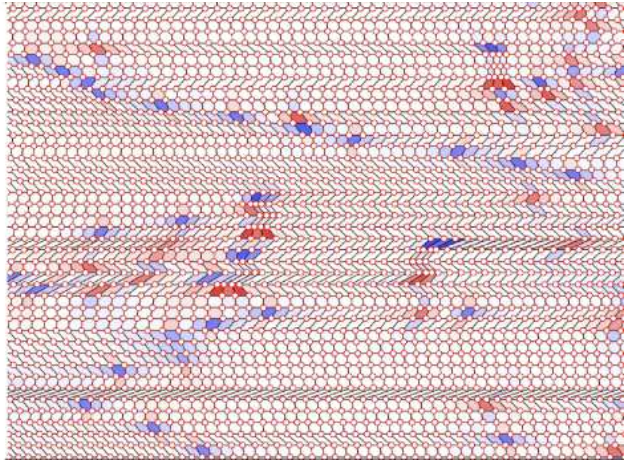}
\end{center}
\caption{ \label{atomicb11} A fragment of a generic dislocational configuration during a steady state simulated plastic flow. red and blue colors indicate stress concentration around dislocational cores of opposite signs. }
\end{figure}

In Fig. \ref{atomicb11}, we show a typical dislocational configuration in a crystal with random quenched disorder subjected to a finite shear. Once again the colors indicate the location of dislocations with different signs.  Here we already see several atomic planes that have experienced either singular or multiple slip. Notice also that the dislocational dipoles have a tendency to concentrate  around  regions with strong inhomogeneity.
\begin{figure}[!h]
\begin{center}
\includegraphics[scale=0.5]{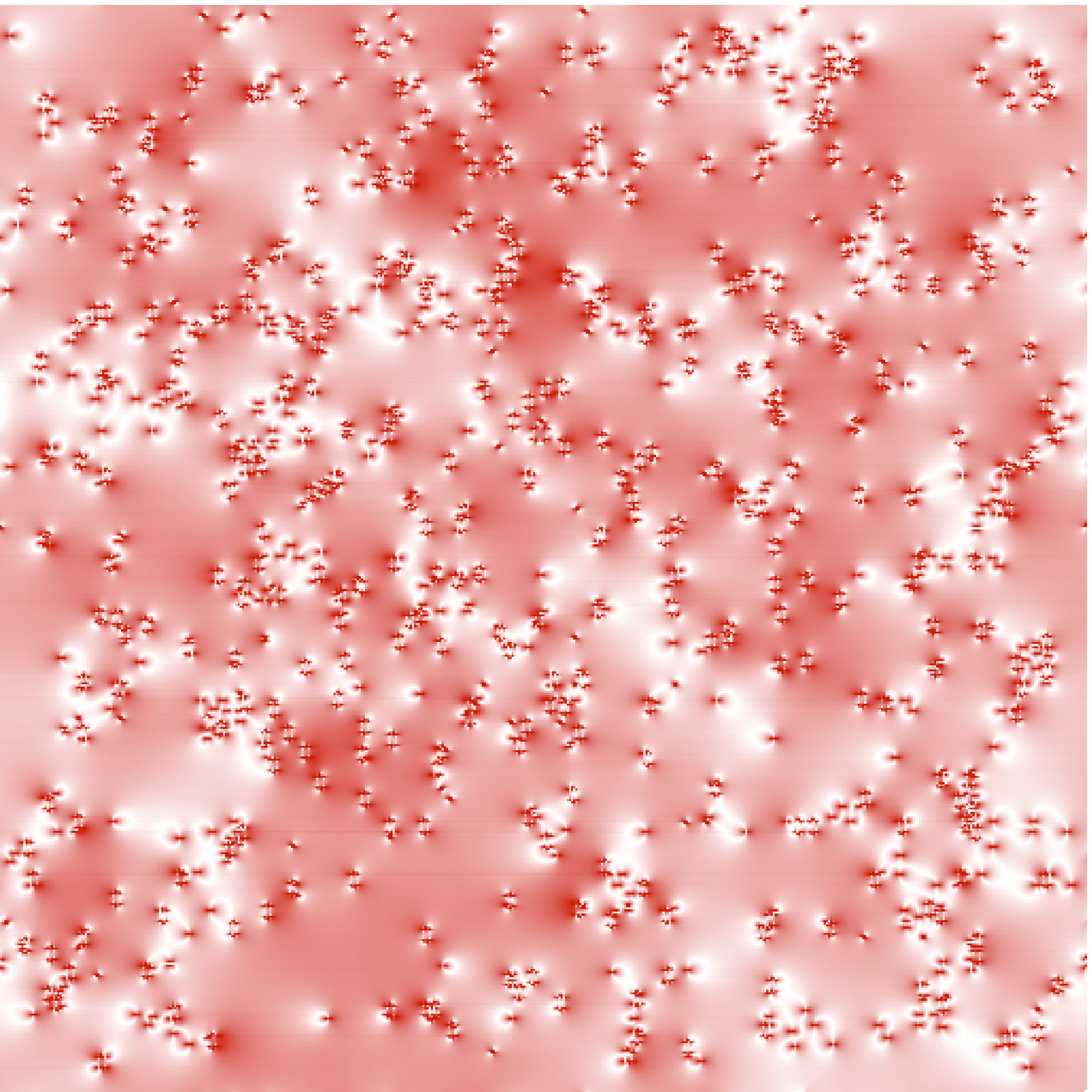}
\end{center}
\caption{\label{stress_field}Snapshot of the total stress field during plastic yielding. Higher color contrast corresponds to higher stress concentration.}
\end{figure}
The corresponding stress field at larger scale is shown in Fig. \ref{stress_field}. We observe that dislocations, which interact through long-range elastic fields, organize themselves into complex microstructures which appear random but turn out to be highly correlated. In our simulations we see that in the steady state regime plastic activity reduces to intermittent dislocational exchanges between stable clusters containing multiple dislocational dipoles. While these clusters remain mostly unchanged from one cycle to another, they have a finite lifetime as in experimental observations reported in \cite{jacobsen_science}.

In Fig. \ref{nucleation},  we show the evolution of the dislocation density during the first six cycles of loading/unloading which we compute by localizing and identifying individual dislocation cores. We observe an initial overshoot and the subsequent stabilization which indicates that the system has reached the shakedown state.  In  Fig. \ref{ss_curves} we present the corresponding macroscopic strain-stress curves exhibiting plastic hysteresis.   The size of the hysteresis loops decreases  with cycling, which indicates that the rate of dissipation diminishes as the system  get stabilized in the shakedown steady state.
 \begin{figure}[h!]
\begin{center}
\subfigure[]{\label{nucleation} \includegraphics[scale=0.24]{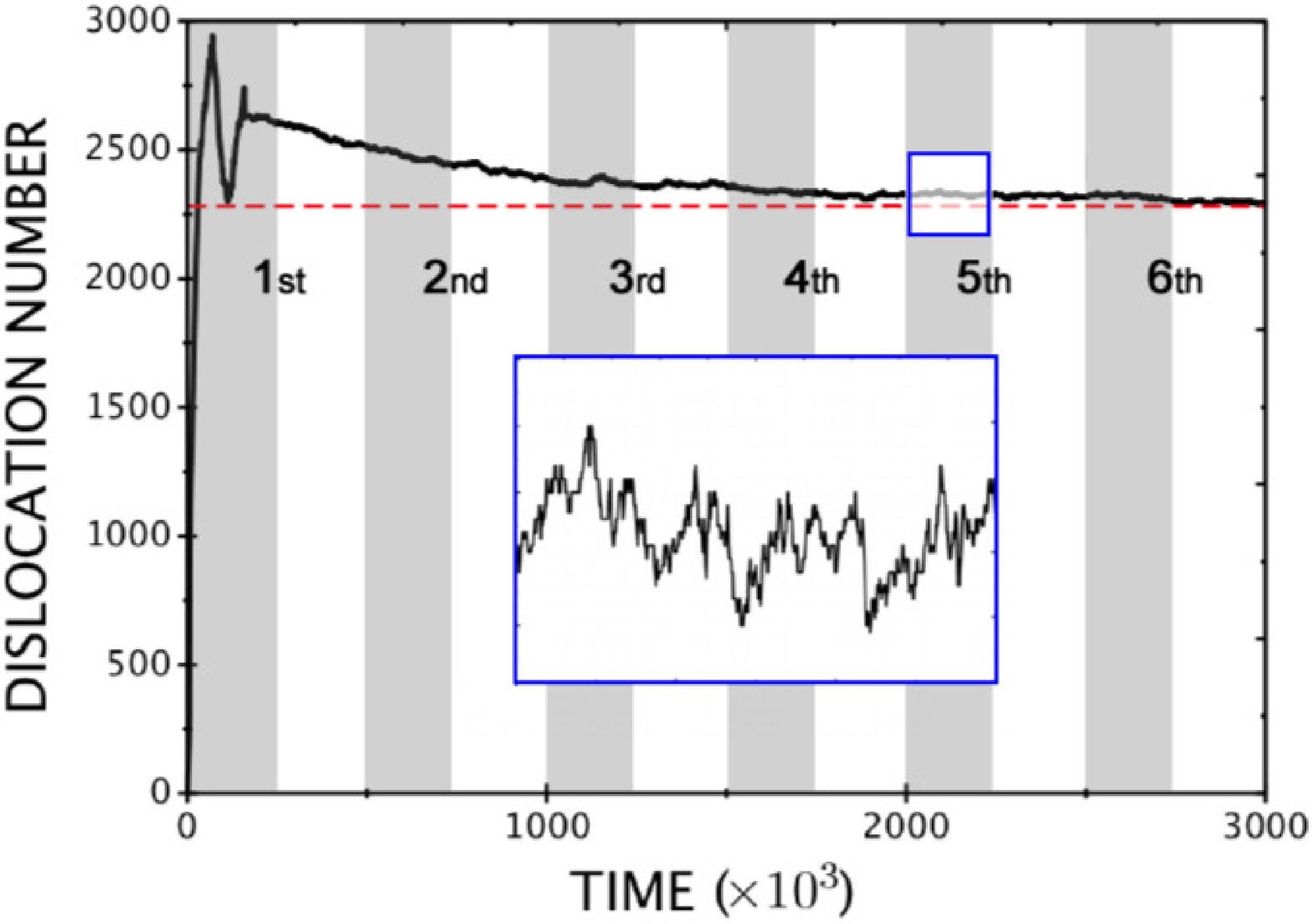}}
\subfigure[]{\label{ss_curves}\includegraphics[scale=0.85]{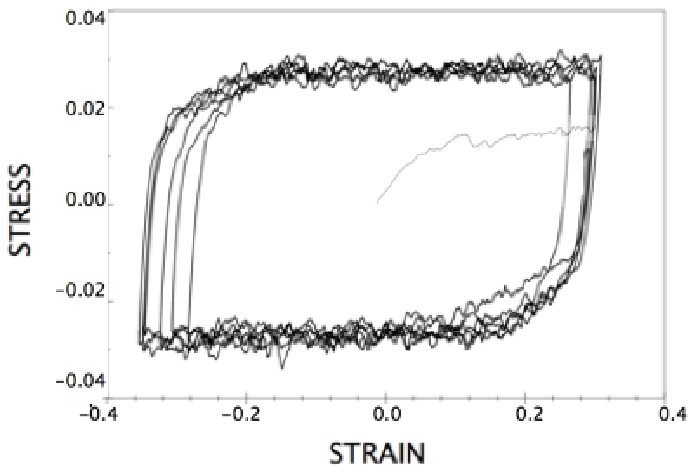}}
\end{center}
\caption{(a) Evolution of the dislocation density in the system subjected to  cyclic loading. The inset shows small scale fluctuations; (b) Macroscopic strain-stress hysteresis during the first six cycles under strain controlled loading conditions.}
\end{figure}

One can see that the macroscopic description of plasticity in our 2D model is quite realistic. At the microscale we see that the initial incipient inhomogeneity is augmented and modified during the cycling. This leads to the formation of statistically stable distribution of residual stress which is largely independent of the initial disorder.
\begin{figure}
\begin{center}
\subfigure[]{\label{dissipenergyfull}\includegraphics[scale=0.20]{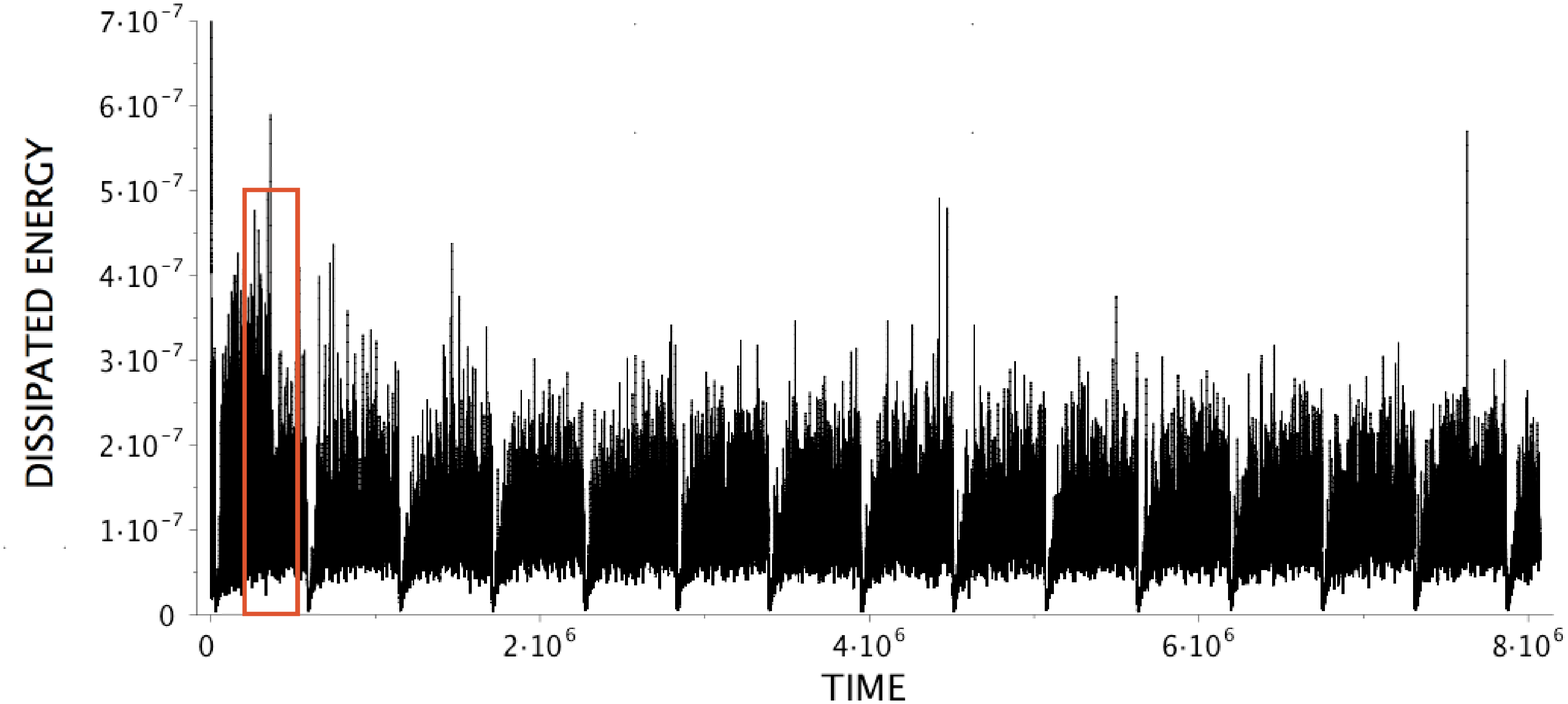}}
\subfigure[]{\includegraphics[scale=0.17]{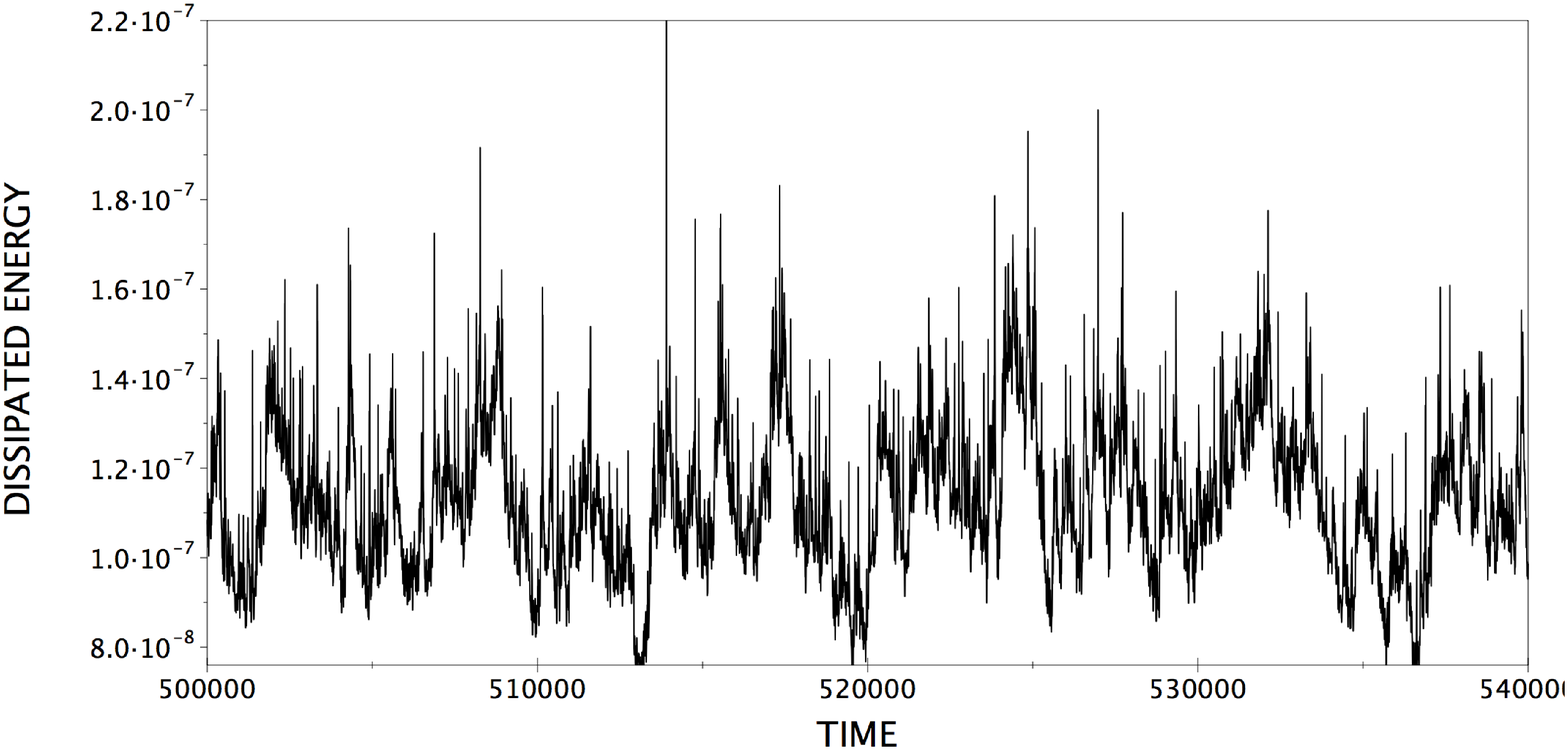}}
\end{center}
\caption{\label{dissip_t}  (a)  Time dependence of the dissipated energy during several cycles; (b) The same signal at shorter time scale (magnification of the highlighted region in (a)).}
\end{figure}

\subsection{Statistics of avalanches in the dynamic model}

The fluctuations associated with intermittent dislocational dynamics are detected in experiment through the acoustic emission (AE) \cite{Weiss:2000kx,PhysRevB.76.224110}. The  of intensity of the AE is accessed by measuring the square of the voltage which is expected to be proportional to the dissipated energy. Therefore by computing the rate of dissipation
\begin{equation}
E(t)=\nu N^{-2} \sum_{i,j}  \dot u(t)_{i,j}^2
\end{equation}
in our  simulations we can make comparison with experimental data. The typical time series for $E(t)$ is shown in Fig. \ref{dissip_t} at two different scales. One can see that the signal has a characteristic spiky appearance and that the complexity of the time series is not reduced with magnification which is an indication of scale free behavior.

To separate individual avalanches we introduce an irrelevant threshold and define the avalanche energy by integrating the dissipation rate over the duration of the avalanche $T$:
\begin{equation}
E = \nu N^{-2}\sum_{i,j} \int_t^{t+T} \dot u_{i,j}^2dt.
\end{equation}
The complexity of the intermittent signal is characterized by
 the probability distribution of avalanches $P(E)$ which in our case stabilizes after few first cycles. In Fig. \ref{in_final_cycle} we show the stationary distribution of avalanches exhibiting  a robust power law pattern for over 4 decades  with exponent $\epsilon\approx1.6\pm0.05$ obtained by the maximum likelihood method \cite{clauset:661}. This value of the exponent is in perfect agreement with experiments in ice crystals and fits the generally accepted range 1.4-1.6 \cite{Dimiduk:2006ys,Zaiser:2008sb,PhysRevLett.100.155502,RichetonBreakdown,Dimiduk:2006ys,Csikor12102007}. It is also consistent with the value obtained for 2D colloidal crystals \cite{Physics:2005zr} and with the value obtained in phase field crystal simulations \cite{ChanThesis}. Curiously, the Gutenberg-Richter law for earthquakes potencies/moments, interpreted as statistics of dissipated energy, gives a very close value of the exponent  $\epsilon\sim1.55-5/3$ \cite{Bak:1991,Benzion:2008}. This is not too surprising, however, in view of the essentially two dimensional nature of the fault friction/plasticity.
\begin{figure}[h!]
\begin{center}
\subfigure[]{\label{in_final_cycle}\includegraphics[scale=0.25]{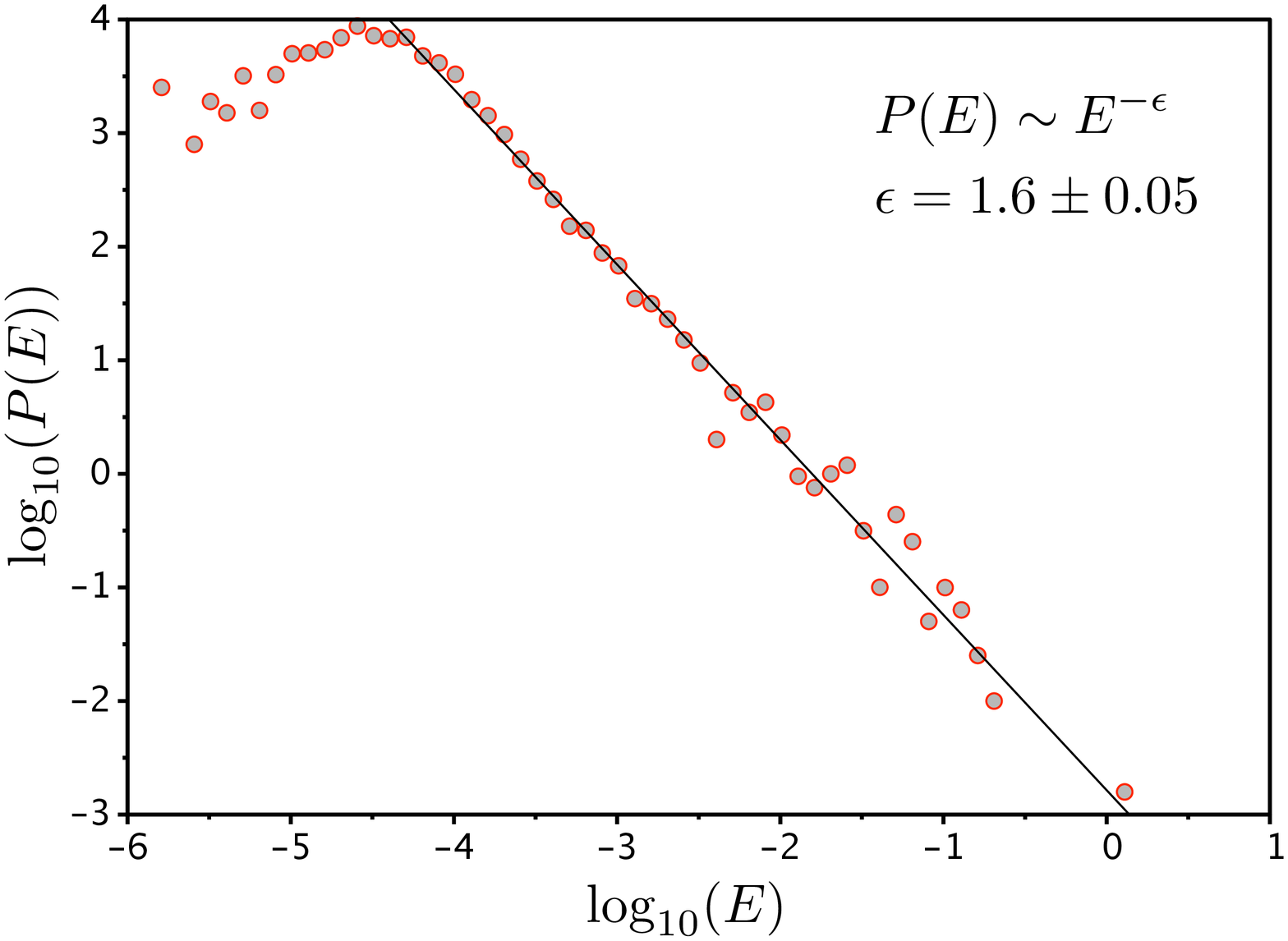}}
\subfigure[]{\label{din_full_cycle}\includegraphics[scale=0.25]{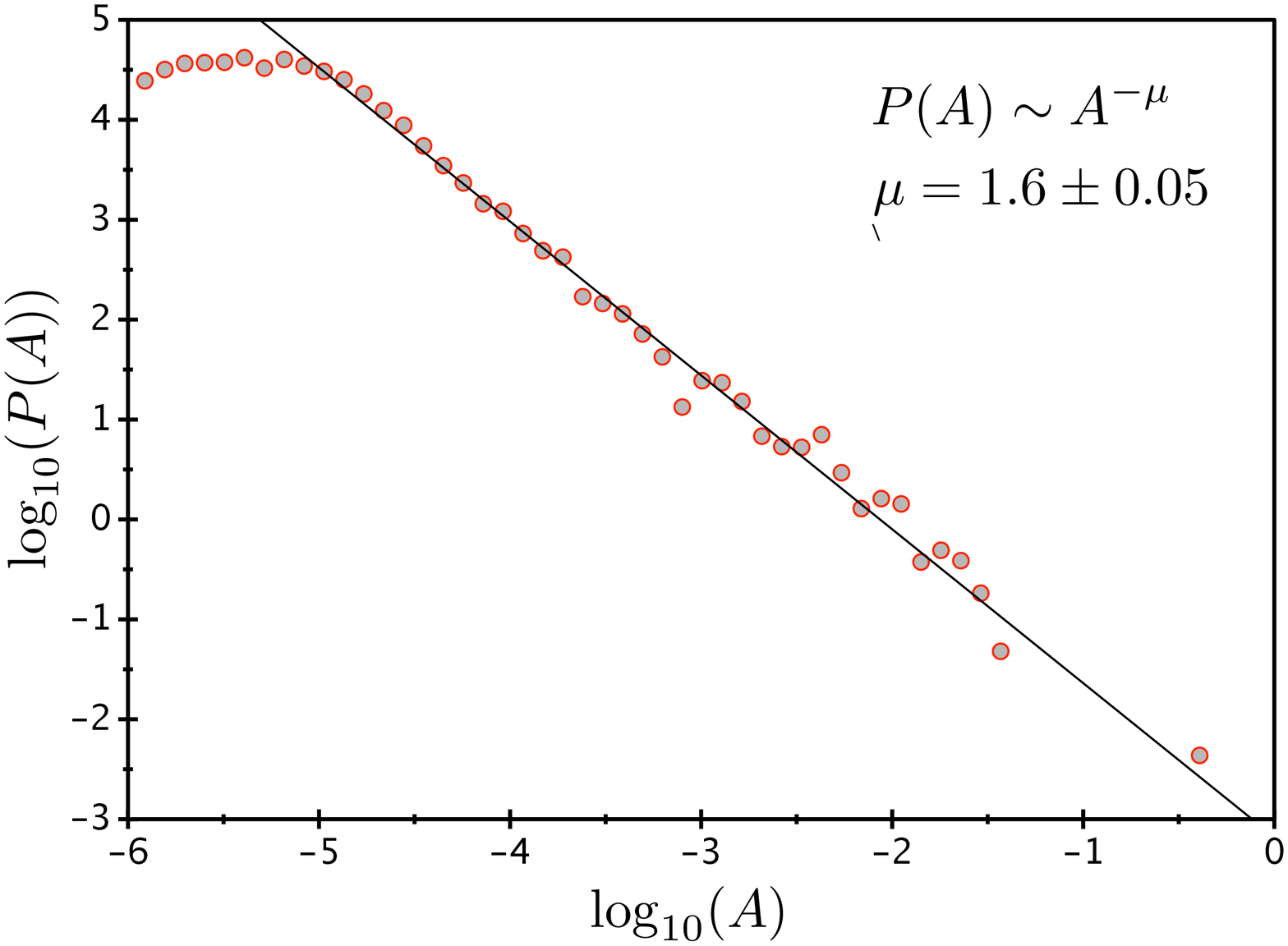}}
\subfigure[]{\label{dur_full_cycle}\includegraphics[scale=0.25]{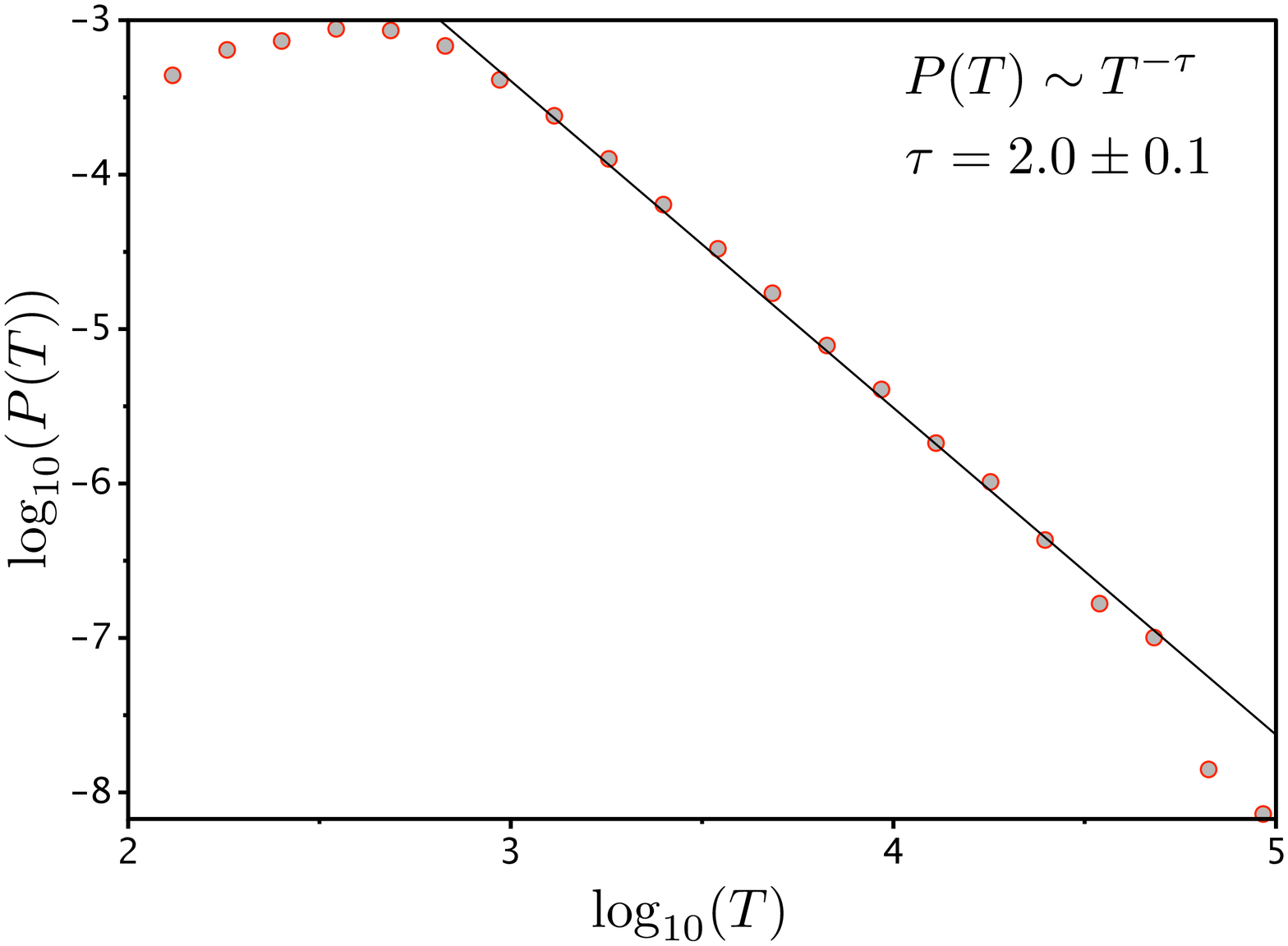}}
\caption{  Log-log plots of the probability density for (a) dissipated energy,  (b)  strain increments, (c) avalanche durations.}
\end{center}
\end{figure}

Due to very small scale of stress fluctuations the dissipated energy during an avalanche is approximately proportional
to the size of the total plastic slip before the system stabilization.  The probability distribution $P(A)$ of the  plastic strain increments
\begin{equation}
A= N^{-2}\sum_{i,j}\int |\dot \xi_{i,j}|dt,
\end{equation}
is shown in Fig. \ref{din_full_cycle} and,  as expected, we again find the power law distribution
 with exponent $\mu=1.6\pm0.05$.  In this graph the abscissa  is normalized by the system size, and therefore one can see that the biggest events traverse the whole system. On the other hand, the lower bound in our scaling region corresponds to the smallest plastic strain increment which scales with the unit cell size (viscous scale is even smaller). This suggests that the cut off have purely geometric nature which is a signature of a scale free regime.

In addition to energy it is also of interest to study the statistics of the avalanche durations \cite{PhysRevLett.77.1182,Perez-Reche:2007et,doi:10.1080/14786431003662572}. The results  shown in  Fig. \ref{dur_full_cycle} suggest a power law scaling  with the exponent $\tau=2.0\pm 0.1$, see Fig. \ref{dur_full_cycle}.  This value can be compared with measurements in  $LiF$ micro-crystals where  the duration exponents were found to be consistently bigger than $2$ \cite{doi:10.1080/14786431003662572}.

We now turn to the study of the power spectrum for dissipated energy $E(t)$ which carries information about temporal correlations not only between avalanches but also inside a single avalanche \cite{Hamon:2002fk}. The log-log plot of the computed power spectrum is shown in Fig. \ref{PSdissip}. It has a characteristic structure of the  $1/f^{\eta}$ noise with $\eta=1$.   Such signals,  also known as pink or flicker noise, are encountered in many  self organized critical systems ranging from astronomy to physiology \cite{PhysRevLett.59.381}, moreover the self organized criticality was originally proposed as an 'explanation of 1/f noise' \cite{Bak:1987}.

It is important to point out that the time series  obtained in our dynamical model result from slightly overlapping  avalanches. This finding  is consistent with the results of the numerical simulations in "running" sand pile models reported in \cite{Kardar.1992}, where the appearance of $1/f$  noise was explicitly linked to the avalanches overlap at finite driving forces and where a very close RG estimate  of the exponent $\eta=5/6$ in 2D case was obtained. Our results also agree with the scaling relation  $\eta=3-\tau$  derived in  \cite{0305-4470-23-9-006} under the assumption of  random superposition of individual avalanches.

\begin{figure}[!h]
\begin{center}
 \includegraphics[scale=0.35]{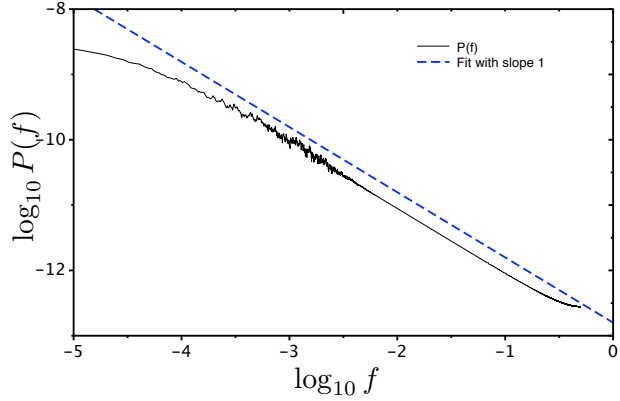}
\end{center}
\caption{\label{PSdissip}Power spectrum of the dissipated energy  in the dynamic model is an example of a $1/f$ noise. }
\end{figure}

Another important signature of a universality class is the average
 shape of avalanches with a given duration $T$. The avalanche shapes  have been studied in different physical systems exhibiting critical behavior \cite{Houston:2001uq,Colaiori:2004fk,Papanikolaou:2011qf}, in particular,  plastic avalanches are known to be  slightly asymmetric (skewed) \cite{Laurson2006}. The average shapes of avalanches  in our dynamic model are shown in Fig. \ref{average_shape}.  We observe that the avalanche shapes in the 2D model show much more variability with respect to durations than in the 1D model. For instance, at longer durations the avalanche in our model are getting more symmetric, which is also a prediction of the mean field model \cite{Papanikolaou:2011qf}, however, the agreement is not complete because some asymmetry always remains. In our calculations, we observed that the number of units close to the marginal stability limit is increasing as the avalanche evolves towards its maximum which may be the origin of the  tail responsible for the observed asymmetry.
 In the dynamic model we were not able to perform the scaling collapse because of the large computational cost associated with explicit resolution of the fast events. This will be done later on in the automaton model.

 \begin{figure}
 \begin{center}
 \includegraphics[scale=0.20]{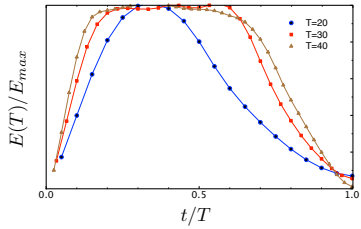}
 \end{center}
 \caption{\label{average_shape} Average avalanche shapes in dynamic model normalized by the magnitude and the duration.}
 \end{figure}

\subsection{Spatial correlations in the dynamic model}

The spatial counterpart of the observed time correlations is the  fractal structure of dislocational patterns. Fig. \ref{atomic_counturen} shows the energy density map $f(\theta, \xi)$ of the system after the fifth cycle. We observe that the energy distribution exhibits spatially separated local peaks associated with dislocation-rich regions which are rather stable in time. The distribution of these peaks is only partially guided by the quenched disorder and is largely a result of self organization.

\begin{figure}[!h]
\begin{center}
\includegraphics[scale=0.5]{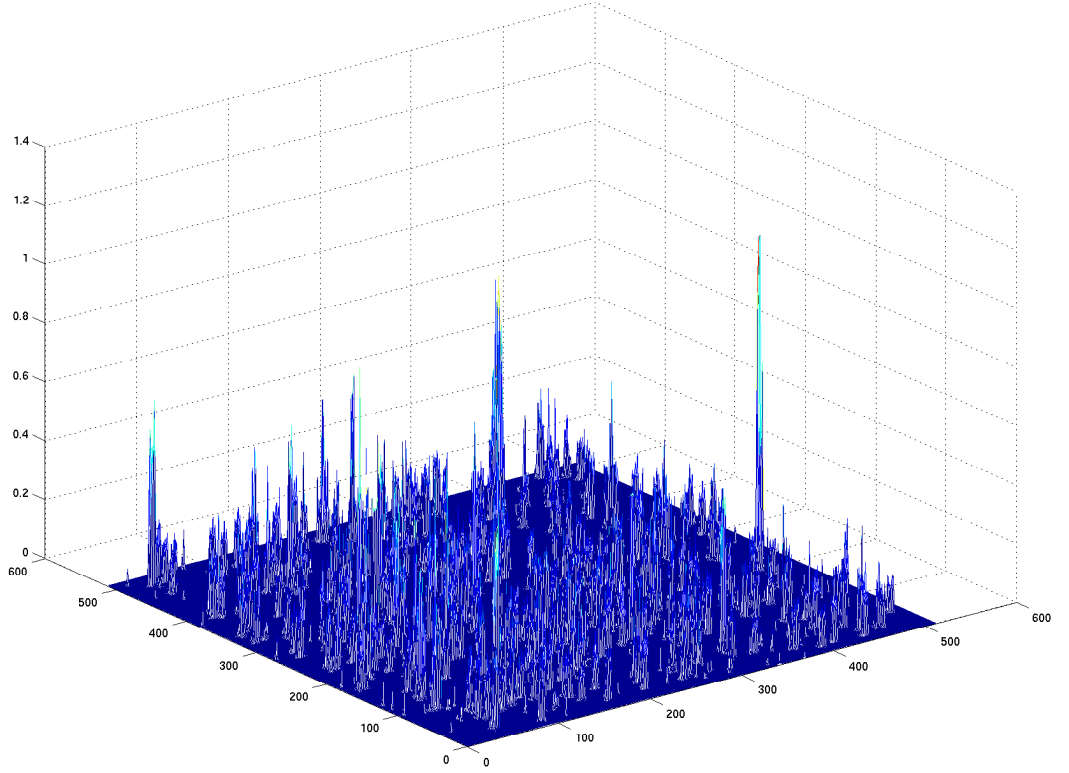}
\end{center}
\caption{ \label{atomic_counturen} Spatial energy density distribution $f(\theta, \xi)$ after the fifth cycle.}
\end{figure}
\begin{figure}[!h]
\begin{center}
\includegraphics[scale=0.35]{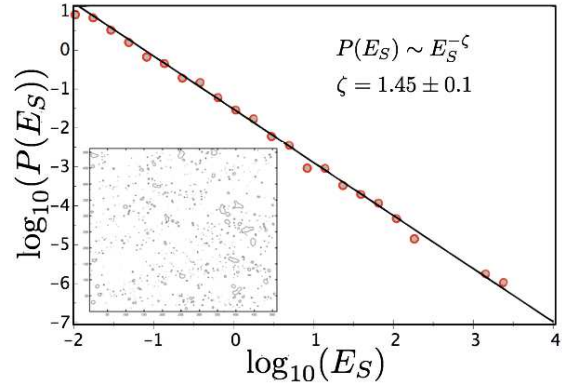}
\end{center}
\caption{\label{contourenergy}The  log-log plot of the probability density for the energies of dislocation rich regions. The insert shows the level set representation of the energy landscape shown in Fig. \ref{atomic_counturen}}
\end{figure}

In order to characterize quantitatively the spatial complexity of this distribution we constructed a level set representation for the energy density and identified the boundaries of the spatially separated local peaks by selecting an irrelevant threshold (see the insert in Fig. \ref{contourenergy}). We then computed the energy density associated with these regions  $E_\Omega=\frac{1}{\Omega}\sum_{\Omega} f(\theta,\xi)$ and obtained its probability distribution which is a power law $P(E_\Omega)\sim E_\Omega^{-\zeta}$  with  exponent $\zeta=1.45$ , see  Fig. \ref{contourenergy}. The obtained scaling is very robust spreading over five decades.
\begin{figure}[!h]
\begin{center}
\includegraphics[scale=0.4]{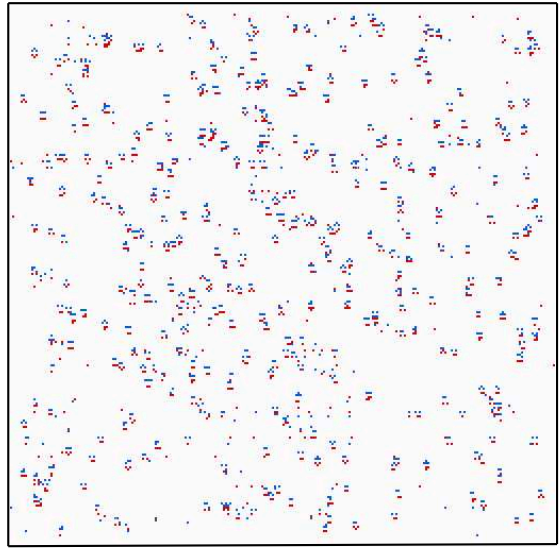}
\end{center}
\caption{\label{confgs111} A fragment of the dislocation  distribution in the dynamic model after the sixth cycle. Red and blue dots correspond to dislocations with positive and negative Burgers vectors, respectively. }
\end{figure}

Another way to quantify the fractal clustering is to compute the correlation function  of the actual dislocation distribution shown in Fig. \ref{confgs111}. We define
 \begin{equation}
\label{corr_int1}
C(r) = 2\frac{\mathscr N_p}{N(N-1)},
\end{equation}
where $N$ is the total number of dislocations and  $\mathscr N_p$ is the number of pairs of dislocations with separation less than $r$ \cite{Hentschel:1982cz}. If
$C(r)\sim r^D$ with non-integer  exponent $D$, this exponent is called  the fractal dimension. In particular, for  a randomly  distributed (not correlated) set of  points, the fractal dimension $D$ is equal to the dimension of space ($D=2$ in our case).

In our numerical simulations we observed  that during the first loading cycle $D\sim2.0$, which is expected given the random nature of the quenched disorder. With cycling, however,  the long range correlations developed progressively and in
 the shakedown regime we recorded $D\sim1.74$ independently of the initial disorder, see Fig. \ref{corr_int}.
 \begin{figure}[!h]
\begin{center}
\subfigure[]{\label{corr_int}\includegraphics[scale=0.29]{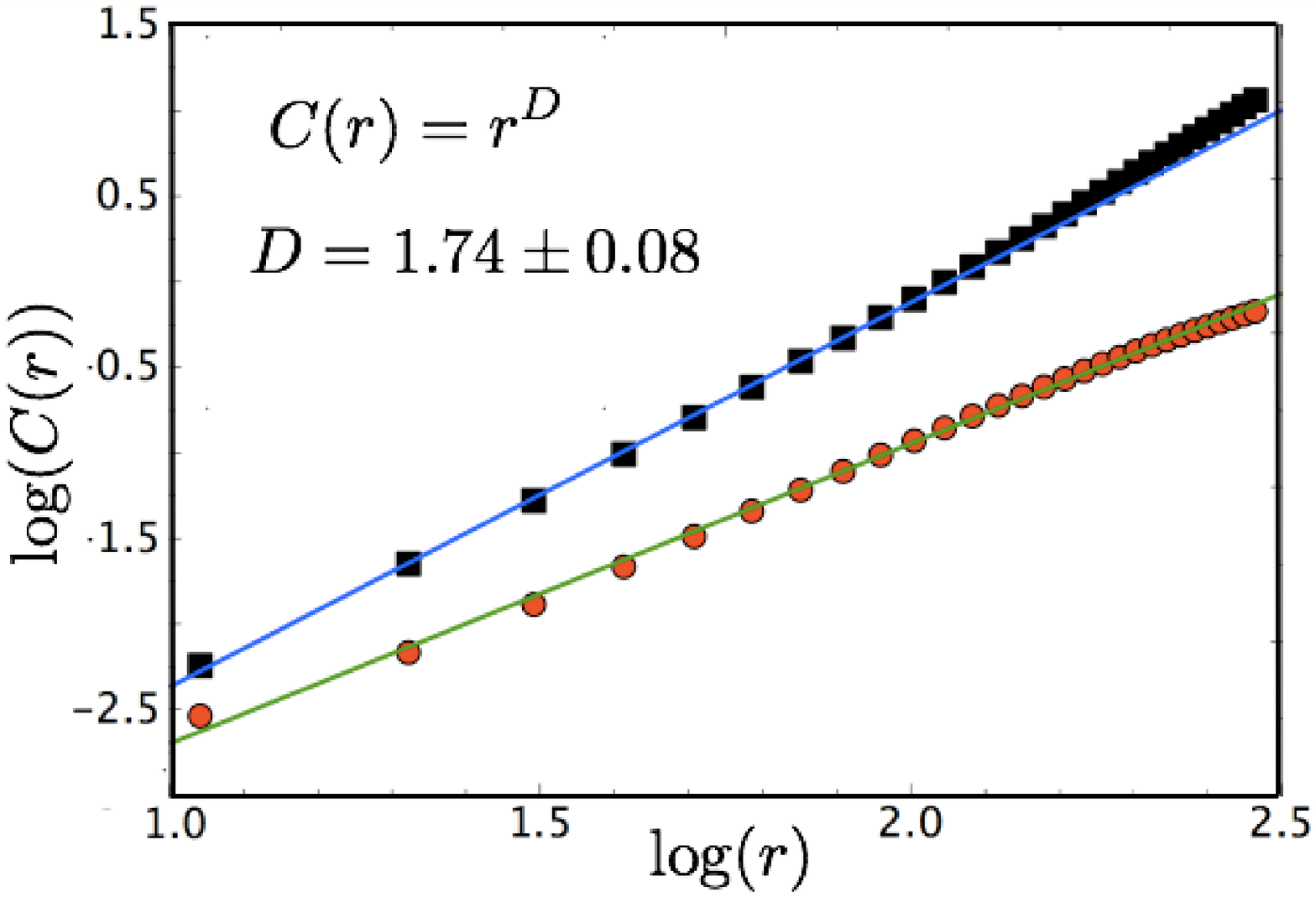}}
\subfigure[]{\label{box_count}\includegraphics[scale=0.29]{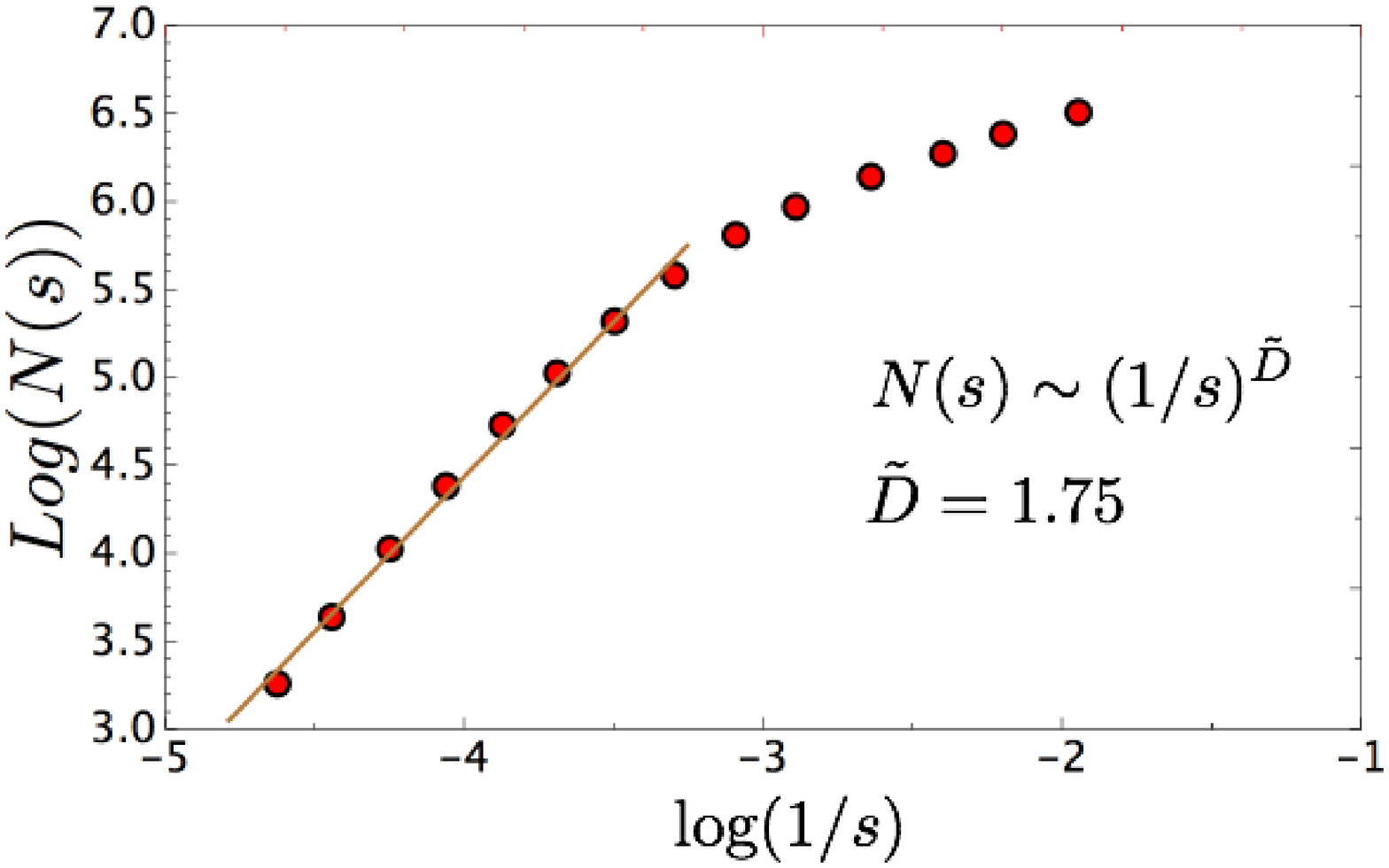}}
\end{center}
\caption{(a) Log-log plot of the correlation function C(r); (b) Log-log plot of the box-counting measure N(s).}
\end{figure}
Interestingly,  the dislocation patterns
 with $D \approx1.64-1.79$ have been observed experimentally in crystals with multiple slip systems;
 in simulations with a single slip system fractal patterning has been previously linked to the possibility of dislocation multiplication \cite{PhysRevLett.84.1487}
 which is operative in our model.

Another  well-known method of identifying the fractal structure of a set is computing its box-counting dimension $\tilde{D}$  \cite{Barnsley:1988fk} . It requires  a covering of a set by boxes of size $s$ and counting the asymptotics of the number of boxes $N(s)$ as $s\rightarrow0$. If  $N(s) =  s^{-\tilde{D}}$ then $\tilde{D}$ is the desired  fractional dimension. Application of this   method to the contour plot shown in the insert in Fig. \ref{contourenergy} gives again $\tilde{D} \approx1.75$,   see Fig.\ref{box_count}  which confirms the consistency of different measures of spatial correlations.

In summary, our 2D dynamical model shows a broad interval of self organized critical behavior characterized by a series of critical exponents revealing temporal and spatial correlations ($\eta=1$, $\alpha=1.6$, $D=1.75$ and $\zeta=1.45$). In the domain of parameters where we studied our system, these exponents do not depend on $K$ and $\xi^0$ and are not affected by the level of disorder $\sigma$. The system self-tunes to criticality by developing a highly correlated structure of persisting inhomogeneities which ensures effective energy dissipation.

\subsection{The automaton model}

Despite the conceptual transparency of the dynamical model, the mechanism of reaching the critical regime remains obscure in this largely computational formulation. In an attempt to obtain a mathematically more transparent version of the model we reduce in this subsection our dynamic equations to an automaton model and compare the results of conceptually similar numerical experiments.  As in 1D case, we utilize the fact that in the quasi-static limit $\nu\rightarrow 0$ viscous relaxation is instantaneous and the driven system can be viewed as almost always residing in the state of mechanical equilibrium.  This dramatically simplifies the tensorial problem because instead of dynamic equations of visco-elasticity one needs to deal only with equilibrium equations of elasto-statics $\partial \Phi / \partial u=0.$

In order to solve the equilibrium equations analytically we again replace the smooth periodic potential shown in  Fig. \ref{Fig_pw} by its piece-wise quadratic approximation  defined in each period $[(d-1/2)(2\xi^0),(d+1/2)(2\xi^0)]$ by
$$
g(\xi_{i,j} ) = \frac{1}{2}(\xi_{i,j}-(2\xi^0)d_{i,j})^2,
$$
Here $d=0, \pm1, \pm2, $ is the integer-valued spin variable describing a quantized slip. In the dynamic version of the model we we always implicitly assuming that $\xi^0=0.5$.

Suppose first that the lattice field $d_{i,j}$ is given. Then
 the  equilibrium equations can be written in the form
 \begin{equation}
\label{eq_mov_oper1}
\begin{aligned}
 &{}K[u_{i+1,j}+u_{i-1,j}-2u_{i,j}] +   [u_{i,j+1}+u_{i,j-1}-2u_{i,j}-2\xi^o(d_{i,j}-d_{i,j-1})]\\
 &{} -[h^1_{i,j}-h^1_{i-1,j}+h^2_{i,j}-h^2_{i,j-1}]=0.
 \end{aligned}
\end{equation}
This linear problem for the displacement field can be mapped into the Fourier space giving an algebraic equation
    \begin{equation}
\label{eq_mov_Fourier1}
[ K\hat s_x^{+}(\bold q )\hat s_x^{-}(\bold q )   +  \hat s_y^{+}(\bold q )\hat s_y^{-}(\bold q )] \hat u(\bold q ) - 2\xi^o \hat s_y^{-}(\bold q )\hat d(\bold q) -\hat H(\bold q)=0.
\end{equation}
where $ \bold q=(q_x,q_y)=(2\pi i/N, 2\pi j /N)$ and all other notations have been already introduced in \ref{subsec1}. By solving (\ref{eq_mov_Fourier1}) we obtain
   \begin{equation}
\hat u(\bold q) = \frac{2\xi^0  s^-_y(\bold q)\hat d(\bold q) + \hat H(\bold q)}{\hat \lambda(\bold q)},
  \end{equation}
where $\hat\lambda(\bold q) = 2K (\cos(q_x)-1) + s^-_y(\bold q)s^+_y(\bold q)$. The  Fourier image of shear strain can be then  calculated as $\hat \xi(\bold{q})=\hat s_y^{-}(\bold q )\hat u(\bold q) $.

It is now straightforward to reformulate the dynamical problem as an integer valued automaton. We first define the residual shear strain   due to quenched disorder  as ${\hat \xi_h(\bold q) =  \hat s_y^{+}(\bold q )(\hat H(\bold q)/\lambda(\bold q))}$. Then we
 observe that the variable
  $$\Delta \xi_{i,j}=\xi_{i,j}-(t+[\hat \xi_h(\bold q)]_{i,j}^{-1}),$$
representing shear strain associated with the slip must be confined between the limits
 $$
-(d_{i,j}-1/2)(2\xi^0)- [\hat \xi_h(\bold q)]_{i,j}^{-1}+t  <\Delta \xi_{i,j}(d) <(d_{i,j}+1/2)(2\xi^0) - [\hat \xi_h(\bold q)]_{i,j}^{-1}+t .
 $$
One can see that the stability of a particular slip configuration $d_{i,j}$ is controlled by two random thresholds which also evolve with the loading (with time).

When one of the thresholds is reached in one lattice point, the integer field  $d_{i,j}$ is updated $$d_{i,j}\rightarrow d_{i,j} + M_{i,j}(d),$$
where
$$
   M_{i,j}(d) =\left\{
  \begin{aligned}
&{}  +1,\hspace{1mm}\text{if} \hspace{1mm} \Delta \xi_{i,j}(d) >-(d_{i,j}-1/2)(2\xi^0) -[\hat \xi_h(\bold q)]_{i,j}^{-1}+t,\\
&{} -1, \hspace{1mm}\text{if}\hspace{1mm} \Delta \xi_{i,j}(d)  <(d_{i,j}+1/2)(2\xi^0) -[\hat \xi_h(\bold q)]_{i,j}^{-1}+t \\
&{}    \hspace{0.5cm} 0 \hspace{1mm}\text{otherwise.}
\end{aligned}
\right.
$$
\begin{figure}[h!]
\begin{center}
\subfigure[]{\label{kernelreal} \includegraphics[scale=0.4]{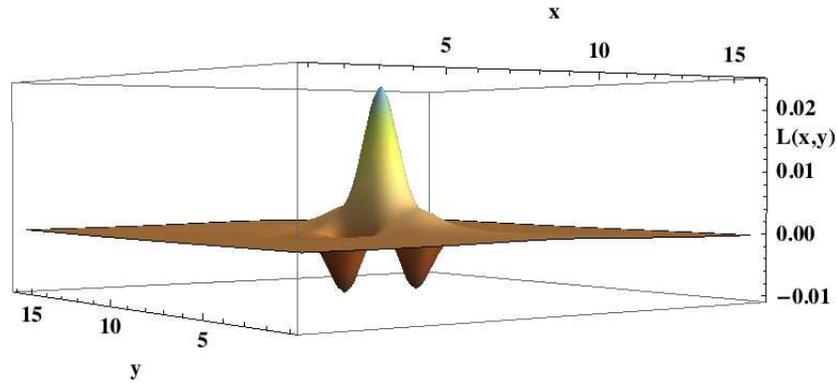}}
\subfigure[]{\label{kernelreal_discrete}\includegraphics[scale=0.45]{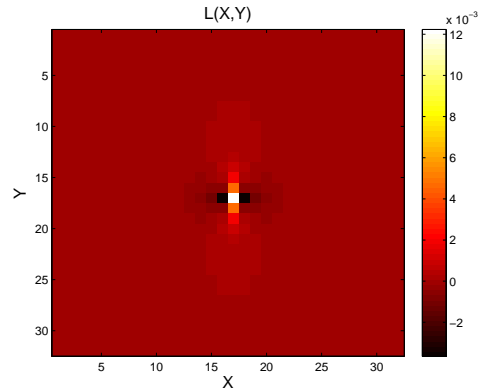}}
\end{center}
\caption{\label{kernelreal} (a) The real space representation and  (b) the discrete map of the kernel (toppling matrix) $  L(x,y)$.}
\end{figure}
After each increment of loading $\Delta t$ one has to check the stability of all units and continue the update until all units are stabilized.

The update of the "slope" field $\Delta \xi_{i,j}$ can be represented in Fourier space as
$$
\hat \Delta \xi(\bold q)  \rightarrow  \hat \Delta \xi(\bold q) + \hat L(\bold q) \hat M(\bold q).
$$
Here the kernel
 $$
\hat L(\bold q) =( 2\xi^0)\frac{\sin^2(q_y/2)}{ K\sin^2(q_x/2)+\sin^2(q_y/2)},\hspace{1mm}\text{for}\hspace{1mm} \bold q \neq 0,
$$
can be viewed as the analog of the toppling matrix in the sand-pile models. The inverse Fourier image of this kernel  in the real space $L(x,y)$ is highly anisotropic, long-range and conservative  (see Fig. \ref{kernelreal}). The conservativeness  is understood here in the sand pile terms and means that after each re-distribution of shear strain (discharge), the total deformation   $\sum_{i,j}\Delta \xi_{i,j}$, which is controlled by the boundary conditions, remains unchanged.

In Fig. \ref{kernelreal_discrete} we show the  discrete map of the real space of the Green's function $L(x,y)$. The discrete map indicates the distribution of the  strain for each discrete  unit (spring) in response to an elementary slip which can also be viewed as a localized force dipole. From this map one can see see that if an element reaches a threshold an strain increase takes place in this unit. Away from this element, in the $x$ direction, the strain is decreased in two neighboring elements but then it again increases reaching zero at infinity. In the $y$ direction, the elastic strain simply decays. Observe that while the structure of our  is  different from  the quadrupolar structure of the 2D elastic kernel in isotropic elasticity \cite{PhysRevE.84.016115}, the main anti-ferromagnetic structure in the direction of the slip is preserved. In the limit $K\rightarrow\infty$ the kernel $\hat L(\bold q)$ vanishes everywhere except $q_x=0$, where $\hat L(0,q_y)= 2\xi^0$ for $q_y\neq0$. One can see that the 2D model reduces in this limit to the 1D model with the kernel (\ref{eq:kernel1d}) (NN version with $\beta=0$).

\subsection{Numerical results for the automaton model}

As we have already noticed, the  automaton representation greatly reduces the complexity of numerical computations which, in particular,  allows one to deal with much bigger domains. In the automaton model fast depinning events are replaced by jumps and outside the jumps the system evolves through a succession of equilibrium states. One can expect that such simplification of the model preserves the basic correlations between the avalanches but may affect the avalanche shapes.  It may also affect the power spectrum particularly at high frequencies. In this section we begin a systematic comparison of the two models.

The stress-strain curves in the automaton model are shown in Fig. \ref{dc}.
\begin{figure}[!h]
\begin{center}
\subfigure[]{\label{hyst_ising_zoom}\includegraphics[scale=0.25]{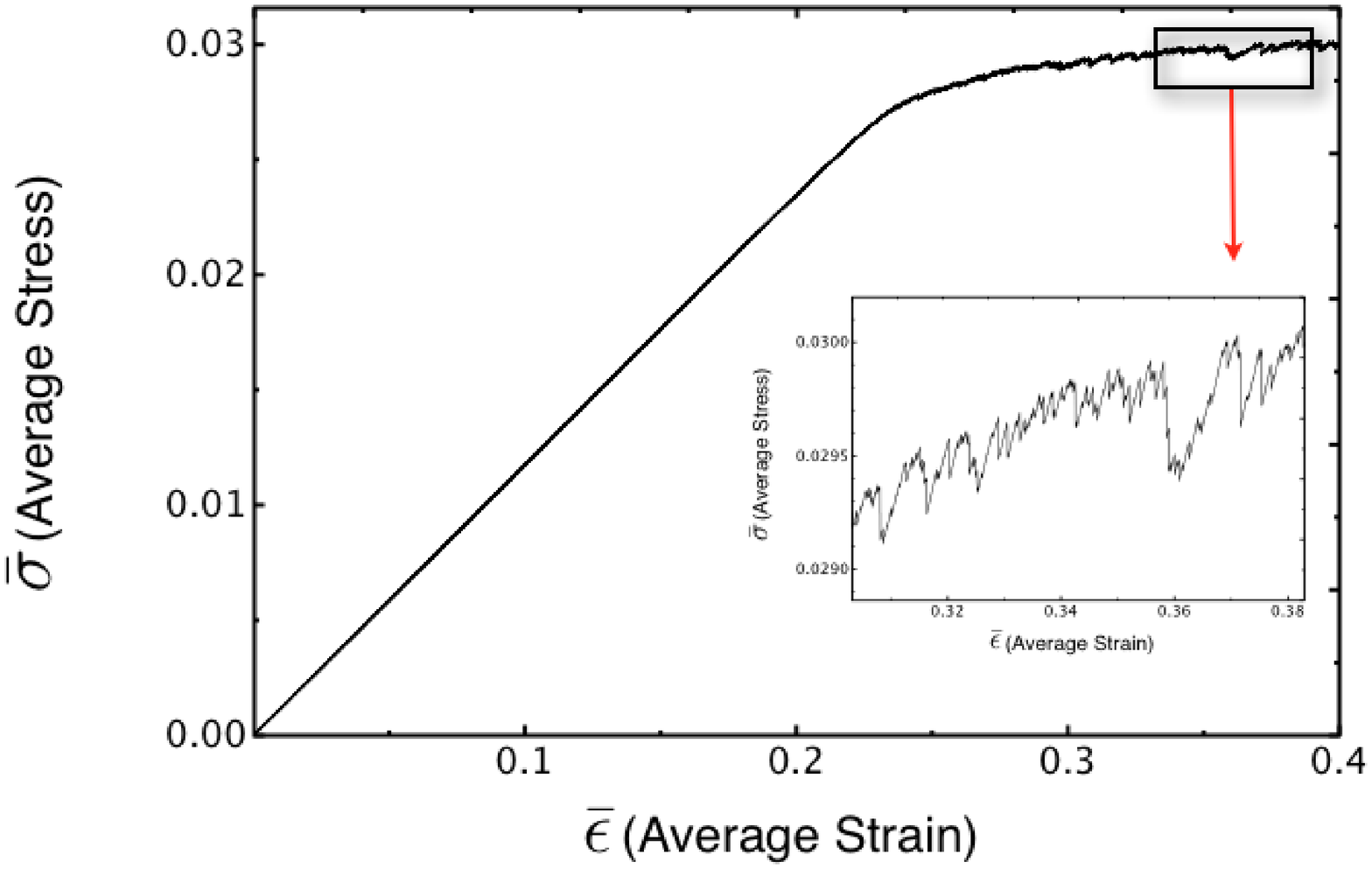}}
\subfigure[]{\label{hyst_ising_cycle}\includegraphics[scale=0.26]{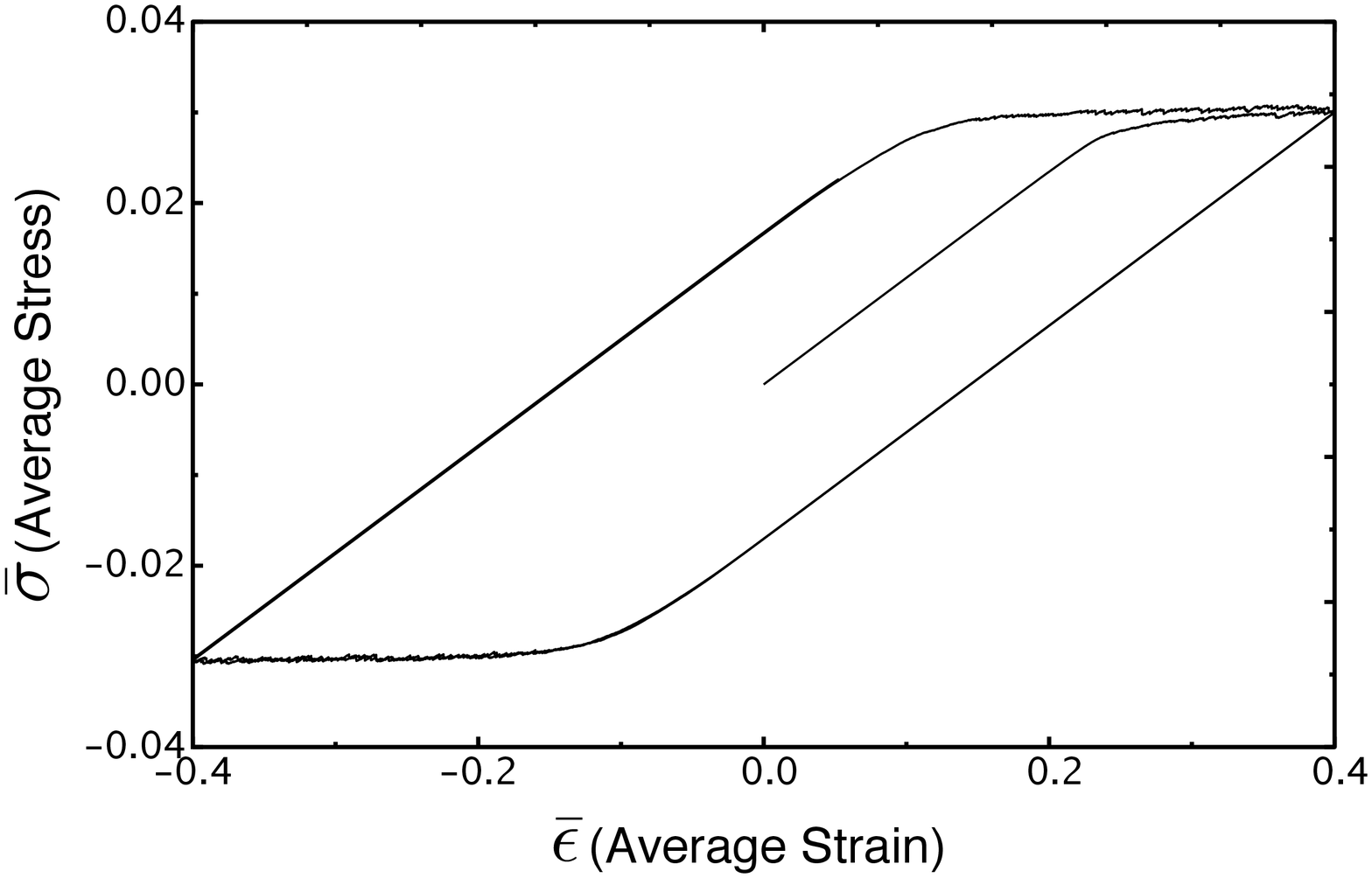}}
\end{center}
\caption{\label{dc}(a) Strain-stress curve for the automaton model in the first cycle; the insert shows  stress fluctuations.  (b) Strain-stress hysteresis  during the first 5 cycles.}
\end{figure}
\begin{figure}[!h]
\begin{center}
\subfigure{ \includegraphics[scale=0.4]{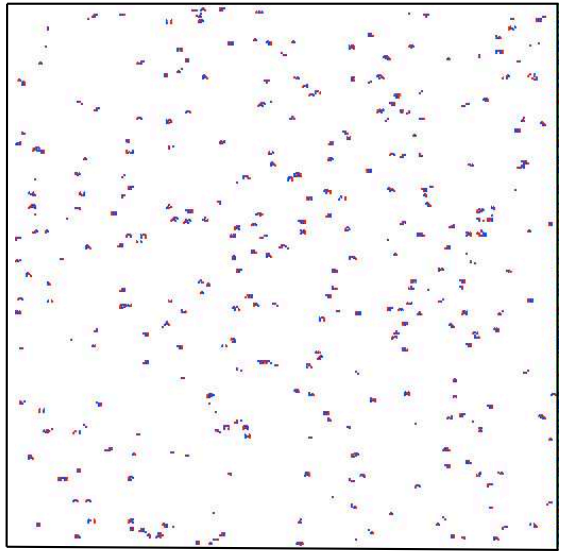}}
\end{center}
\caption{Distribution of dislocation cores obtained in the automaton model; red and blue colors correspond to dislocations of different signs.}
\label{dislo_distr1}
\end{figure}
We observe that the macroscopic response remains basically the same as in the dynamical model. The small difference in the configuration of the hysteresis loops  is due to the fact that in both models we compute  the  stress-strain relation only at some selected time steps. For instance, in the automaton model we  know exactly the silent intervals and therefore   can capture the strain/stress curve with higher resolution. Instead, in the dynamic model we randomly chose our test points which leads to stronger fluctuations.

A typical steady dislocation pattern in the automaton model is shown in Fig.\ref{dislo_distr1}. One can see that the dislocations again form clusters without an obvious scale. A computation of the correlation function for this set of points gives a slightly bigger fractal dimension $D\approx 1.9\pm0.1$ than in the dynamic model.

The study of the temporal correlations in the automaton model can be based on the time series representing the dissipated energy (\ref{eq:dissip1dg}). The typical signal in the shakedown state  is shown in Fig. \ref{energy_bursts}.
\begin{figure}[!h]
\begin{center}
\subfigure { \includegraphics[scale=0.25]{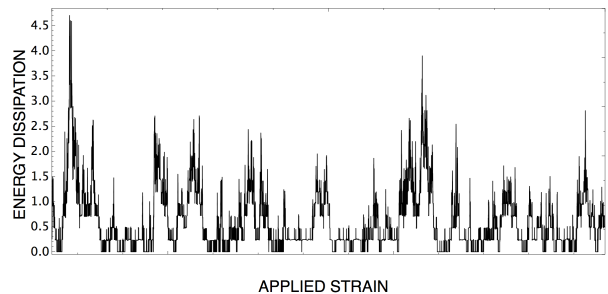}}
\end{center}
\caption{Energy dissipation in the automaton model. }
\label{energy_bursts}
\end{figure}
Since now we have access to data for different domain sizes we can perform the finite-size scaling (FSS) analysis of the avalanche distribution  \cite{E.Fisher:1971fk}.  In the critical state the power law probability distribution is expected to have a size dependent cut-off
\begin{equation}
P(E) = E^{-\epsilon}\varphi\left(\frac{E}{E_c}\right),
\end{equation}
where $\varphi$ is a universal shape function and  $E_c$ is the cut-off  energy with the scaling $E_c\sim N^\delta$.  According to the FSS hypothesis both, the exponents ($\delta$ and $\epsilon$), and the shape function $\varphi $, define the universality class.
\begin{figure}[h!]
\begin{center}
\subfigure[]{\label{q_simgaq}\includegraphics[scale=0.26]{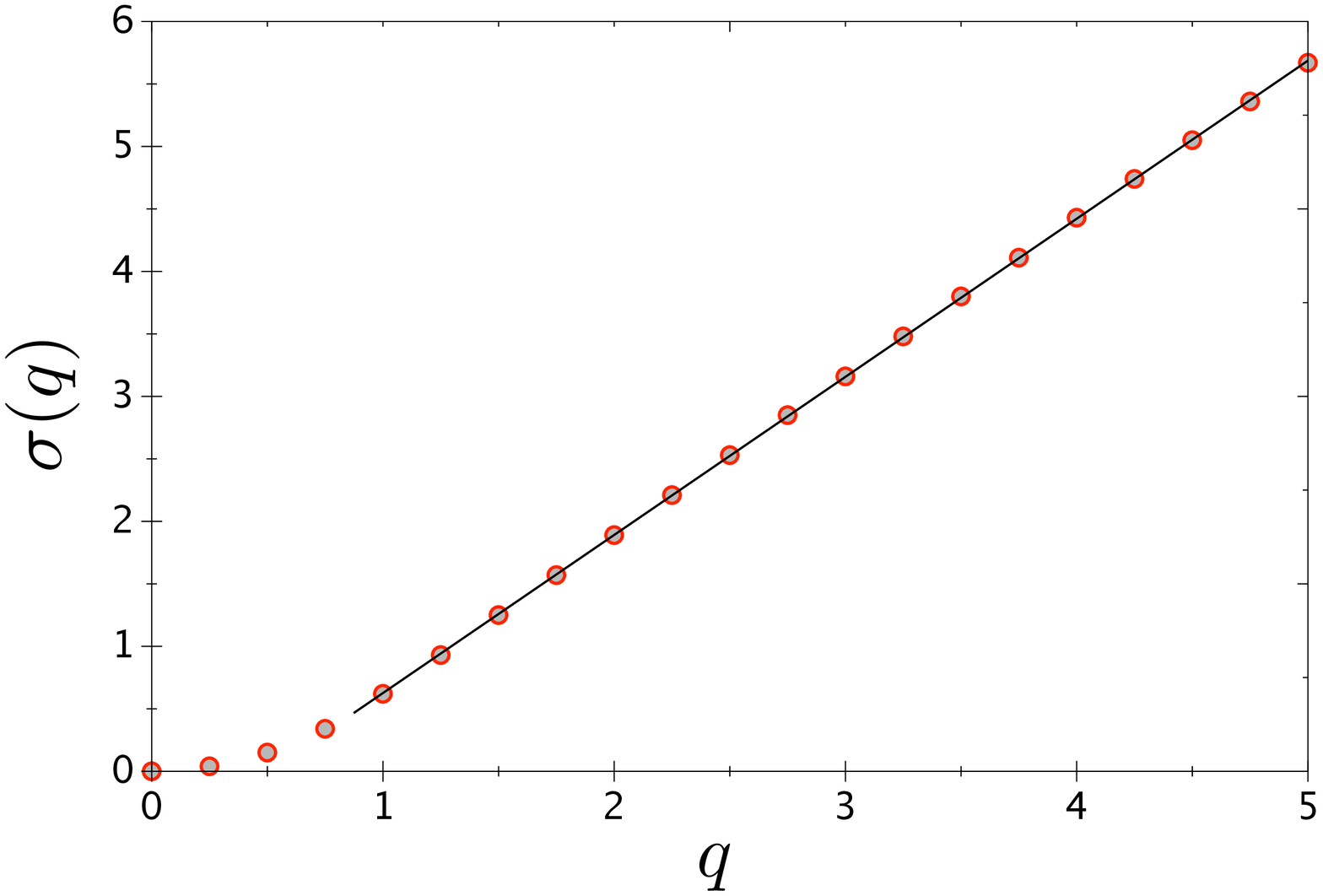}}
\subfigure[]{\label{data_collapse} \includegraphics[scale=0.26]{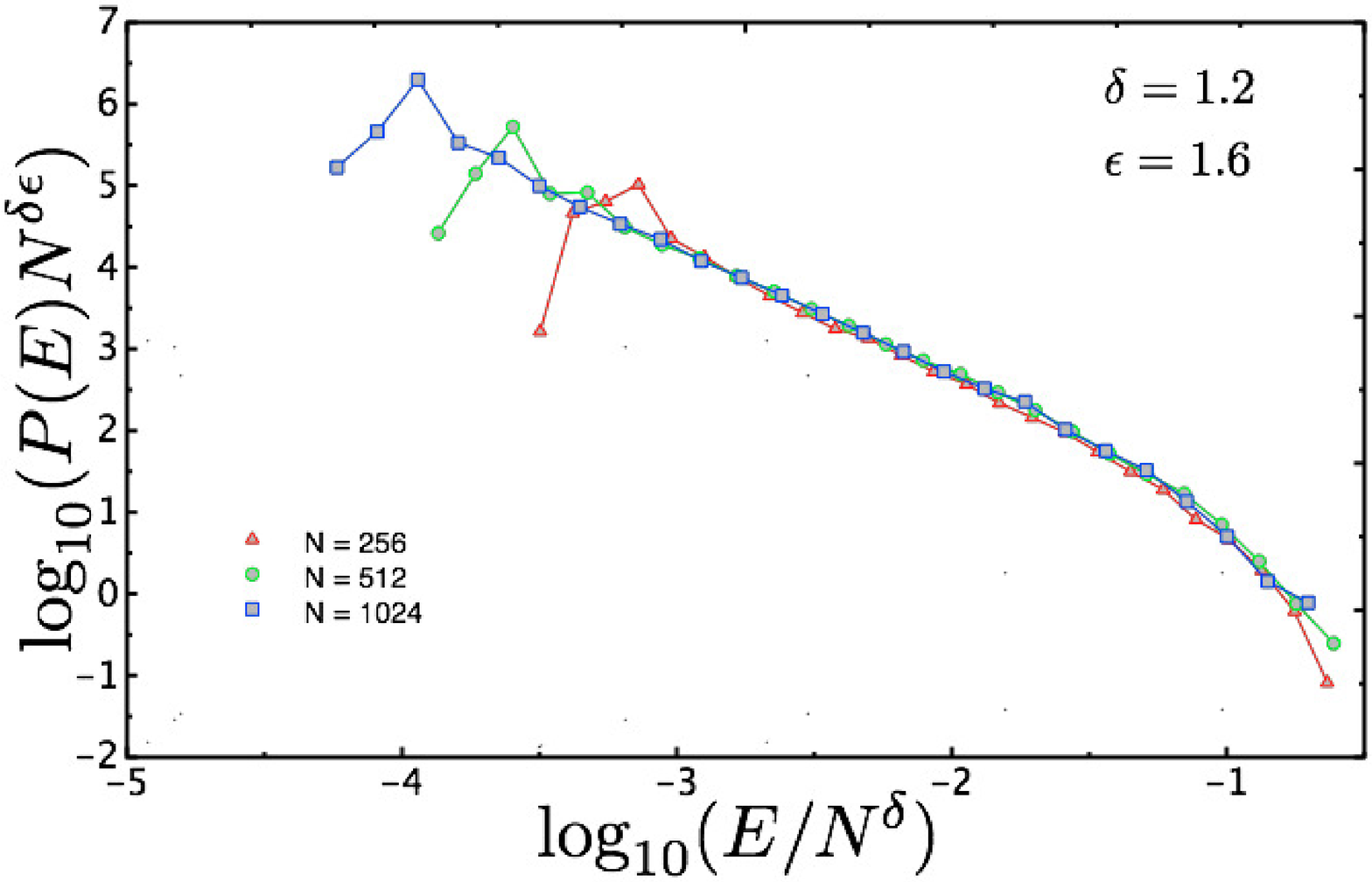}}
\end{center}
\caption{(a) Log-log plot of $\sigma(q)$ versus $q$. The slope in the linear regime is $1.2$. (b) Scaling collapse of the dissipated energy in the automaton model for different system sizes $N$. }
\end{figure}
If the FSS hypothesis is valid we must obtain for large $E$   that $<E^q> = \int E^q P(E)dE \sim N^{\sigma(q)}$,
where $\sigma(q) = \delta(q-\epsilon+1)$. The exponent $\sigma(q)$ can be computed from  the slope of the log-log plot of  $<E^q>$ vs. $N$ and the slope of $\sigma(q)$ with respect to q is the cut-off exponent $\delta$. As we see from Fig. \ref{q_simgaq}, the linear behavior starts from $q\approx0.6$ with a slope $\delta = 1.2\pm0.1$.  The exponent  $\delta \approx 1.2\pm0.1$  is close to the value $\delta \approx 1$ obtained for plastic strain increments in \cite{Miguel:2001dk, 1742-5468-2007-04-P04013}.
 Once $\delta$ is calculated, using $<E> \sim N ^{\sigma(1)}$, one finds the scaling relation $(2-\epsilon)\delta = \sigma(1)$ that leads to $\epsilon\approx1.6$.  The data collapse in the coordinates $ E/ N^{\delta}$ and $ P(E)N^{\delta\epsilon}$, presented in  Fig. \ref{data_collapse}, shows an excellent agreement with the criticality hypothesis.
 The knowledge of the exponent $\delta$ allows one to introduce a correlation length and study the size effect associated with collective interaction of dislocations \cite{JMPS.44.1996,Beato:2011dq}.

As we have already seen, one can also access temporal correlations through the analysis of the power spectrum for the dissipated energy.
\begin{figure}[!h]
\begin{center}
\label{PS} \includegraphics[scale=0.3]{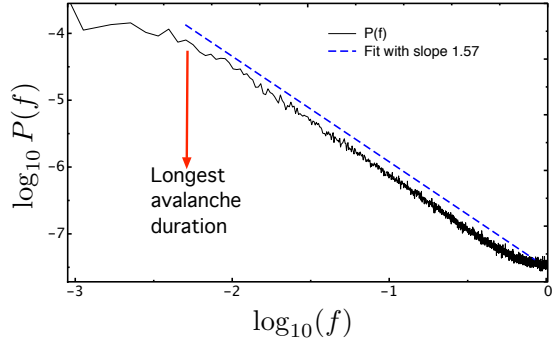}
\end{center}
\caption{ Power spectrum in the automaton model has a power law structure $P(f)\sim f^{-\eta}$ with $\eta=1.57$. }
\end{figure}
The power spectrum for our automaton is shown in Fig. \ref{PS}. Observe that the 2D power spectrum is different from the 1D spectrum shown in Fig. \ref{PS1D1}. It is, however,  similar to power spectrum one obtained in our dynamic model although with a larger exponent  $\eta \approx 1.57$ (which is, interestingly,  very close to the value $1.59\pm 0.05$ obtained for the 2D BTW sand pile \cite{Laurson2006}).  The scaling part of the power spectrum is limited on the low frequency side  by the duration of the longest avalanche. The absence of correlations in the low frequency region  shows that  the power spectrum  reflects correlations  within a single  avalanches while missing correlations between different  avalanches  (see also Ref. \cite{1742-5468-2005-11-L11001}).

We now move to the study of the avalanche shapes. Recall that in a critical state one can expect to see avalanches of all sizes and the average avalanche shapes must scale with their durations and magnitudes in a  universal way \cite{Kuntz:2000fk}. This scaling  hypothesis can be thoroughly tested in the automaton model.

The first step is to average  the avalanches of a given duration $T$ and plot the resulting function  $<E(t)_T>$ against the rescaled time $t/T$. In the critical state one should see the following scaling  \cite{PhysRevE.73.056104}
\begin{equation}
<E(t)_T> = T^{\gamma-1}e(t/T),
\label{scaling_E}
\end{equation}
where $e$ is a universal scaling function and $\gamma$ is the exponent in the power law relation linking the average avalanche size $<E(T)> = < \int_t^{t+T} E(t)dt>$ with its duration $T$
\begin{equation}
<E(T )>\sim T^{\gamma}.
\label{scaling_s}
\end{equation}
The relation (\ref{scaling_s}) is tested in Fig. \ref{STTT}, which  confirms the power law scaling and gives the value of exponent $\gamma \approx 1.3$.
\begin{figure}[!h]
\begin{center}
\includegraphics[scale=0.6]{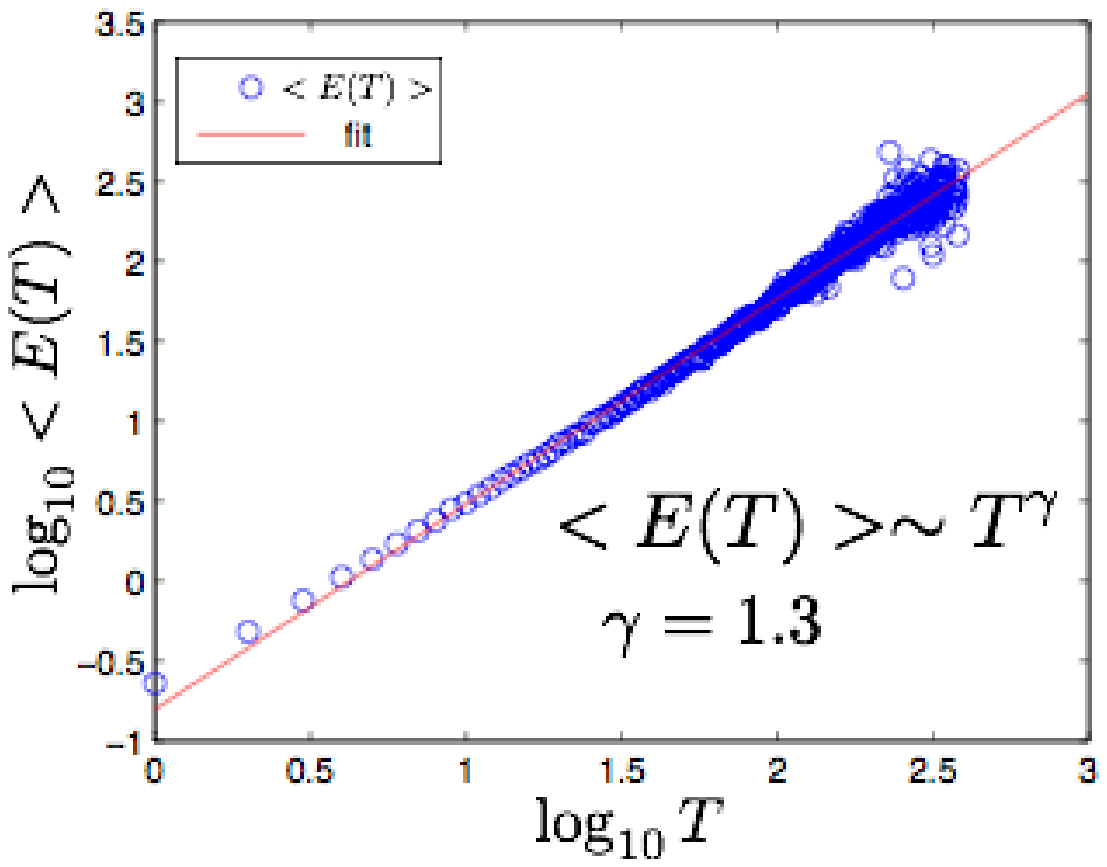}
\end{center}
\caption{ \label{STTT} Log-log plot of $<E(T)>$ vs T.}
\end{figure}
In Fig. \ref{q_simgaq2}, we present the average avalanche shapes $E(t)$ for different durations before data collapse. 
The data collapse based on the relation (\ref{scaling_s}) is shown in  Fig. \ref{q_simgaq3}. The slightly asymmetric close to parabolic shape of avalanches is in agreement with the results of DDD simulations \cite{Laurson2006}.  Notice that we presented only avalanches with durations in the limited range because below $T=20$ the average avalanche shapes deviate considerably from the 'parabolic' pattern,  while above $T=50$ the data become noisy and the scaling collapse is less satisfactory.
 \begin{figure}[!h]
\begin{center}
\subfigure[]{\label{q_simgaq2}\includegraphics[scale=0.33]{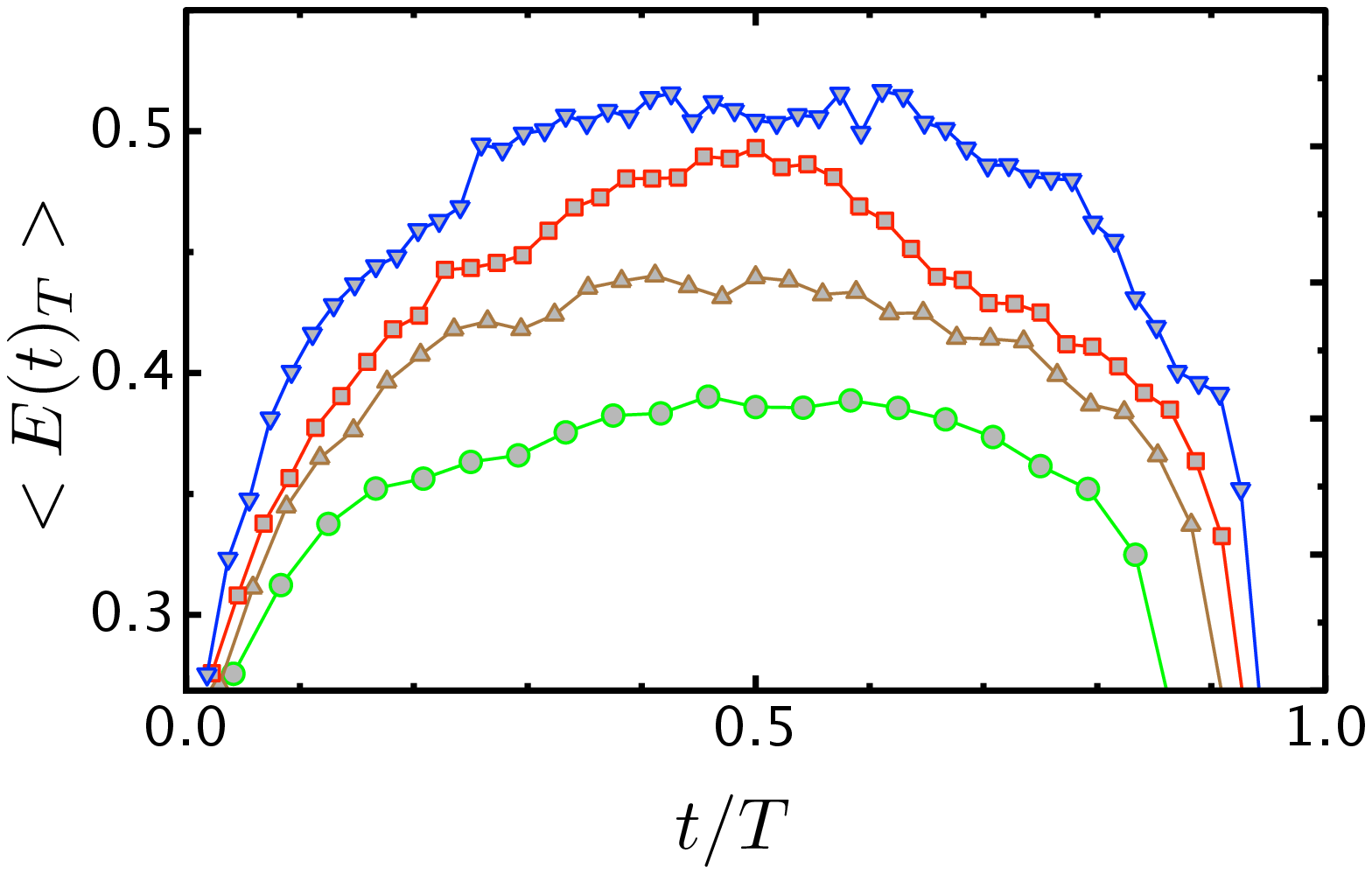}}
\subfigure[]{\label{q_simgaq3}\includegraphics[scale=0.35]{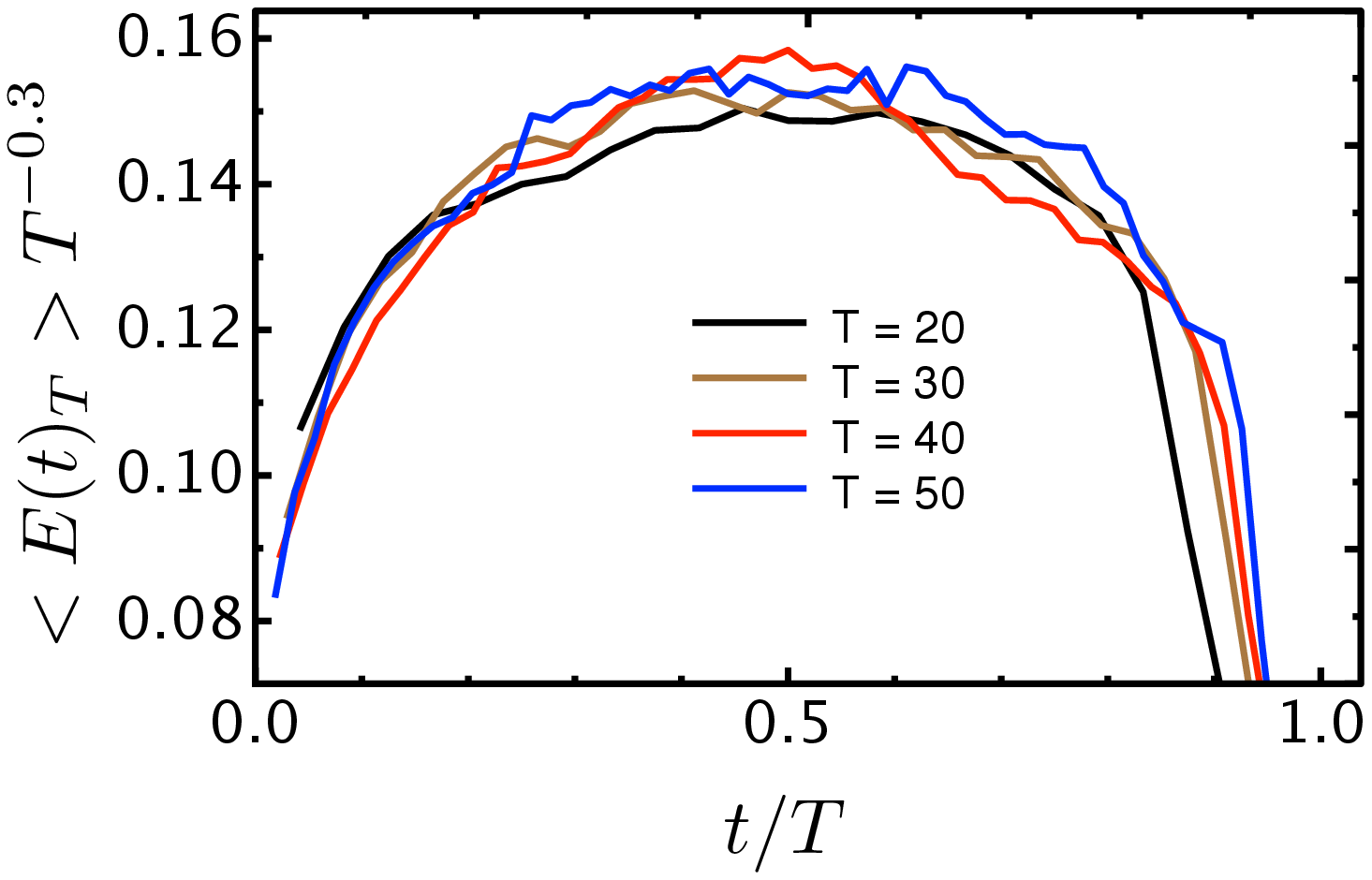}}
\end{center}
\caption{\label{q_simgaq}Average avalanche shapes of a given duration in the automaton model:  (a)  before and (b) after the scaling collapse.}
\end{figure}

Another way to reveal the scaling of the of avalanche shapes is to compare avalanches with a given total dissipation  $E_0$. In this case, one can anticipate the following scaling relation
\begin{equation}
\label{shapeE}
<E(t)_{E_0}> = E_0^{1-\alpha}\tilde{e}(t/E_0^\alpha).
\end{equation}
where $\tilde{e}$ is again a universal cut off function. The relation  (\ref{scaling_s}) suggests that the total dissipated energy scales with the avalanche duration as $T\sim E_0^{\alpha}$, where $\alpha=1/\gamma\approx0.76$.

In Figs. \ref{ESC1} and \ref{ESC2} we present  our data for the average avalanche shapes at a given total energy before and after the scaling collapse. One can see that the scaling hypothesis works rather well. Interestingly, our average shapes (at a given total dissipation $E_0$) are qualitatively similar to the average earthquake profiles at a given total moment \cite{Houston:2001uq}. The skewed shape is a manifestation of the fact that avalanches with a given total dissipation can have different durations. One can except that the maximum amplitude of the shorter avalanches is higher than of the longer avalanches with the same total energy. Then in the average shape the shorter avalanches will mostly contribute to the peak at small time while the long avalanches will be mostly responsible for the tail at large times which  leads to the observed asymmetry.
\begin{figure}[!h]
\begin{center}
\subfigure[]{\label{ESC1} \includegraphics[scale=0.35]{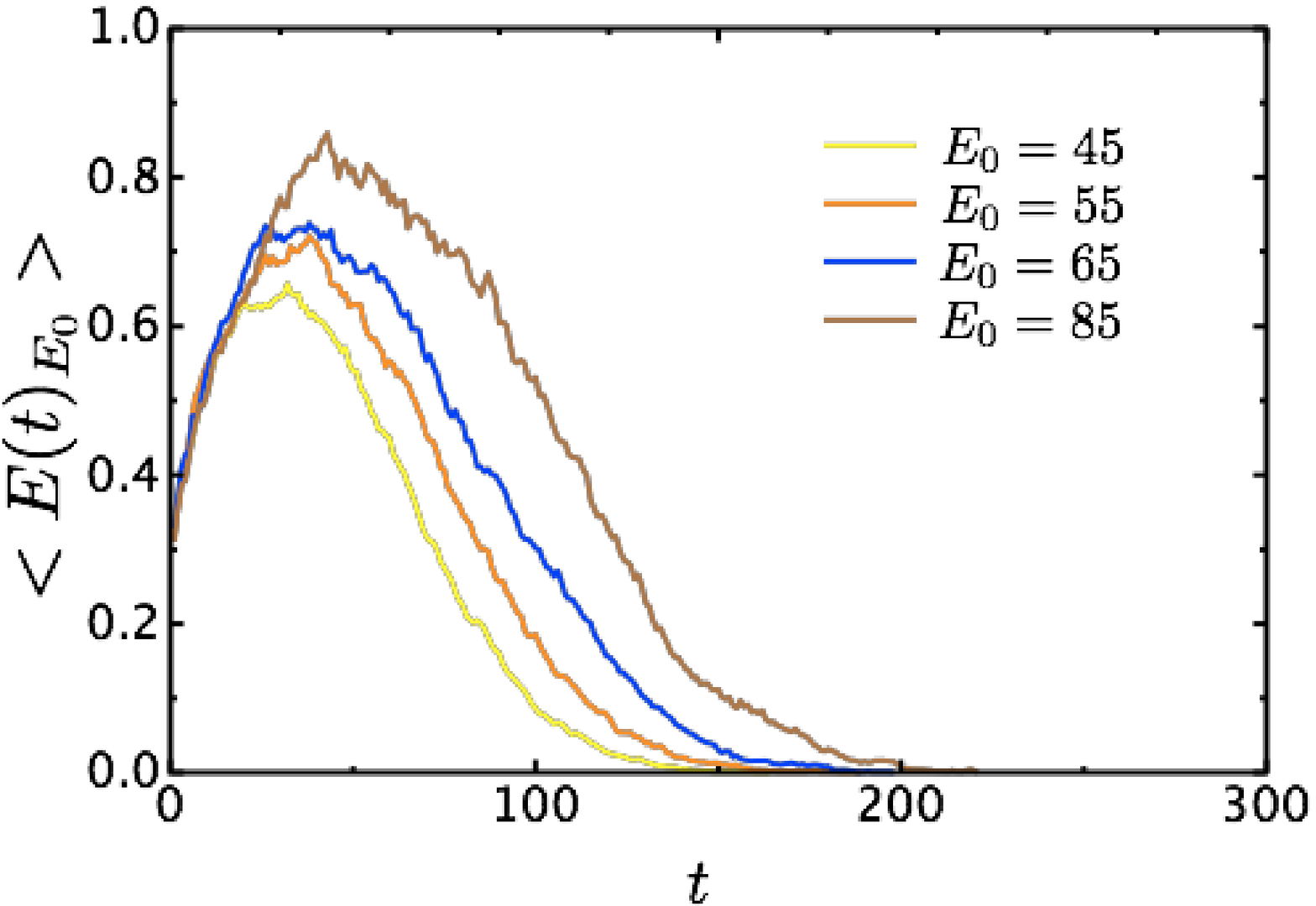}}
\subfigure[]{\label{ESC2} \includegraphics[scale=0.35]{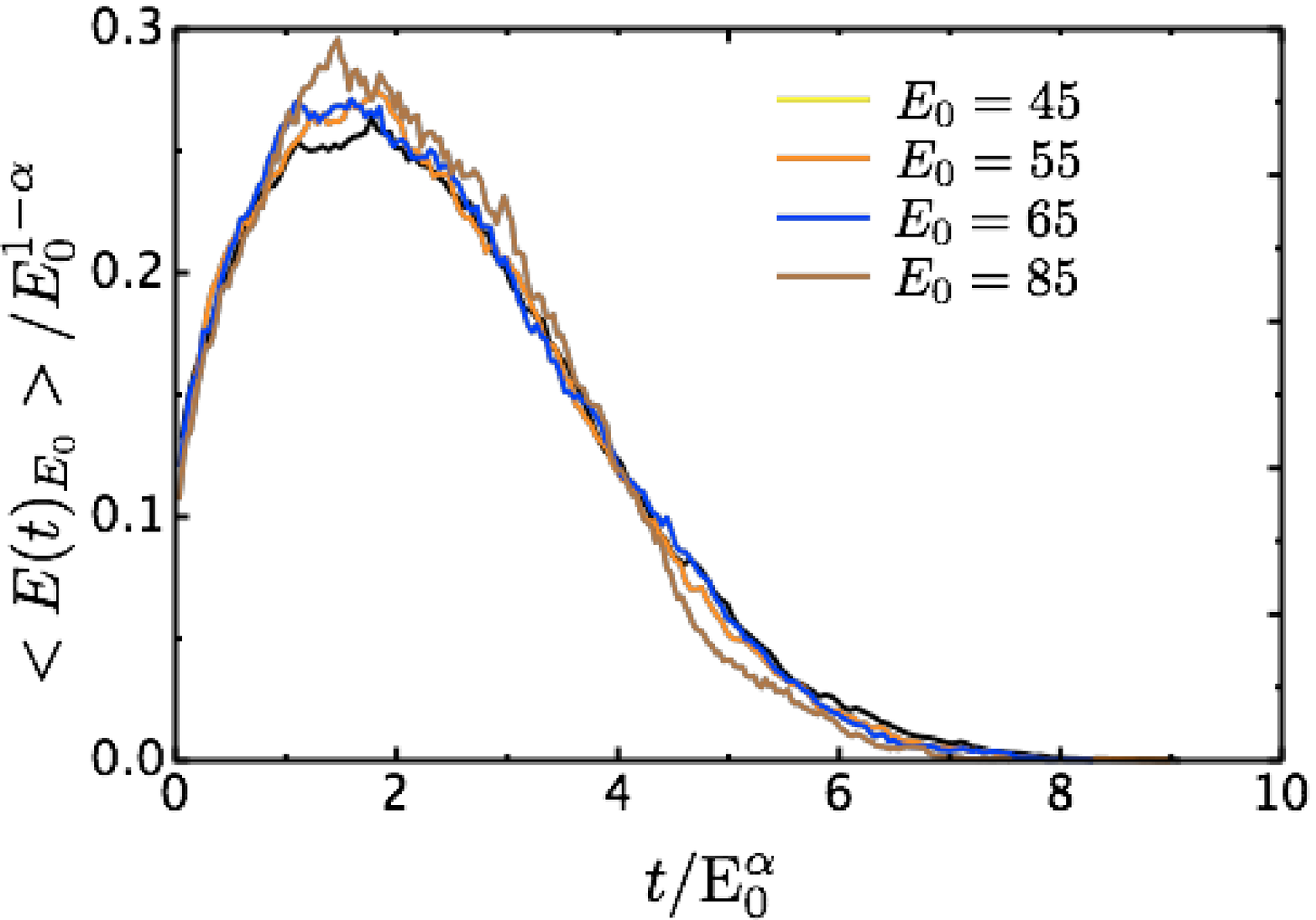}}
\end{center}
\caption{ (a) Average avalanche shapes at a given dissipated energy $E_0$ before (a) and after  (b) the scaling collapse.}
\end{figure}
We observe the same asymmetry in the mean field theory of depinning where the universal scaling functions $e$ and $\tilde e $ are known exactly for some classes of periodic potentials, for instance, in the  piece-wise quadratic case
  $ e(x)\sim x(1-x)$  and $\tilde e(x)\sim x e^{-x^2}$ \cite{Fisher:1998fk,PhysRevB.53.14872}, with the energy duration scaling  $T\sim E^{1/2}$ \cite{Fisher:1998fk}. A direct comparison shows that neither the exponent $\alpha$  nor  the average avalanche shapes match the predictions of the mean field theory. Similar analysis of the scaling functions for avalanches with given durations and magnitudes in the cases of earthquakes and magnets, which also questions the validity of the mean field theory, can be found in \cite{dahmen:2009}.

Finally, one can check the consistency of the obtained data by varifuing the universal relations between the exponents $\epsilon$, $\tau$ and $\gamma$.  Thus, given that the probability distributions $P(E)$ and $P(T)$ are related through
$P(T)dT\sim P(E)dE
$ we can use the  scaling relation $<E(T)>\sim T^\gamma$ to obtain
$
\gamma(1-\tau) = 1-\epsilon
$\cite{0305-4470-23-9-006}.
This scaling relation works fairly well for our model showing the consistency of  the obtained numerical data.

\section{Conclusions}

In this paper we developed a highly idealized model of crystal plasticity allowing one to study
cooperative effects in dislocation dynamics.  Quite remarkably, our toy  model, based on a simple system of coupled ODEs, has shown an excellent agreement with experimental observations of fluctuations during plastic yielding 
 and with previous much more elaborate numerical studies of avalanche statistics. The model confirmed robust temporal intermittency and spatial fractality of the shakedown regime and provided the first evidence towards the possibility of a flicker noise which remains to be checked in experiment.
 
To make the occurrence of scaling in our model
 more transparent we further simplified the formulation by reducing the continuous dynamics to an integer automaton. More precisely, we replaced the fast  dissipative stages of dynamics by jump discontinuities and then linearized the elastic behavior inside individual energy wells. The legitimacy of such model reduction has been proved rigorously only in 1D case where the fluctuations have a Gaussian structure \cite{A.:2011kx}, and the generalization to higher dimensions appear to be nontrivial in view of the observed long range correlations. The derived automaton model, however,  is much lighter numerically than the dynamic model which allowed us to perform the scaling collapse and to reveal the size effect (correlation length divergence) associated with  collective interaction of dislocations.

Based on the computed value of the  critical exponents one has a temptation to conjecture that both models, the one with continuous dynamics and smooth potential, and the one with discrete dynamics and piece-wise quadratic potential belong to the same universality class. This conclusion would be markedly different from the prediction of the mean field theory where smooth and cusped potentials lead to different universality classes  \cite{PhysRevB.48.7030,PhysRevB.31.1396}. In fact, the dynamic and the automaton models show different exponents in the power spectrum. Also the fractal dimensions of the dislocation pattern in the two models are slightly different. Given that the automaton model presents a drastic simplification of the discrete model it may be expected to capture only a limited interval of temporal and spatial scales. Different behavior of the two models outside this interval may affect the maximum likelihood fit of the exponent and this may be the source of the observed disagreement. For instance, the remaining disagreement may be due to the small but finite overlap of the avalanches in the dynamic model while they are perfectly separated in the automaton model.
 More detailed studies are needed to resolve this issue definitively.

An important remaining question is whether the toppling rules in the automaton are Abelian meaning that the outcome of the instability in multiple sites does not depend on the toppling order. We compared numerically the conventional updating strategies such as simultaneous updates of all unstable units,  updates involving only the unit with the smallest value of strain or updates of randomly chosen units. We found that the microscopic configuration shows some small dependence on the choice of the strategy while the macroscopic manifestations, including the shakedown hysteresis loop and the statistics of avalanches (critical exponents) remain unaffected. This means that the model exhibits a weak (statistical) form of Abelian symmetry. Another important property of the model is that the energy is lowered after each avalanche, which is explicit in the dynamic formulation and can be inferred for the automaton formulation.  These two features, additional internal symmetry and the dissipative structure, are crucial for the possibility of the mathematical study of the origin of scale free behavior.

To make our model more realistic and useful in mechanical engineering applications it is of interest to develop a tensorial formulation allowing one to simulate the critical behavior in the configurations with arbitrary geometry and with generic loadings. In particular, the complex inhomogeneous setting will give rise to interesting manifestations of the size effect associated with the growth of the correlation length from the microscopic values to the scale of the whole structure. On materials science side, it would be of interest to include the effects of  micro-cracking during cyclic loading associated with extreme dislocational pile-ups. Such coupling of plasticity and damage will open the way to the study of the crossover from   plastic to 'spinodal' criticality associated with the formation of the macroscopic crack and the ultimate failure of a solid.

More generally, the obtained reduction of a complex multi-particle dynamics to an integer automation suggests that similar development can be attempted for other distributed systems exhibiting criticality.  The revealed inherent temporal discretness can be a  tool bringing dramatic simplification and transparency to otherwise impenetrable problems ranging from turbulent flows to the rheology of cytoskeleton.

\section{Acknowledgements}
The authors would like to thank  E. Vives for helpful discussions.This work was supported by the French ANR-2008, grant EVOCRIT.

\bibliographystyle{model3-num-names}

\end{document}